%% file: paper_arxiv.tex
\documentclass[11pt]{article}
\usepackage[margin=1in]{geometry}

\usepackage{url}
\usepackage{algorithm}
\usepackage{algpseudocode}
\algnewcommand{\INPUT}{\Require}
\algnewcommand{\OUTPUT}{\Ensure}
\algnewcommand{\STATE}{\State}
\algnewcommand{\FOR}{\For}
\algnewcommand{\ENDFOR}{\EndFor}
\algnewcommand{\IF}{\If}
\algnewcommand{\ELSE}{\Else}
\algnewcommand{\ENDIF}{\EndIf}

\usepackage{amsmath,amssymb}

\usepackage[hidelinks]{hyperref}

\usepackage[round]{natbib}

\usepackage{amsthm}
\newtheorem{theorem}{Theorem}
\newtheorem{lemma}{Lemma}
\newtheorem{proposition}{Proposition}

\usepackage{authblk}
\usepackage{array}      
\usepackage{multirow}   
\usepackage{ragged2e}   
\usepackage{booktabs}   

\usepackage{booktabs,longtable,caption}
\usepackage{placeins}
\usepackage{xcolor}
\definecolor{darkgreen}{rgb}{0.0, 0.5, 0.0}

\input{definitions.tex}

\begin{document}

\title{Distribution-Free Selection of Low-Risk Oncology Patients for Survival Beyond a Time Horizon}

\author[1]{Matteo Sesia\thanks{CONTACT Matteo Sesia. Email: sesia@marshall.usc.edu}}
\author[2]{Vladimir Svetnik}

\affil[1]{Departments of Data Sciences and Operations, and of Computer Science\\
University of Southern California, California, USA}

\affil[2]{Biostatistics and Research Decision Science, Merck \& Co., Inc., Rahway, New Jersey, USA}

\date{}

\maketitle

\begin{abstract}
\input{abstract.tex}
\end{abstract}

\thispagestyle{empty}   

\input{body.tex}

\bibliographystyle{plainnat} 
\bibliography{bibliography}

\clearpage

\appendix

\renewcommand{\thesection}{A\arabic{section}}
\renewcommand{\theequation}{A\arabic{equation}}
\renewcommand{\thetheorem}{A\arabic{theorem}}
\renewcommand{\thecorollary}{A\arabic{corollary}}
\renewcommand{\theproposition}{A\arabic{proposition}}
\renewcommand{\thelemma}{A\arabic{lemma}}
\renewcommand{\thetable}{A\arabic{table}}
\renewcommand{\thefigure}{A\arabic{figure}}
\renewcommand{\thealgorithm}{A\arabic{algorithm}}
\setcounter{figure}{0}
\setcounter{table}{0}
\setcounter{theorem}{0}
\setcounter{algorithm}{0}

\input{appendix.tex}

\end{document}

%% file: definitions.tex
\usepackage{booktabs}
\usepackage{multirow,amssymb,amsmath,amsfonts,color}
\usepackage{mathtools}
\newtheorem{definition}{Definition}

\usepackage{enumitem} 

\renewcommand{\P}[1]{\mathbb{P}\left[#1\right]}
\newcommand{\EV}[1]{\mathbb{E}\left[#1\right]}
\newcommand{\EVc}[2]{\mathbb{E}\left[#1 \mid #2\right]}

\newcommand{\I}[1]{\mathbb{I}\left[#1\right]}
\newcommand{\Dtrain}{\mathcal D_{\mathrm{train}}}
\newcommand{\Dcal}{\mathcal D_{\mathrm{cal}}}
\newcommand{\Dtest}{\mathcal D_{\mathrm{test}}}
\newcommand{\Shat}{\widehat S}
\newcommand{\Ghat}{\widehat G}

\newcommand{\Lamhat}{\widehat\Lambda}

\newcommand\indep{\protect\mathpalette{\protect\independenT}{\perp}}\def\independenT#1#2{\mathrel{\rlap{$#1#2$}\mkern2mu{#1#2}}}

%% file: abstract.tex
We study how to select a subset of patients who are unlikely to experience an adverse event within a given time horizon, by calibrating a screening rule based on the output of any survival model. We consider two complementary frameworks. The first extends the classical idea of estimating the event rate among selected patients using a hold-out dataset, integrating it with the Learn-Then-Test method. This provides approximate high-probability guarantees that are comparable to those obtainable from simultaneous confidence bands while being often less conservative. The second takes a different perspective by reformulating the problem in terms of multiple hypotheses testing, enabling false discovery rate (FDR) control via the Benjamini–Hochberg procedure applied to selective conformal p-values. This provides approximate guarantees in expectation. We clarify the theoretical relationship between these approaches, explain how both can handle right-censored data and be made doubly robust via augmented inverse probability of censoring weighting, and compare them empirically using simulations and oncology data from the Flatiron Health Research Database. Our results reveal a practical trade-off: FDR-based screening is typically more powerful, while high-probability calibration is more conservative but offers stronger guarantees, especially when few patients are selected. We also provide practical guidance on implementation and tuning.


%% file: body.tex
\section{Introduction}

\subsection{Background and Motivation}

A common statistical question in medicine is how to identify a subset of patients who are sufficiently unlikely to experience an adverse event within a certain time horizon. This problem has many applications, including determining eligibility for less intensive management strategies such as active surveillance instead of surgery, and optimizing the allocation of limited clinical resources toward patients most likely to benefit.
This problem can be seen as one of large-scale testing: among many candidates, the goal is to select as many as possible among those who are truly low-risk, while controlling the rate of type-I errors—patients who are selected but nonetheless experience the event within the horizon of interest.
In practice, this task is further complicated by right-censoring, as follow-up may end before the event occurs for many patients, leaving their event status by the time horizon of interest unobserved.

As a concrete example arising from oncology care, consider a phase II de-escalation study that aims to enroll only patients who are sufficiently low risk, with a protocol target of no more than 10\% event probability within 3 months among participants. This means that approximately 90\% of enrolled patients should remain event-free for at least 3 months. 
Proper calibration of the screening rule is essential and requires balancing competing considerations: an overly liberal rule may admit too many high-risk patients, raising safety concerns or prompting early study termination, whereas an overly conservative rule may exclude many eligible patients, resulting in missed opportunities to improve patient well-being and make scientific findings.

In modern clinical practice, patient screening is typically data-driven and guided by survival probabilities estimated from predictive models that leverage rich demographic information and detailed clinical histories. 
Classical approaches include exponential or Weibull models and the semi-parametric Cox proportional hazards model \citep{cox1972regression}, while in large-scale, high-dimensional settings more flexible machine-learning methods \citep{bertsimas2020machine}—including random survival forests \citep{Ishwaran2008}, deep neural networks \citep{katzman2018deepsurv}, and gradient-boosting models \citep{barnwal2022survival}—often provide more accurate risk stratification.

Yet, all such models have limitations: classical survival models rely on assumptions that may be difficult to validate (e.g., parametric forms or proportional hazards), whereas machine-learning approaches typically prioritize point predictions over principled uncertainty estimation. As a result, it is natural to treat the predictive model itself as a ``black box'' and seek statistical guarantees directly at the level of the downstream screening rule, which selects patients whose estimated survival probabilities exceed a threshold that needs to be \emph{calibrated} after the model has been trained.

A standard way to \emph{evaluate} the event rate among patients selected by a
given screening rule is to apply it to an independent hold-out dataset. Under
right-censoring, this is commonly done using inverse probability of censoring
weighting (IPCW), which yields consistent estimates of time-dependent event rates
provided the censoring mechanism is correctly modelled
\citep{robins1993information}. An augmented variant (AIPCW)
recovers information from censored observations and is \emph{doubly robust} \citep{robins1992recovery},
remaining consistent if either the censoring model or the survival model is
correct. Either estimator can be paired with asymptotic or resampling-based
inference to construct confidence intervals for the event rate. When evaluating a
fixed screening rule, pointwise inference over the threshold is sufficient,
although more conservative simultaneous confidence bands can also be constructed.

The main challenge is that screening rules are typically tuned by evaluating
performance across multiple candidate thresholds and choosing the one that
maximizes yield subject to an acceptable event rate. Pointwise confidence
intervals do not account for this data-dependent selection, and so come with no
guarantee for the chosen rule, while uniform confidence bands, although valid,
are often overly conservative and lead to inefficient screening. These
limitations motivate the need for principled methods that can account for data-dependent
tuning while providing meaningful statistical guarantees, all while accommodating
right-censoring in the data.

\subsection{Contributions}

We make three main contributions.
First, we show how to calibrate screening rules by applying fixed-sequence
testing \citep{bauer1991multiple} via the Learn–Then–Test (LTT) framework
\citep{angelopoulos2025learn} to pointwise risk bounds, computed under
right-censoring from IPCW or AIPCW estimates of the event rate. This yields approximate
{\em high-probability} (HP) guarantees for data-driven screening rules that are comparable in
interpretation to those obtained from uniform confidence bounds, but often are
achieved less conservatively. To the best of our knowledge,
this represents the first application of the LTT framework with right-censored
time-to-event data.

Second, we study an alternative calibration strategy based on conformal prediction \citep{vovk2005algorithmic} and false discovery rate (FDR) control, building on \citet{sesiavsvetnik2025}. We recast screening in terms of testing random hypotheses on the event times of future patients, construct ``selective'' conformal p-values \citep{jin2023selection} that account for censoring via IPCW, and apply the Benjamini--Hochberg (BH) procedure \citep{benjamini1995controlling}. This typically selects more patients then the LTT approach while approximately controlling the event rate {\em in expectation} rather than with high probability. We further develop an augmented (AIPCW) variant of this method that enjoys doubly robust asymptotic FDR guarantees.

Third, we provide a direct theoretical and empirical comparison of HP and FDR-based calibration. Although these methods have different targets, they essentially address the same practical problem and can be meaningfully related. We clarify the interpretation of their inferences, identify regimes in which their behavior diverges, and characterize the resulting trade-offs. A key theoretical contribution is Theorem \ref{thm:fdr-risk}, which explicitly connects the event rate among selected patients to the FDR targeted by the conformal method, formalizing how the two concepts diverge only when selection events are rare. We also compare both approaches empirically against classical calibration strategies based on pointwise and uniform confidence intervals \citep{park2003estimating,zheng2008time,zheng2010semiparametric,fleming2013counting}.

\begin{table*}[!htb]
\small
\centering
\caption{Summary of screening calibration methods compared in this paper.
The first four methods, which are studied both theoretically and empirically, differ in their inferential objectives and in whether they operate from an inductive (cohort-independent) or transductive (cohort-adaptive) perspective. The last two
rows correspond to benchmarks included in the empirical comparison: a model-based baseline that treats predicted survival probabilities
as exact, and an oracle that assumes access to the true population-level survival
function.}
\label{tab:methods-summary}
\begin{tabular}{lllll}
\toprule
 {Method} & {Calibration} & {Perspective} & {Yield} & {Where} \\
\midrule
HP-Pointwise 
& Heuristic
& Inductive 
& High 
& Section~\ref{subsec:pointwise} \\

HP-Uniform 
& High-probability
& Inductive 
& Low 
& Section~\ref{subsec:uniform} \\

HP-LTT 
& High-probability
& Inductive 
& Medium 
& Section~\ref{subsec:ltt} \\

FDR-Conformal 
& In expectation 
& Transductive 
& High 
& Section~\ref{sec:fdr} \\

\midrule
Model-based baseline 
& None
& Inductive
& High 
& Section~\ref{sec:experiments} \\

Oracle benchmark 
& Exact 
& Inductive
& Optimal
& Section~\ref{sec:experiments} \\
\bottomrule
\end{tabular}
\end{table*}

\subsection{Related Works} \label{sec:related}

This paper integrates and builds on ideas from several areas of research, including multiple testing \citep{benjamini1995controlling,storey2004strong}, IPCW and AIPCW for right-censored time-to-event data \citep{robins1992recovery,robins1993information,vock2016adapting}, distribution-free risk control via the LTT framework \citep{angelopoulos2025learn}, and conformal prediction \citep{vovk2005algorithmic,angelopoulos2024theoretical}.

The LTT approach adopts an \emph{estimation-oriented} perspective, aiming to control the event rate among selected patients with \emph{high-probability}. This perspective is closely aligned with standard calibration practices that rely on IPCW estimates of event rates among selected patients as a function of the screening threshold parameter. Despite the natural connection, and to the best of our knowledge, the LTT framework has not been used before in the analysis of censored time-to-event data.

Conformal prediction offers a complementary approach from a \emph{predictive
inference} perspective, aiming to control in expectation a related notion of event rate among selected patients at the cohort rather than population level.
Conformal prediction has been the subject of substantial
recent research and has seen growing use in medical applications
\citep{vazquez2022conformal,olsson2022estimating}, including oncology
\citep{bergquist2022uncertainty} and survival analysis
\citep{candes2023conformalized,gui2024conformalized,qin2025conformal,davidov2025conformalized}.
In this paper, we apply conformal prediction to calibrate cohort screening rules,
using {\em selective conformal p-values} \citep{jin2023selection} combined with
BH \citep{benjamini1995controlling}.

While conformal prediction originally focused on fully observed and exchangeable data, 
these assumptions can be relaxed \citep{tibshirani2019conformal,barber2023conformal}.
\citet{jin2023model} extend the selective conformal p-values of
\citet{jin2023selection} to handle covariate shift through covariate-based likelihood ratio reweighting.
Censoring, however, induces a different form of distribution shift
\citep{candes2023conformalized} that requires time-dependent weights. 
We build on the IPCW-based
conformal approach of \citet{sesiavsvetnik2025}, augmenting it to
make it doubly robust. Doubly robust conformal methods are gaining attention
more broadly \citep{paul2025multiply}. In survival analysis they have been
proposed before by \citet{sesia2024doubly} and \citet{farina2025doubly}, where they are used instead for constructing prediction intervals for survival times.

The problem studied in this paper is also related to, but distinct from, time-dependent receiver operating characteristic (ROC) analysis, which focuses on assessing how well a prognostic score ranks patients according to their risk of experiencing an event by a given time horizon \citep{heagerty2000time,heagerty2005survival,blanche2013package,yuan2018threshold}. Discrimination metrics such as the area under the curve quantify ranking accuracy but do not ensure that any specific decision threshold yields an acceptable event rate among patients classified as low or high risk. In contrast, clinical screening decisions depend on \emph{calibration at the operating point}—that is, ensuring that the event rate among selected (or unselected) patients meets a predefined target. This paper addresses the calibration problem directly.

\section{Preliminaries}

\subsection{Notation, Data, and Main Assumptions} \label{sec:notation}

For each individual indexed by $i\in\mathbb N$, let $X_i\in\mathcal X\subseteq
\mathbb R^d$ denote a vector of covariates, $T_i>0$ the event time of interest,
and $C_i>0$ the censoring time. We assume individuals are independent 
and identically distributed (i.i.d.)\ samples from
an unknown population.
For each individual $i$ in the available data we observe only
\begin{align*}
  & O_i = (X_i, \tilde T_i, E_i),
  & \tilde T_i =  T_i \land C_i,
  && E_i = \I{T_i<C_i},
\end{align*}
where $a \land b := \min\{a,b\}$.
In words, \( \tilde{T}_i \) is the observed time and \( E_i \in \{0,1\} \) indicates whether the event occurred before censoring.
When the index is unimportant we write $(X,T,C)$ and $O=(X,\tilde{T},E)$ for a generic draw from the population.

Let $S$ and $G$ denote the population-level survival and censoring functions,
\[
  S(t\mid x) := \P{T>t\mid X=x} , \qquad G(t\mid x) := \P{C>t\mid X=x} ,
\]
and let $\Lambda^{C}(\cdot\mid x)$ denote the cumulative hazard of the censoring
time, defined by $d\Lambda^{C}(u\mid x) := - dG(u\mid x) / G(u^{-}\mid x)$, where
$G(u^{-}\mid x)$ denotes the left limit. If $C$ is continuously distributed,
this reduces to $\Lambda^{C}(t\mid x) = -\log G(t\mid x)$.

We assume access to estimates $\Shat$ and $\Ghat$ obtained from ``black box''
survival and censoring models. In practice the censoring model is trained
analogously to the survival model but with the event indicators reversed.
The survival model is used to compute covariate-based scores
$\hat z(x)\in [0,1]$ ranking patients by their estimated likelihood of surviving
beyond a fixed horizon $t_0>0$. A simple choice is
\begin{align} \label{eq:s-scores}
  \hat z(x) = \Shat(t_0\mid x) ,
\end{align}
although other options are possible; e.g., see Section~\ref{sec:fdr-bh}.

The following assumptions are used throughout the paper.

\noindent \textbf{Assumptions}
\begin{enumerate}
\item[(A1)] $T\indep C\mid X$, and the calibration data set $\Dcal=\{(X_i,\tilde T_i,E_i)\}_{i=1}^n$ are
i.i.d.\ draws from the population. Moreover, $T$ and $C$ are continuously distributed given $X$.
\item[(A2)] $\Shat$ and $\Ghat$ are fitted on a separate training data set $\Dtrain$, independent of $\Dcal$, and are treated as fixed. There exist $c,s\in(0,1]$ with
$G(u\mid x)\wedge\Ghat(u\mid x)\ge c$ and $S(u\mid x)\wedge\Shat(u\mid x)\ge s$
for all $x\in\mathcal X$ and $u\le t_0$.
\item[(A3)] The cohort $\Dtest=\{(X_{n+j},T_{n+j},C_{n+j})\}_{j=1}^m$ consists of
i.i.d.\ draws from the same population, independent of $\Dcal$ and $\Dtrain$.
\end{enumerate}
The conditional independence in Assumption (A1) says that censoring is uninformative given the covariates and
is standard in the analysis of time-to-event data.
Assumption (A2) implies that every individual has a nonzero probability of remaining uncensored up to $t_0$,
and that property is reflected by the fitted censoring model.

\subsection{Problem Statement and Outline of Approaches} \label{sec:problem-formulation}

We consider a monotone family of screening rules indexed by $\lambda\in[0,1] \cup \{+\infty\}$,
\begin{align} \label{eq:def-screening-rule}
  A_\lambda(x) := \I{\hat z(x) \geq \lambda} .
\end{align}
Throughout the paper, we treat $\hat z$ as fixed and focus on selecting a data-driven
threshold $\hat\lambda$ using the calibration data $\Dcal$.
Intuitively, the screening rule indexed by \( \hat{\lambda} \) should
select as many patients as possible while ensuring that the selected patients are truly
``low risk'' within the time horizon \( t_0 \), as made precise below.

Given any (possibly data-dependent) screening rule $A:\mathcal X\to\{0,1\}$, define
the \emph{selected-set risk} and the expected \emph{yield} at horizon $t_0$,
conditional on $\Dcal$, as
\begin{align} \label{eq:def-r}
  r(A;\Dcal) &:= \P{T\le t_0 \,\middle|\, A(X)=1,\ \Dcal} , \\
  \label{eq:def-mu}
  \mu(A;\Dcal) &:= \P{A(X)=1 \mid \Dcal} .
\end{align}
In words, $r(A;\Dcal)$ is the expected event rate before $t_0$ among selected patients,
and $\mu(A;\Dcal)$ is the expected proportion of cohort patients who are selected.
The explicit dependence of $r$ and $\mu$ on $\mathcal{D}_{\mathrm{cal}}$ emphasizes that both quantities are random if the selection rule depends on a data-driven threshold $\hat{\lambda}$ (e.g., $A = A_{\hat{\lambda}}$). We will sometimes suppress this dependence in notation when it is clear from context.

Given a target event rate $\alpha\in(0,1)$, the goal is to {\em approximately} solve
\begin{align} \label{equation:optim}
\begin{split}
\text{maximize}_{\lambda\in[0,1] \cup \{+\infty\}} \quad & \mu(A_\lambda;\Dcal) \\
\text{subject to} \quad & r(A_\lambda;\Dcal)\le\alpha .
\end{split}
\end{align}
Since $\mu(A_\lambda;\Dcal)$ is non-increasing in $\lambda$, this is equivalent
to minimizing $\lambda$ subject to the risk constraint.
However, the constraint still depends on an unknown quantity, $r(A_\lambda; \mathcal{D}_{\mathrm{cal}})$, hence why this problem can be solved only  approximately. 

In Section~\ref{sec:methodology-hp}, we present methods designed to approximately
solve~\eqref{equation:optim} from an \emph{inductive} perspective, allowing the
solution \( \hat{\lambda} \) to depend only on \( \mathcal D_{\mathrm{cal}} \) and enforcing the risk constraint with high
probability.
In Section~\ref{sec:fdr}, we introduce an alternative approach from a
\emph{transductive} (or \emph{cohort-adaptive}) perspective, which additionally allows
\( \hat{\lambda} \) to depend on the observed covariate data from the specific
cohort to which the screening rule is applied, and seeks to satisfy the risk
constraint in expectation.
Both families of methods rely on IPCW or its augmented variant to handle right-censoring in \( \mathcal D_{\mathrm{cal}} \).

\section{Calibration via High-Probability Risk Control} \label{sec:methodology-hp}

The methods in this section approximately solve~\eqref{equation:optim} while allowing the solution $\hat{\lambda}$ to depend only on the calibration data $\mathcal D_{\mathrm{cal}}$, without looking at the covariates in $\Dtest$, and aiming to satisfy the risk constraint with high probability; i.e.,
\begin{align} \label{eq:hp}
  \mathbb{P}[r(A_{\hat{\lambda}};\Dcal) > \alpha] < \delta,
\end{align}
where the probability is taken over $\Dcal$ and $\delta$ is a significance threshold specified by the practitioner.
These methods require confidence bounds for $r(A_{\lambda}; \mathcal{D}_{\mathrm{cal}})$, as a function of $\lambda \in [0,1]$.
We describe three such approaches:
(i) Pointwise calibration, which constructs upper confidence bounds at fixed values of $\lambda$ and then greedily selects the most liberal threshold meeting the constraint;
(ii) Uniform calibration, which employs simultaneous confidence bounds across a grid of $\lambda$ values to protect against data-dependent threshold tuning; and (iii) Learn–then–Test (LTT) or fixed-sequence calibration, which performs pointwise estimation sequentially over ordered thresholds.

\subsection{Pointwise Risk Estimation} \label{subsec:pointwise-estimation}
 
For any {\em fixed} $\lambda$ such that $\mu(\lambda)>0$, the selected-set risk~\eqref{eq:def-r} does not depend on $\Dcal$ and may be written as
\begin{align} \label{eq:r-theta-mu}
  & r(\lambda) = \frac{\theta(\lambda)}{\mu(\lambda)},
  & \theta(\lambda) = \EV{\I{T\le t_0}A_\lambda(X)},
  && \mu(\lambda) = \EV{A_\lambda(X)}.
\end{align}
In the special case of $\mu(\lambda)=0$, the risk becomes meaningless and we may set it to zero by convention.
Since $\mu(\lambda)$ involves only the covariates, it is simply estimated by
$\hat\mu(\lambda)=n^{-1}\sum_{i=1}^n A_\lambda(X_i)$. The main difficulty is estimating the numerator $\theta(\lambda)$,
which depends on the event times and thus requires correcting for censoring.

\subsubsection{Inverse Probability of Censoring Weighting (IPCW)}
The standard \emph{event-time} IPCW estimator of $\theta(\lambda)$ upweights observed events before $t_0$ by the inverse probability of remaining uncensored at the event time:
\begin{align} \label{eq:ipcw-et}
  & \hat r_{\mathrm{et}}(\lambda)
  = \frac{\hat{\theta}_{\mathrm{et}}(\lambda)}{\hat{\mu}(\lambda)},
  & \hat{\theta}_{\mathrm{et}}(\lambda)
  = \frac{1}{n} \sum_{i=1}^n A_\lambda(X_i)\, \frac{\I{\tilde T_i \le t_0,\, E_i=1}}{\hat G(\tilde T_i\mid X_i)}.
\end{align}
This reweighting approximates what would have been observed in a fully followed data set, by relying on Assumptions (A1)--(A2) as well as on consistent estimation of $G$.
A related approach is \emph{fixed-time} IPCW, which instead upweights individuals known to be event-free at $t_0$ by $1/\hat G(t_0\mid X_i)$, as described in Supplement Section~\ref*{app:high-probability}.
We adopt the event-time weighted (IPCW--ET) form throughout, as it aligns most closely with the FDR method of Section~\ref{sec:fdr}.
Moreover, we follow the standard practice of applying mild winsorization \citep{cole2008constructing} to the weights $\hat w_i = 1 / \hat G(\tilde T_i\mid X_i)$. 

A simple calculation shows that the bias of the IPCW estimator is
\begin{equation} \label{eq:ipcw-bias}
  \mathbb E\big[\hat{\theta}_{\mathrm{et}}(\lambda)\big] - \theta(\lambda)
  =
  \mathbb E\!\left[
    A_\lambda(X)\,\mathbb I\{T\le t_0\}
    \left(\frac{G(T\mid X)}{\hat G(T\mid X)} - 1\right)
  \right],
\end{equation}
which is \emph{first order} in the relative error of $\hat G$ and does not vanish as $n$ grows. 
Note that, since the variable inside the expectation is nonzero only for selected patients who experience the event before $t_0$, the bias 
is only affected by the accuracy of $\hat G$ in the corresponding region of the covariate space, not on average over the entire population.

\subsubsection{Augmented IPCW (AIPCW)} \label{sec:conformal-aipcw}

The first-order bias of the IPCW estimator can be reduced to second order by
augmenting~\eqref{eq:ipcw-et} with a term that recovers information from censored
observations \citep{robins1992recovery}. 
This gives the AIPCW estimator
\begin{align} \label{eq:aipcw}
  & \hat r_{\mathrm{dr}}(\lambda)
  = \frac{\hat{\theta}_{\mathrm{dr}}(\lambda)}{\hat{\mu}(\lambda)},
  & \hat \theta_{\mathrm{dr}}(\lambda)
  = \frac{1}{n} \sum_{i} A_\lambda(X_i)\, \hat{\Psi}^{\mathrm{dr}}(O_i, t_0),
\end{align}
which uses the pseudo-outcomes
\begin{align} \label{eq:aipcw-pseudo}
\begin{split}
  \hat{\Psi}^{\mathrm{dr}}(O_i, t_0)
  & := \frac{E_i\,\mathbb I(\tilde T_i \le t_0)}{\hat G(\tilde T_i\mid X_i)}
  + \frac{(1-E_i)\,\mathbb I(\tilde T_i \le t_0)\,\hat Q(\tilde T_i, t_0 \mid X_i)}{\hat G(\tilde T_i \mid X_i)} \\
  & \qquad - \int_{0}^{\tilde T_i \wedge t_0} \frac{\hat Q(u, t_0 \mid X_i)}{\hat G(u \mid X_i)}\, d\hat\Lambda^{C}(u \mid X_i) ,
\end{split}
\end{align}
for each individual $i \in \mathbb{N}$, where
\begin{equation} \label{eq:qhat}
  \hat Q(u, t_0 \mid x) := \frac{\hat S(u^{-} \mid x) - \hat S(t_0 \mid x)}{\hat S(u^{-} \mid x)}
  \;\approx\; \mathbb P\big(T \le t_0 \mid T \geq u,\, X = x\big),
  \qquad u \le t_0 ,
\end{equation}
and $\Lamhat^{C}$ denotes the cumulative hazard implied by $\Ghat$. 
If $\Ghat(\cdot\mid x)$ is a step function on a grid
$0 = u_0 < \cdots < u_K = t_0$, $\Lamhat^{C}(\cdot\mid x)$
is the discrete measure placing mass
$1 - \Ghat(u_k\mid x)/\Ghat(u_{k-1}\mid x)$ at each $u_k$, so that the integral
in~\eqref{eq:aipcw-pseudo} is a finite sum. 
We refer to Supplement Section~\ref*{app:aipcw-implementation} for  
computational details on how to efficiently evaluate these pseudo-outcomes in practice.

The second and third terms in~\eqref{eq:aipcw-pseudo} depend on the subjects
censored before $t_0$, which are disregarded by IPCW~\eqref{eq:ipcw-et}. 
The second term credits each censored
subject with the estimated conditional probability that the event would still
have occurred by $t_0$, while the third subtracts the estimated expected value of
this correction. Together they form a stochastic integral with respect to the
censoring martingale \citep{fleming2013counting}. Two special cases are worth
noting: in the absence of censoring $\hat{\Psi}^{\mathrm{dr}}(O_i, t_0)$ reduces
to $\mathbb I(T_i \le t_0)$; and setting $\hat Q \equiv 0$ recovers IPCW.

The main advantage of AIPCW over IPCW is that its bias is multiplicative in the errors of $\hat S$ and $\hat G$, so it may disappear even if $\hat{G}$ is not consistent for $G$. This makes $\hat\theta_{\mathrm{dr}}$ {\em doubly robust}.
One disadvantage of AIPCW is that $\hat r_{\mathrm{dr}}(\lambda)$, unlike $\hat r_{\mathrm{et}}(\lambda)$, may be negative, potentially complicating its interpretation.

\begin{proposition} \label{prop:double-robustness}
Under Assumptions (A1)--(A2), writing
$Q(u, t_0\mid x) = \{S(u\mid x) - S(t_0\mid x)\}/S(u\mid x)$,
for every $\lambda \in [0,1]$,
  \begin{equation} \label{eq:exact-bias}
    \mathbb E\big[\hat \theta_{\mathrm{dr}}(\lambda) \big] - \theta(\lambda)
    = \mathbb E\!\left[
        A_\lambda(X) \int_{0}^{t_0}
          \frac{G(u)}{\hat{G}(u)}\, S(u)\, \big(\hat Q - Q\big)(u)\,
          \big(d\Lambda^{C} - d\hat\Lambda^{C}\big)(u)
      \right].
  \end{equation}
  Here every function inside the integral is conditional on $X$, and $\hat Q$, $Q$
  are evaluated at $(u, t_0 \mid X)$; no continuity is required of $\hat S$ or
  $\hat G$. In particular, if either $\hat G = G$ or $\hat S = S$ almost surely,
  then $\mathbb E[A_\lambda(X)\hat{\Psi}^{\mathrm{dr}}(O,t_0)] = \theta(\lambda)$
  for all $\lambda$.
\end{proposition}
The proof of this result is in Supplement Section~\ref{appendix:proofs}.

\subsubsection{Upper Confidence Bounds}

In addition to point estimates, the calibration procedures considered in this paper require one-sided upper confidence bounds (UCBs) for \( r(\lambda) \) at a prescribed significance level \( \delta \in (0,1) \), so that, for example, a \(95\%\) confidence bound corresponds to \( \delta = 0.05 \). Several approaches are available.
 
A classical approach that tends to work well in moderate-to-large samples is the delta-method, which approximates $\hat r(\lambda)$ as asymptotically Gaussian and computes its large-sample variance using a Taylor expansion, leading to an upper bound of the form
\[
  \mathrm{UCB}^{\mathrm{pt}}(\lambda;\delta)
  = \hat r(\lambda) + z_{1-\delta}\,\hat{\mathrm{SE}}\big(\hat r(\lambda)\big),
\]
where $z_{1-\delta}$ is the $(1-\delta)$ quantile of the standard normal distribution. We refer to Supplement Section~\ref*{app:high-probability} for additional details, including an expression for $\hat{\mathrm{SE}}(\hat r(\lambda))$. An alternative approach with similar characteristics is the nonparametric bootstrap; see Algorithm~\ref*{alg:bootstrap-ub} in Supplement Section~\ref*{app:bootstrap-ub}.

Both the IPCW estimator~\eqref{eq:ipcw-et} and its augmented
counterpart~\eqref{eq:aipcw} are of the form
$\hat\theta(\lambda) = n^{-1}\sum_i A_\lambda(X_i)\hat{\Psi}(O_i,t_0)$,
differing in how the pseudo-outcomes are defined: in the first case
$\hat{\Psi}(O_i,t_0) = \mathbb I(\tilde T_i \le t_0,\, E_i = 1)/
\hat G(\tilde T_i\mid X_i)$, and in the second $\hat{\Psi}(O_i,t_0)$ is given
by~\eqref{eq:aipcw-pseudo}. As shown in Supplement
Section~\ref*{app:high-probability}, the delta-method variance depends
on the estimator only through the influence contributions
\begin{equation} \label{eq:influence-generic}
  \phi_i(\lambda)
  = \frac{A_\lambda(X_i)\big\{\hat{\Psi}(O_i,t_0) - \hat r(\lambda)\big\}}{\hat\mu(\lambda)},
\end{equation}
so the same expressions apply to either choice, with different numerical values. 
Another difference is in how well the delta-method bound is justified. Both treat $\hat S$ and $\hat G$ as
fixed, which is asymptotically harmless for the augmented estimator, whose bias
is second order in the nuisance errors, but which for the IPCW estimator ignores
a first-order contribution from the estimation of $\hat G$.

When very few patients are selected (e.g., $\le 15$), the asymptotic approximations underlying both the delta-method and bootstrap are no longer justified, requiring instead the use of finite-sample bounds. In our context, a relatively tight finite-sample upper confidence bound for $r(\lambda)$ can be derived by combining an empirical-Bernstein inequality \citep{maurer2009empirical} for the numerator with an exact binomial lower bound for the denominator; see Supplement Section~\ref*{app:ipcw-fs} for further details.

\subsection{Greedy Pointwise Calibration} \label{subsec:pointwise}

Given a grid $\Lambda=\{\lambda_1,\dots,\lambda_L\}$, the simplest calibration strategy
computes $\hat r(\lambda)$ and a pointwise UCB for each $\lambda\in\Lambda$ and
chooses
\begin{align} \label{eq:cal-greedy}
  \hat\lambda_{\mathrm{greedy}} = \min\ \big\{ \lambda\in\Lambda : \mathrm{UCB}^{\mathrm{pt}}(\lambda;\delta)\le\alpha\big\},
\end{align}
with the convention that $\hat\lambda_{\mathrm{greedy}} = +\infty$ if the set is empty.
Algorithm~\ref{alg:greedy} in Supplement Section~\ref*{app:hp-greedy} summarizes this procedure.

A limitation of this strategy is that it offers no guarantee for
the risk of the tuned screening rule, $r(\hat\lambda_{\mathrm{greedy}})$, even if
the pointwise bounds $\mathrm{UCB}^{\mathrm{pt}}(\lambda;\delta)$ are themselves
valid in finite samples. This is because $\mathrm{UCB}^{\mathrm{pt}}(\lambda;\delta)$ 
is valid only at a pre-specified threshold $\lambda$, and
searching across $L$ candidates could in the worst case inflate the
undercoverage probability from $\delta$ to $\min\{1,L \delta\}$. While this worst case is
pessimistic in practice, because $\hat r(\lambda)$ is positively correlated across
nearby thresholds, the numerical experiments in Section~\ref{sec:experiments}
show that pointwise calibration does not reliably achieve~\eqref{eq:hp}.

\subsection{Uniform Calibration} \label{subsec:uniform}

One way of ensuring valid inference after data-driven threshold selection is to replace the pointwise bound with a more conservative \emph{uniform} bound $\mathrm{UCB}^{\text{unif}}(\lambda; \delta)$  satisfying:
\begin{equation*}
\P{ r(\lambda) \le \mathrm{UCB}^{\text{unif}}(\lambda; \delta), \; \forall\,\lambda \in \Lambda } \ge 1 - \delta.
\end{equation*}
The calibrated threshold can then be safely selected as
\begin{equation*}
\hat\lambda_{\text{unif}} = \min \big\{\lambda\in\Lambda  :\
\mathrm{UCB}^{\text{unif}}(\lambda;\delta)\le \alpha\ \big\},
\end{equation*}
with the convention that $\hat\lambda_{\mathrm{unif}} = +\infty$ if the set is empty; see Algorithm~\ref{alg:uniform} in Supplement Section~\ref*{app:uniform}. This reliably achieves~\eqref{eq:hp}.

The simplest approach to obtain uniform confidence bounds applies the pointwise methods from Section~\ref{subsec:pointwise-estimation} with a Bonferroni correction, replacing $\delta$ with $\delta / |\Lambda|$. However, this tends to be overly conservative.
A more efficient alternative is the Gaussian multiplier bootstrap \citep{chernozhukov2013gaussian}, which simulates correlated fluctuations in $\hat{r}(\lambda)$ across the $\lambda$ grid. This accounts for dependencies between estimates and can produce much tighter bands than Bonferroni when $\hat{r}(\lambda)$ varies smoothly in $\lambda$.
We refer to Supplement Section~\ref*{app:uniform} for further details on both approaches.

\subsection{Learn–Then–Test (LTT) Calibration} \label{subsec:ltt}
An intermediate approach to achieve~\eqref{eq:hp} that sits between pointwise and uniform
calibration is provided by the LTT procedure
\citep{angelopoulos2025learn}. 

Proceeding as in Section~\ref{subsec:pointwise-estimation}, LTT constructs pointwise upper confidence bounds $\mathrm{UCB}^{\mathrm{pt}}(\lambda; \delta_\star)$ for the risk $r(\lambda)$ over a grid  $\Lambda$ of candidate thresholds, using a significance level $\delta_\star$ defined below. To determine the calibrated threshold $\hat{\lambda}_{\mathrm{LTT}}$, LTT does not search greedily over all $\lambda \in \Lambda$ as in~\eqref{eq:cal-greedy}. Instead, it evaluates whether $\mathrm{UCB}^{\mathrm{pt}}(\lambda; \delta_\star) \leq \alpha$ sequentially over $\lambda$ along one or more pre-specified, typically monotone paths through $\Lambda$—for instance, decreasing values of $\lambda$ corresponding to increasingly liberal screening rules—and stops immediately before the first ``unsafe'' threshold where the risk upper confidence bound exceeds $\alpha$. When multiple ($M>1$) paths are considered, the final $\hat{\lambda}_{\mathrm{LTT}}$ is the most liberal (i.e., smallest) among the corresponding stopping points.

This is an instance of fixed-sequence testing \citep{bauer1991multiple}, ensuring high-probability risk control without excessive conservativeness. Formally, if the risk upper bounds $\mathrm{UCB}^{\mathrm{pt}}(\lambda; \delta_\star)$ are {\em pointwise} valid, then $\mathbb{P}[r(\hat{\lambda}_{\mathrm{LTT}}) > \alpha] \le \delta$.
Intuitively, by setting $\delta_\star = \delta/M$, LTT operates as if only $M$ candidate thresholds were being tested. 
The Bonferroni adjustment in $\delta_\star = \delta/M$ is typically mild, since a small number of well-designed paths ($M \ll |\Lambda|$) is usually sufficient to find a good calibrated threshold.

Concretely, we implement LTT using threshold paths defined as follows. We begin by fixing an \emph{anchor} threshold $\lambda^{(1)}_0 = 1 - \alpha$, motivated by the intuition that if the survival model $\hat{S}$ is well-calibrated, then $\hat{z}(x) \approx S(t_0 \mid x)$, so selecting individuals with $\hat{z}(x) \ge \lambda$ should result in an event rate near $1 - \lambda$, implying $r(\lambda^{(1)}_0) \approx \alpha$. Then, we define a decreasing sequence $\lambda^{(1)}_0 > \lambda^{(1)}_1 > \lambda^{(1)}_2 > \cdots > 0$ to explore progressively more liberal thresholds. Intuitively, if the anchor is too conservative ($r(\lambda^{(1)}_0) < \alpha$), relaxing the threshold can safely increase yield.
To guard against the possibility that the anchor itself may already be too liberal, we include a second decreasing path $\lambda^{(2)}_0 > \lambda^{(2)}_1 > \lambda^{(2)}_2 > \cdots \geq \lambda^{(1)}_0$ starting from a more conservative anchor, e.g., $\lambda^{(2)}_0 = 1 - \alpha/2$.

For each candidate $\lambda$, we compute $\mathrm{UCB}^{\mathrm{pt}}(\lambda; \delta_\star)$ for $r(\lambda)$ as described in Section~\ref{subsec:pointwise-estimation}, using $\delta_\star = \delta/2$ since we consider $M = 2$ paths. 
In practice, we recommend using delta–method or bootstrap confidence bounds, resorting to finite–sample safeguards when the selected set is very small (e.g., $\leq 15$).
Along each path, the algorithm begins at the anchor and moves sequentially toward smaller $\lambda$. It stops when the upper confidence bound first exceeds $\alpha$, recording the last accepted value as $\hat{\lambda}_{\mathrm{LTT}}^{(1)}$ or $\hat{\lambda}_{\mathrm{LTT}}^{(2)}$, depending on the path. If no threshold is accepted, we set $\hat{\lambda}_{\mathrm{LTT}}^{(l)} = +\infty$, corresponding to no selections. Finally, the calibrated threshold is $\hat{\lambda}_{\mathrm{LTT}} = \hat{\lambda}^{(1)}_{\mathrm{LTT}} \land  \hat{\lambda}^{(2)}_{\mathrm{LTT}}$.
The complete procedure is summarized by Algorithm~\ref{alg:ltt} in Supplement Section~\ref*{app:ltt}.

Compared to greedy pointwise and uniform calibration, LTT seeks a middle ground between efficiency and conservativeness.
It accounts for the data-driven choice of threshold while being often less conservative than the 
uniform approach, though not always: the search
halts at the first threshold whose bound exceeds $\alpha$, which may happen too
early if the pointwise estimates are noisy, as shown empirically in
Section~\ref{sec:experiments}.

In the next section, we turn to a complementary framework based on conformal prediction and FDR control, which removes the need to specify a significance level $\delta$.

\section{Calibration via FDR Control and Conformal Inference} \label{sec:fdr}

\subsection{Cohort-Specific Screening Rules and Hypothesis Testing}

We now turn to a \emph{transductive} perspective, allowing the threshold to depend
on the covariates of the test cohort $\Dtest$, and reformulating the calibration problem in the language of hypothesis testing
and FDR. 
A practical advantage of this approach is that it requires neither a pre-specified threshold grid nor a significance level $\delta$; instead, the user only needs to specify the target risk level $\alpha$ (for example, $10\%$). 

For each cohort patient $j \in [m]$, consider the null and alternative ``hypotheses''
\begin{align} \label{eq:null-hyp}
H^{0}_{j}:\ T_{n+j} \le t_0
\qquad \text{vs.} \qquad
H^{a}_{j}:\ T_{n+j} > t_0,
\end{align}
where \( H^{0}_{j} \) states that patient \( j \) experiences the
event before the time horizon \( t_0 \).
Departing from classical hypothesis testing, however, these ``hypotheses'' are
not statements about fixed population parameters but rather correspond to random variables,
more precisely written as \( H^{0}_{j} = \I{T_{n+j} \le t_0} \).

For any cohort-specific and possibly data-dependent screening rule $A$, let $\hat{\mathcal S}\subseteq[m]$ denote the selected cohort patients and $\mathcal H_0 := \{j\in[m]: T_{n+j}\le t_0\}$ the cohort patients who experience the event before \( t_0 \), corresponding to the true nulls.
In this context, we define the \emph{false discovery proportion} (FDP) and the FDR as:
\begin{align} \label{eq:FDR}
  & \mathrm{FDP} := \frac{|\hat{\mathcal S}\cap\mathcal H_0|}
                       {\max\{1,|\hat{\mathcal S}|\}} ,
  & \mathrm{FDR} := \EV{\mathrm{FDP}} ,
\end{align}
with the expectation taken over both $\Dcal$ and $\Dtest$.
The goal is to select a large subset $\hat{\mathcal S}$ of cohort patients while controlling the FDR below $\alpha$.

\subsection{Screening with the Benjamini--Hochberg Procedure} \label{sec:fdr-bh}

For each cohort individual $j \in [m]$ we compute a \emph{nonconformity score}
$Z_{n+j}$ quantifying how plausible it seems that the individual survives beyond
$t_0$, so that larger values indicate stronger evidence against
$H^{0}_{j}: T_{n+j} \le t_0$. We take these to be the fitted survival curve
evaluated at a {\em shifted} horizon,
\begin{align} \label{eq:def-scores-test}
  Z_{n+j} = \hat{S}(t_0 + \gamma \mid X_{n+j}), \qquad j \in [m] ,
\end{align}
where the time shift $\gamma \ge 0$ is a free parameter.

The yield of this procedure is sensitive to the choice of $\gamma$, and a sufficiently large value is typically needed. This is reminiscent of
the role of $M$ in the uncensored regression setting of
\citet{jin2023selection}, where scores of the form
$V(x,y) = M \cdot y - \hat\mu(x)$ are used and $M$ must be large enough to
avoid excessive conservativeness. The most intuitive choice, $\gamma = 0$, performs very poorly:
 in the large-sample limit it induces an all-or-nothing
behavior in which either every subject is selected or none, depending only on
whether the mean risk of the entire cohort meets the target; larger values
of $\gamma$ are needed to isolate a low-risk subgroup.
Supplement Section~\ref{app:tuning-gamma} explains why, and
Algorithm~\ref{alg:tuning-gamma} therein gives a practical procedure for tuning
$\gamma$ on data independent of $\Dcal$. All results below hold conditionally
on $\gamma$, which we now treat as fixed.

To assess the evidence that $Z_{n+j}$ provides against $H^{0}_{j}$ in a
model-agnostic manner, we compare it to analogous scores computed on the
calibration data, using the observed event or censoring times in place of
$t_0$.
Specifically, define
\begin{align} \label{eq:def-scores}
Z_i = \hat{S}(\tilde{T}_i + \gamma \mid X_i), \qquad i \in [n].
\end{align}
Then, each $Z_{n+j}$ is ranked against $Z_{1},\ldots,Z_{n}$ by the {\em IPCW conformal p-value} function
\begin{align}\label{eq:ipcw-conformal-p}
& \hat{p}(z)
= 1 \land \frac{1 + \sum_{i=1}^{n} E_i\,\I{\tilde{T}_i \le t_0} \, \hat{w}_i \,\I{Z_i \ge z}}{1 + n},
& \hat{w}_i = \frac{1}{\hat{G}(\tilde{T}_i \mid X_i)} .
\end{align}
Intuitively, $p_j := \hat{p}(\hat{S}(t_0 + \gamma \mid X_{n+j}) )$ is the IPCW-adjusted empirical
rank of \( Z_{n+j} \) among the calibration scores of individuals
who experienced the event before \( t_0 \).

We call $p_1,\ldots,p_m$ \emph{selective conformal p-values}, consistent with \citet{jin2023selection},
because a smaller value of \( p_j \) provides stronger evidence against $H^{0}_{j}$ and these statistics are used as input for BH, 
even though they are not p-values in the classical sense.
Recall that applying BH means computing an adaptive p-value threshold 
\[
\hat{q}(\alpha)
:= \max\left\{ \tau \in [0,1] :
\#\{j : p_j \le \tau\} \ge \frac{m}{\alpha}\,\tau
\right\}
\]
and then defining the set $\hat{\mathcal{S}}$ of selected patients as:
\begin{align} \label{eq:sel-pvalue}
\hat{\mathcal{S}}
= \{ j \in [m] : p_j \le \hat{q}(\alpha) \}.
\end{align}
Algorithm~\ref{tab:alg-fdr} summarizes this method, including two extensions explained below.

This approach can be viewed as a special case of the general screening
framework from Section~\ref{sec:problem-formulation}, where the
selected set is written as
\(
\hat{\mathcal{S}} = \{ j \in [m] : A_\lambda(X_{n+j}) = 1 \}
\)
for screening rules of the form
\( A_\lambda(x) := \I{\hat{z}(x) \ge \lambda} \).
Indeed, by defining the score \( \hat{z}(x) := 1 - \hat{p}(\hat{S}(t_0 + \gamma \mid x)) \) and the
data-adaptive threshold \( \hat{\lambda} := 1 - \hat{q}(\alpha) \), the BH rule
can be expressed equivalently as
\(
\hat{\mathcal{S}} = \{ j \in [m] : \hat{z}(X_{n+j}) \ge \hat{\lambda} \}.
\)
A key difference relative to the inductive methods of
Section~\ref{sec:methodology-hp} is that both components of this screening rule are
data-driven: the score \( \hat{z} \) depends on the calibration data
\( \mathcal D_{\mathrm{cal}} \), while the threshold \( \hat{\lambda} \) further
adapts to the specific cohort through the BH procedure.

In the absence of censoring (i.e., $E_i \equiv 1$, $\tilde T_i = T_i$, and $G\equiv1$) the
conformal p-values~\eqref{eq:ipcw-conformal-p} reduce to those of \citet{jin2023selection}, with which BH controls the FDR in finite samples.
While in our censored setting exact FDR control is not guaranteed, it continues to hold well in practice and can be guaranteed in the large-sample limit provided that $\hat G$ is consistent \citep{sesiavsvetnik2025}. In Section~\ref{sec:fdr-augmented} we introduce an AIPCW extension that makes Algorithm~\ref{tab:alg-fdr} doubly robust, and in Section~\ref{sec:fdr-asymptotic} we compare its asymptotic FDR properties to those of the IPCW implementation.
Section~\ref{sec:fdr-relation} studies the relation between FDR and HP risk control, motivating the optional abstention step introduced in the last line of Algorithm~\ref{tab:alg-fdr}.

\begin{algorithm*}[!htb]
\caption{FDR-conformal patient screening.}
\label{tab:alg-fdr}
\begin{algorithmic}[1]
\Require Pre-trained survival model $\hat{S}$ and censoring model $\hat G$;
right-censored calibration data $\mathcal D_{\mathrm{cal}}=\{(X_i,\tilde T_i,E_i)\}_{i=1}^n$;
test cohort covariates $\{X_{n+1},\dots,X_{n+m}\}$;
target risk level $\alpha \in (0,1)$;
time shift parameter $\gamma\ge 0$;
selection stability threshold $\nu \in [0,1]$;
flavor $\texttt{dr} \in \{\textsc{false}, \textsc{true}\}$.

\State For $i \in [n]$, compute $\hat{w}_i = 1 / \hat G(\tilde T_i\mid X_i)$, with winsorization.

\State For $i \in [n]$ and $j \in [m]$, compute $Z_i$ and $Z_{n+j}$ as in~\eqref{eq:def-scores}, with time shift $\gamma$.

\If{$\texttt{dr} = \textsc{false}$}
  \State For $j \in [m]$, compute $p_j = \hat{p}(Z_{n+j})$ with the p-value function $\hat{p}$ in~\eqref{eq:ipcw-conformal-p}.
\Else
  \State For $j \in [m]$, compute $p_j = \hat{p}(Z_{n+j})$ with the p-value function $\hat{p}$ in~\eqref{eq:aipcw-conformal-p}.
\EndIf

\State Apply BH at level $\alpha$ to $(p_1,\ldots,p_m)$ and obtain rejection set $\hat{\mathcal{S}} \subseteq [m]$.

\State \textit{(Optionally abstain)} Compute $\hat{\mathbb P}(|\hat{\mathcal{S}}|>0 \mid \mathcal D_{\mathrm{cal}})$ by nonparametric bootstrap over test rows (see Section~\ref{sec:fdr-relation});
set $\hat{\mathcal{S}} = \emptyset$ if $\hat{\mathbb P}(|\hat{\mathcal{S}}|>0 \mid \mathcal D_{\mathrm{cal}}) < \nu$.

\Ensure Selected cohort subset $\hat{\mathcal S} \subseteq [m]$.
\end{algorithmic}
\end{algorithm*}

\subsection{Augmented Conformal p-Values} \label{sec:fdr-augmented}

The p-value function in~\eqref{eq:ipcw-conformal-p} can be made doubly robust 
with an AIPCW approach similar to Section~\ref{subsec:pointwise-estimation}.
To highlight the connection with Section~\ref{subsec:pointwise-estimation}, it is helpful 
to rewrite~\eqref{eq:ipcw-conformal-p} as follows.
Define the
non-conformity score function
\begin{equation} \label{eq:conformal-score-R}
  R(x,t) = \hat S(t + \gamma \mid x)\,\I{t \le t_0},
\end{equation}
and the random variable $Z = \hat{S}(\tilde{T} + \gamma \mid X)$.
Note that
$\mathbb{I}[\tilde{T} \le t_0]\,\mathbb{I}[Z \ge z] = \mathbb{I}[R(X,\tilde{T}) \geq z]$ almost
surely for any $z>0$, so that $\hat{p}(z)$
in~\eqref{eq:ipcw-conformal-p} becomes
\begin{equation} \label{eq:pvalue-as-score}
  \hat{p}(z)
  \;=\;
  1 \land \frac{1 + \sum_{i=1}^{n} E_i \, \hat{w}_i \,\I{R(X_i,\tilde{T}_i) \ge z}}{1 + n},
\end{equation}
which is essentially the IPCW estimator of $H(z) := \P{R(X,T) \ge z}$.

Moreover, since $R(x,t)$ is by construction non-increasing in $t$, the
equivalence $\{R(X,T) \ge z\} = \{T \le b_z(X)\}$ holds for any $z > 0$, where
\begin{equation} \label{eq:def-bz}
  b_z(X) := \sup\big\{t \ge 0 : R(X,t) \ge z \big\} \leq t_0.
\end{equation}
This allows writing $H(z) = \P{T \le b_z(X)}$, which highlights that $H(z)$ has a 
functional form similar to that of $\theta(\lambda)$ in
Section~\ref{subsec:pointwise-estimation}, with $t_0$ replaced
by the covariate-dependent horizon $b_z(X)$. Therefore, we can reuse the same AIPCW construction,
replacing $t_0$ with $b_z(X_i)$ in the pseudo-outcomes in~\eqref{eq:aipcw-pseudo}, 
which gives $\hat H(z) := n^{-1}\sum_{i=1}^{n}\hat{\Psi}(O_i, b_z(X_i))$ as the AIPCW estimator of $H(z)$.

While it may be tempting to define the augmented conformal p-value directly as
$[1+n\hat H(z)]/(1+n)$, note that $\hat H(z)$ need not be non-increasing in $z$,
nor take values in $[0,1]$. Therefore, we replace
$\hat H(z)$ by its least non-increasing majorant and clip the conformal p-value to the unit interval,
computing:
\begin{align} \label{eq:aipcw-conformal-p}
  & \hat{p}_{\mathrm{dr}}(z) := \Pi \left( \frac{1 + \sum_{i=1}^{n}\hat{\Psi}(O_i, b_z(X_i)) }{1+n} \right),
  & \Pi f(z) := \max\{0, \min\{1,\ \sup_{z'\ge z} f(z')\}\}.
\end{align}
Then, cohort patients are selected by applying BH
to the augmented conformal p-values $\hat{p}_{\mathrm{dr}}(\hat{S}(t_0 + \gamma \mid X_{n+1})), \ldots, \hat{p}_{\mathrm{dr}}(\hat{S}(t_0 + \gamma \mid X_{n+m}))$.

\subsection{Asymptotic FDR Analysis} \label{sec:fdr-asymptotic}

We now study the asymptotic behaviour of the FDR when BH is applied to the
selective conformal p-values~\eqref{eq:ipcw-conformal-p} or their augmented
counterparts~\eqref{eq:aipcw-conformal-p}. 
We let $n\to\infty$ with the training sample growing alongside it, so that 
the fitted survival and censoring models are also indexed by $n$.
The results depend on model accuracy through
\begin{align} \label{eq:def-err}
\begin{split}
  \mathcal G_n &:= \Big\{\EV{\big|\hat G(T\mid X)-G(T\mid X)\big|^2}\Big\}^{1/2} , \\
  \mathcal S_n & := \Big\{\EV{\sup_{u\le t_0}\big|\hat S(u\mid X)-S(u\mid X)\big|^2}\Big\}^{1/2} , \\
  \mathcal L_n &:= \left\{\EV{\left(\int_0^{t_0}
    \big|d\Lambda^{C}(u\mid X)-d\hat\Lambda^{C}(u\mid X)\big|\right)^{2}}\right\}^{1/2} ,
\end{split}
\end{align}
which measure the error of the censoring model on the CDF and hazard scales and
that of the survival model on the CDF scale. Each term averages over the
covariate distribution and is
therefore of the kind that existing consistency results for flexible survival estimators
can deliver \citep[e.g.,][]{ishwaran2010consistency,chen2019nearest}. 
Note, however, that $\mathcal L_n$ measures the error of $\hat G$ on the
hazard scale, so it depends on the accuracy of a derivative that may be estimated at a slower rate then;
see Remark~2.9 in \citet{farina2025doubly} for a discussion of this point in a
related setting.

As $n\to\infty$, we consider a cohort size $m$ that may remain fixed or also
grow. If the cohort grows, we track the reciprocal
of the realized selection proportion, $m/\max(|\hat{\mathcal S}|,1)$, where
$|\hat{\mathcal S}|$ is the number of hypotheses rejected by BH at level
$\alpha$; in particular, we track
\begin{align} \label{eq:def-kappa}
  & \kappa_{n,m} := \EV{\frac{m}{\max(|\hat{\mathcal S}|,1)}},
  & \kappa^{(2)}_{n,m} := \EV{\bigg(\frac{m}{\max(|\hat{\mathcal S}|,1)}\bigg)^{\!2}}^{1/2},
\end{align}
which satisfy $\kappa_{n,m}\le\kappa^{(2)}_{n,m}\le m$.
If $m$ grows, $\kappa^{(2)}_{n,m}$ may still remain bounded as long as the selected set contains a
non-vanishing fraction of the cohort.

\begin{theorem} \label{thm:fdr-asymptotic}
Suppose Assumptions \textup{(A1)}--\textup{(A3)} hold along a sequence of
problems indexed by $n\to\infty$, with $c$, $s$ not varying along the sequence.
Then BH applied at level $\alpha$ to the conformal p-values defined above
satisfies
\begin{align*}
  \mathrm{FDR} & \le \alpha + O\big(\mathcal G_n + n^{-1}\big) \cdot \kappa_{n,m} + O\big(n^{-1/2}\big) \cdot \kappa^{(2)}_{n,m},                     
   && \text{under IPCW~\eqref{eq:pvalue-as-score},} \\
  \mathrm{FDR} & \le \alpha + O\big(\mathcal S_n\mathcal L_n + n^{-1}\big) \cdot \kappa_{n,m} + O\big(n^{-1/2}\big) \cdot \kappa^{(2)}_{n,m},                     
  && \text{under AIPCW~\eqref{eq:aipcw-conformal-p},}
\end{align*}
with implicit constants depending only on $c$ and $s$. 
\end{theorem}

The proof of Theorem~\ref{thm:fdr-asymptotic} is in Supplement Section~\ref{app:fdr-risk}. 
This result implies that $\limsup\mathrm{FDR}\le\alpha$ if $\kappa^{(2)}_{n,m}=O(1)$ and
$\mathcal G_n=o(1)$ under IPCW, or $\mathcal S_n\mathcal L_n=o(1)$ under AIPCW.
The difference between these two asymptotic FDR bounds thus
 mirrors the difference between the biases of the IPCW and AIPCW
estimators in Section~\ref{subsec:pointwise-estimation}. 
If both errors are $o(n^{-1/4})$, the AIPCW procedure attains
$\mathrm{FDR}\le\alpha+O(n^{-1/2})\cdot\kappa^{(2)}_{n,m}$, which matches the parametric rate
whenever $\kappa^{(2)}_{n,m}=O(1)$.

\subsection{Relation to Risk Control and the Need for Abstention} \label{sec:fdr-relation}

Controlling the FDR is closely related to controlling the selected-set risk~\eqref{eq:def-r} in expectation.
This connection makes FDR-based screening directly comparable to the
HP calibration methods of Section~\ref{sec:methodology-hp} and motivates the abstention step of Algorithm~\ref{tab:alg-fdr}.

\begin{theorem}\label{thm:fdr-risk}
Let \( (p_j,H^{0}_{j})_{j=1}^m \) be pairs of random variables, with
\( p_j\in[0,1] \) and \( H^{0}_{j}\in\{0,1\} \), that are i.i.d.\ conditional
on \( \mathcal D_{\mathrm{cal}} \). For any \( \tau\in[0,1] \), define
\[
\bar r(\tau; \mathcal D_{\mathrm{cal}})
:=
\mathbb{P}\!\left(H^{0}_1=1 \,\middle|\, p_1\le \tau,\ \mathcal D_{\mathrm{cal}}\right).
\]
Run BH at level \( \alpha \) on \( (p_1,\ldots,p_m) \) and let
\( \hat{\mathcal S} \subseteq [m] \) denote the corresponding rejection set.
Then the FDR, defined as in~\eqref{eq:FDR} with
\( \mathcal{H}_0 = \{j \in [m] : H^{0}_{j} = 1\} \), satisfies
\[
\mathrm{FDR}
=
\mathbb{E}\!\left[
\mathbb{P}\big(|\hat{\mathcal S}| > 0 \mid \mathcal D_{\mathrm{cal}}\big)
\cdot
\EV{ \bar r\!\left(\frac{\alpha |\hat{\mathcal S}|}{m}; \mathcal D_{\mathrm{cal}}\right) \mid |\hat{\mathcal S}|>0, \mathcal D_{\mathrm{cal}}}
\right],
\]
with the convention that the product is zero when $\mathbb P(|\hat{\mathcal S}|>0\mid\mathcal D_{\mathrm{cal}})=0$.
\end{theorem}
The proof of Theorem~\ref{thm:fdr-risk} is in Supplement Section~\ref{app:fdr-risk}.
To see how this result connects to the risk \( r \) in
\eqref{eq:def-r}, note that a BH threshold \( \alpha k / m \)
corresponds to an adaptive threshold
\( \hat{\lambda} = 1 - \alpha k / m \) for scores
\( \hat{z}(X_{n+j}) = 1 - \hat{p}(\hat{S}(t_0 + \gamma \mid X_{n+j})) \). Thus,
\[
\bar{r}\!\left(\frac{\alpha k}{m}; \mathcal D_{\mathrm{cal}}\right)
= r\left(A_{\hat{\lambda}}; \mathcal D_{\mathrm{cal}}\right),
\]
where \( A_{\hat{\lambda}} \) denotes the cohort-specific screening rule
\( A_{\hat{\lambda}}(X_{n+j}) := \mathbb{I}\{\hat{z}(X_{n+j}) \ge \hat{\lambda}\} \).

It follows that in regimes where
\( \mathbb{P}(|\hat{\mathcal S}| > 0 \mid \mathcal D_{\mathrm{cal}}) \) is close to
one---corresponding to cohorts that are not too small and screening rules with
non-negligible yield---FDR control is effectively equivalent to controlling the
selected-set risk \( r \) in expectation.
On the other hand, if $\mathbb{P}(|\hat{\mathcal S}|>0 \mid \mathcal D_{\mathrm{cal}})$
is small, the FDR may fall well below $\alpha$ while the event rate $\bar r$ among
selected patients is larger. This motivates the abstention step introduced in the last line of
Algorithm~\ref{tab:alg-fdr}, which we now discuss in detail.
 
For a fixed calibration data set and pre-trained pair of survival and censoring
models, we estimate
$\mathbb{P}(|\hat{\mathcal S}|>0 \mid \mathcal D_{\mathrm{cal}})$ via nonparametric
bootstrap. We resample the $m$ cohort individuals with replacement,
recompute the p-values, reapply BH, and compute the fraction of resamples
selecting at least one cohort patient. If the estimate thus obtained falls below
a predetermined threshold $\nu$ (we use $\nu = 0.9$), Algorithm~\ref{tab:alg-fdr} abstains and selects no cohort patients.
Abstention never increases the FDR because 
the selected set is either the original $\hat{\mathcal S}$ or $\emptyset$, 
and the empty set always has zero FDP by definition~\eqref{eq:FDR}.
 
The effect of abstention is to make Algorithm~\ref{tab:alg-fdr} more
conservative, addressing the gap between marginal FDR validity
and the per-cohort reliability clinicians need in practice. 
Consider for example a scenario with 100 distinct
cohorts drawn from the same population; in 90 cohorts no patients are
selected, while in the remaining 10 all selected patients experience the event
before the horizon $t_0$. The average FDP is $0\cdot0.9 + 1\cdot0.1 = 0.1$, so
the FDR is controlled at $\alpha = 0.1$. Yet the event
rate among selected patients is $100\%$ in every cohort for which any selection
is made, making the method practically unreliable.
This is consistent with Theorem~\ref{thm:fdr-risk}: here
$\mathbb{P}(|\hat{\mathcal S}|>0 \mid \mathcal D_{\mathrm{cal}}) \approx 0.1$,
$\bar r \approx 1$, and their product is the nominal FDR level. The abstention step is
designed precisely for this situation. As long as the estimated selection probability
falls below $\nu$, no selections are made in any of the 100 cohorts.

In practice, an abstention indicates that the data are not sufficient
to identify a low-risk subgroup for this cohort at this horizon. 
The practitioner may react by shortening the time horizon, relaxing the target survival level,
collecting more data, or deferring to clinical judgement. 
In any case, abstention seems preferable to making unreliable and potentially misleading 
selections when the evidence is weak.

Finally, we remark that, even with optional abstention, Algorithm~\ref{tab:alg-fdr}
still only targets the marginal FDR, which says little
about the event rate among selected patients when few are selected. The
HP methods of Section~\ref{sec:methodology-hp} are easier to
interpret in that regime, as they bound the selected-set risk 
regardless of yield, and should therefore be preferred when few selections are expected,
either because the cohort is small or the yield is low.
Otherwise the two approaches essentially target the same quantity, by
Theorem~\ref{thm:fdr-risk}, and BH has the advantage of not requiring a
significance level $\delta$.

\section{Numerical Experiments with Semi-Synthetic Data} \label{sec:experiments}

We compare the calibration methods described above using semi-synthetic data designed to
mimic the structure of a real oncology study while retaining perfect knowledge of
the underlying event times, which enables full evaluation.
Specifically, we begin with real patient covariates and fit models for both the
event-time and censoring-time distributions conditional on these covariates. These models are then
treated as generative mechanisms to simulate realistic synthetic event and censoring times.

\subsection{Data Description} \label{sec:data}

We examine data from the Flatiron Health Research Database (FHRD), an electronic health record-derived database comprising data from over 280 oncology practices at 800+ unique sites of care in the US, spanning the period from December 2010 to May 2023 \citep{Ma2020,Birnbaum2020,Becker2023,Johnson2025}. All data are de-identified and subject to technical and administrative safeguards 
to protect patient confidentiality.
We focus on a cohort of 7{,}000 patients diagnosed with advanced 
non--small cell lung cancer (aNSCLC). 
For each patient, 43 covariates describe demographic and socioeconomic 
characteristics, smoking status, vital signs, functional status, comorbidity index, 
tumor histology, number and location of metastatic sites, treatment history, 
indicators of prior laboratory assessments, and a wide range of clinical laboratory 
measurements. 
The outcome measures include a binary event indicator and the corresponding 
time to event or censoring.

\subsection{Semi-Synthetic Data Generation} \label{subsec:semi-synth}

Let $X = \{X_i\}_{i=1}^{N_0}$ denote the FHRD covariate matrix, with $N_0=7{,}000$ rows and $43$ columns. These covariates are treated as fixed design points representing the target population. Using all available observations to approximate this population as closely as possible, we fit two {\em generative} models, respectively denoted as $(\hat{S}^\star, \hat{G}^\star)$, to estimate the conditional distribution of the survival and censoring times given the covariates.
We use a generalized random forest (GRF) implemented in the \texttt{grf} \textsf{R} package for both $\hat{S}^\star$ and $\hat{G}^\star$; the latter is fit similarly to the former after flipping the event indicators.

For each individual $i$, we generate independent uniform random variables $U_i, V_i \sim \text{Uniform}(0,1)$ and define event and censoring times:
\[
T_i = \inf\{t : \hat{S}^\star(t \mid X_i) \le U_i\}, \qquad 
C_i = \inf\{t : \hat{G}^\star(t \mid X_i) \le V_i\}.
\]
Thus, $T \indep C \mid X$ holds by construction. The observation for the $i$-th semi-synthetic individual is $(X_i,\widetilde{T}_i, E_i)$, where $\widetilde{T}_i = T_i \land C_i$ and $E_i = \mathbb{I}\{T_i < C_i\}$. This gives a new right-censored dataset with realistic dependence patterns and known ground-truth event times. The full resampling procedure is independently repeated $100$ times.

\subsection{Analysis Protocol} \label{subsec:analysis-protocol}

Each replicate of the semi-synthetic data is randomly split into a training set ($n_{\mathrm{tr}}=5{,}000$), a calibration set ($n = 1{,}000$), and a test cohort ($m=1{,}000$). We consider four screening horizons, $t_0 \in \{2, 3, 6, 9\}$ months, and aim to select a low-risk subgroup with a target event-rate of $\alpha = 0.10$, corresponding to survival probability approximately $0.90$ by time $t_0$. The training data are used to fit survival and censoring models, using one of three approaches: a generalized random forest, the Cox proportional hazards model, and an accelerated failure time gradient boosting model implemented in the \texttt{xgboost} \textsf{R} package. This setup allows studying method robustness under both correctly specified and misspecified settings. In each case, the calibration data are used to calibrate the screening rule, and the data from the test cohort to evaluate performance.

We compare all methods summarized in Table~\ref{tab:methods-summary}, including
three HP calibration methods, the FDR-based conformal approach,
and two reference benchmarks, all evaluated using the same fitted survival and
censoring models.

The HP methods are: \emph{HP-Pointwise} (Section~\ref{subsec:pointwise}), 
which greedily selects the smallest feasible
threshold using pointwise upper confidence bounds; \emph{HP-Uniform} 
(Section~\ref{subsec:uniform}), which constructs simultaneous confidence bounds
via Gaussian-multiplier perturbations with 1000 draws; and \emph{HP-LTT}
 (Section~\ref{subsec:ltt}). These methods share a common threshold
grid defined by 200 quantiles of the calibration score distribution and rely on
pointwise risk estimates based on asymptotic delta-method approximations, with a
fallback to empirical Bernstein bounds when fewer than 15 patients are selected.
Unless stated otherwise, all HP methods use a significance level of
$\delta = 0.10$, and we also examine sensitivity to this choice.

The \emph{FDR-Conformal} method (Section~\ref{sec:fdr}) tunes the time shift
parameter $\gamma$ using the training data as described in Supplement
Section~\ref*{app:tuning-gamma}. To align with the other calibration approaches and
the interpretation of Theorem~\ref{thm:fdr-risk}, this method abstains whenever a
bootstrap estimate of $\mathbb{P}(|\hat{\mathcal{S}}|>0)$ falls below $0.9$.

By default, all four calibration methods are applied using AIPCW pseudo-outcomes for double robustness; we also report comparisons against the simpler event-time IPCW alternatives. In both cases the weights are winsorized at their 99th percentile.

For reference, we also report results for two benchmarks: the \emph{model-based}
baseline, which treats predicted survival probabilities as exact and selects the
largest subset with average predicted survival exceeding $0.90$ (using both
training and calibration data for model fitting), and an \emph{oracle} benchmark
that uses the true generative survival function $S^\star(t \mid x)$ to select the
largest subset achieving 90\% average survival, representing the best attainable
performance under perfect knowledge of the data-generating process.

\subsection{Evaluation Metrics} \label{subsec:metrics}
For each replicate and horizon $t_0$ we record: the \emph{yield}, or number of
cohort patients selected; the FDP~\eqref{eq:FDR}, namely the proportion of
selected patients who experience the event by $t_0$, counted as zero when none
are selected; the \emph{event rate}, which is the same proportion but averaged
only over replicates with at least one selection; and whether any selection was
made at all. The empirical distribution of each metric is summarized across the 100
replicates.

\subsection{Results}

\subsubsection{Performance at Different Time Horizons}

\begin{figure*}[!htbp]
\centering
\includegraphics[width=0.9\textwidth]{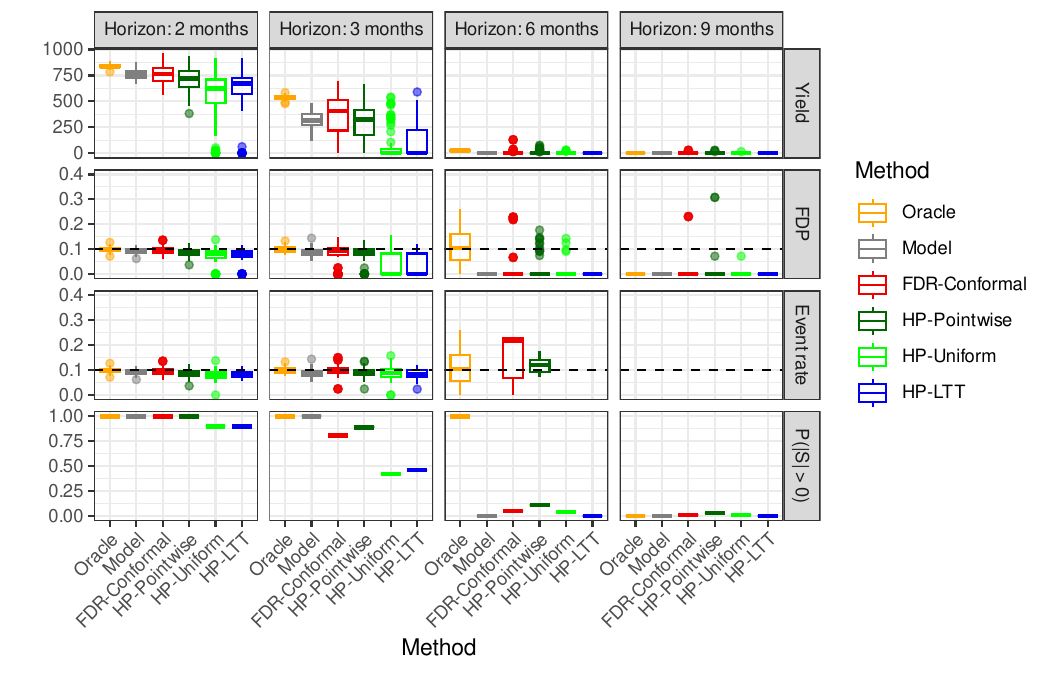}
\caption{
Summary of low-risk screening results on semi-synthetic data, over 100 independent replicates.
Top row: yield.
Second row: false discovery proportion, counted as zero when no patients are selected.
Third row: event rate among selected patients, evaluated only on replicates with at least one selection; it is not reported if selections occur in fewer than 10\% of replicates.
The dashed horizontal line denotes the target event rate, 10\%.
Fourth row: proportion of replicates in which at least one patient is selected.
All methods use the same random forest models. High-probability (HP) methods are applied at significance level $\delta=0.1$.
}
\label{fig:1}
\end{figure*}

Figure~\ref{fig:1} summarizes the results for a well-specified setting, where
both the data-generating and fitted models are generalized random forests.
The methods are mostly distinguished by yield: FDR-Conformal selects the most
patients, followed by HP-Pointwise, HP-LTT, and HP-Uniform. Yield declines at
longer horizons for every method, as fewer patients can be certified low-risk.
The average FDP (the empirical FDR) remains below $\alpha = 0.10$ for all
methods, even where the event rate among selected patients is higher; this
happens especially for FDR-Conformal and HP-Pointwise when the yield is low.
It is unsurprising, because FDR-Conformal does not control the event rate in
expectation when selections are very rare (Theorem~\ref{thm:fdr-risk}), and
HP-Pointwise does not account for the data-driven selection of the screening
threshold.

Table~\ref{tab:semi-synthetic_grf} in Supplement
Section~\ref*{app:numerical-sim} summarizes these results numerically and adds
an estimate of the risk exceedance probability $\P{r(A;\Dcal) > \alpha}$, which
the HP methods promise to bound by $\delta$. This is closely related to the fraction of
replicates whose event rate falls above the dashed line in
Figure~\ref{fig:1}, but because each realized event rate is itself measured on
a potentially small selected set, it is better estimated using a beta-binomial
model \citep{skellam1948probability}, as detailed in Supplement
Section~\ref{app:numerical-risk-estim}. The table shows that 
$\P{r(A;\Dcal) > \alpha} \leq \delta = 0.1$ holds throughout for HP-Uniform and
HP-LTT but not for HP-Pointwise, which reaches $\approx 0.19$ at $t_0 = 3$
months.

\subsubsection{The Effect of the HP Significance Level $\delta$}

\begin{figure*}[!htbp]
\centering
\includegraphics[width=0.85\textwidth]{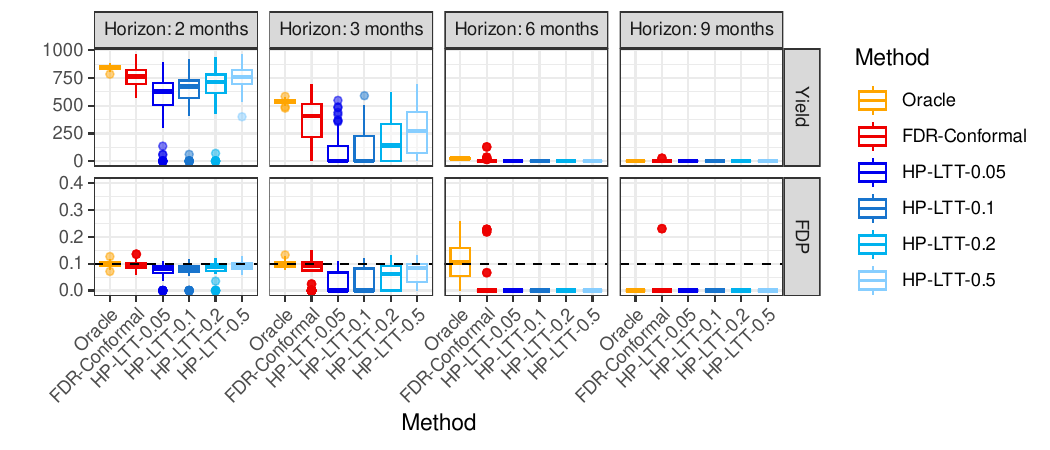}
\caption{
Summary of low-risk screening results on semi-synthetic data, as in Figure~\ref{fig:1}.
Here, HP-LTT is applied at different significance levels $\delta$, ranging from $\delta = 0.05$ (more conservative) to $\delta=0.5$ (more liberal).
}
\label{fig:2}
\end{figure*}

Figure~\ref{fig:2} illustrates how the significance level~$\delta$ affects the behavior of HP-LTT, in the same setting as Figure~\ref{fig:1}. 
More conservative significance levels (e.g.,~$\delta=0.05$) substantially reduce yield. In contrast, when applied with very liberal significance levels (e.g.,~$\delta=0.5$), HP-LTT performs more similarly to FDR-Conformal. The interpretation, however, differs. 
FDR-Conformal approximately controls the event rate among selected patients below~10\% on average, while HP-LTT with~$\delta=0.5$ aims to keep the event rate below~10\% in at least half of the replicates, saying nothing about how large it may be in the other half. Because the event rate is bounded below by~0\% but can substantially exceed the 10\% target, controlling its \emph{expected value} is more meaningful than controlling it in probability at a very liberal significance level such as~50\%.
Tables~\ref{tab:fig2-details-23} and~\ref{tab:fig2-details-69} in Supplement
Section~\ref*{app:numerical-sim} report numerical summaries at each
$\delta$ for all three HP approaches, including estimates of
 $\P{r(A;\Dcal) > \alpha}$. This is controlled
below $\delta$ by HP-Uniform and HP-LTT but not by HP-Pointwise.

\subsubsection{The Effect of Survival Model Misspecification}

\begin{figure*}[!htbp]
\centering
\includegraphics[width=0.8\textwidth]{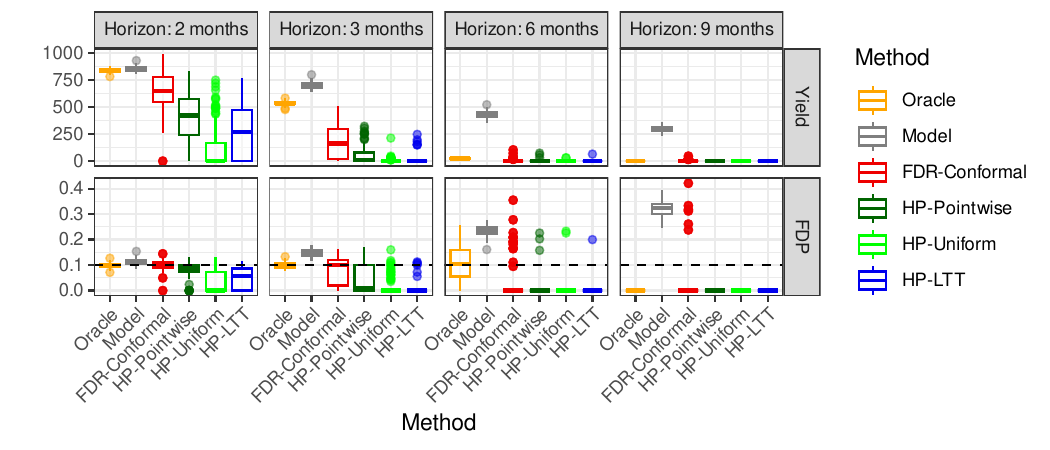}
\caption{
Summary of low-risk screening results obtained by applying different calibration methods to semi-synthetic data simulated using a random forest generative model, at different screening horizons. All methods use the same mis-specified gradient boosting survival model, which leads to higher-than-expected event rates at long horizons if applied to select low-risk patients without calibration. Other details as in Figure~\ref{fig:1}.
}
\label{fig:3}
\end{figure*}

Figure~\ref{fig:3} presents results from experiments analogous to those in
Figure~\ref{fig:1}, except that the survival model is now misspecified: a
gradient boosting model is fitted on data generated from a random forest-based
generative model. As before, the HP calibration methods use a
significance level of $\delta = 0.1$. In this setting, the uncalibrated model
selects too many patients and exceeds the target event rate at longer horizons.
By contrast, all calibrated methods keep the FDP below the 10\% target, though
their yields decline markedly as the time horizon $t_0$ increases; as in
Figure~\ref{fig:1}, the event rate among selected patients can be higher where
the yield is low. The relative performance ordering among valid calibration
methods remains consistent with the well-specified case, with FDR-Conformal
selecting the most patients on average. A numerical summary is 
in Table~\ref*{tab:semi-synthetic_xgb} in Supplement
Section~\ref*{app:numerical-sim}. Figure~\ref*{fig:semi-synthetic-n1000-cox-delta0.1}
and Table~\ref*{tab:semi-synthetic_cox} summarize qualitatively similar results
obtained using the Cox model instead of gradient boosting.

\subsubsection{The Effect of the Calibration Sample Size}

\begin{figure*}[!htbp]
\centering
\includegraphics[width=0.8\textwidth]{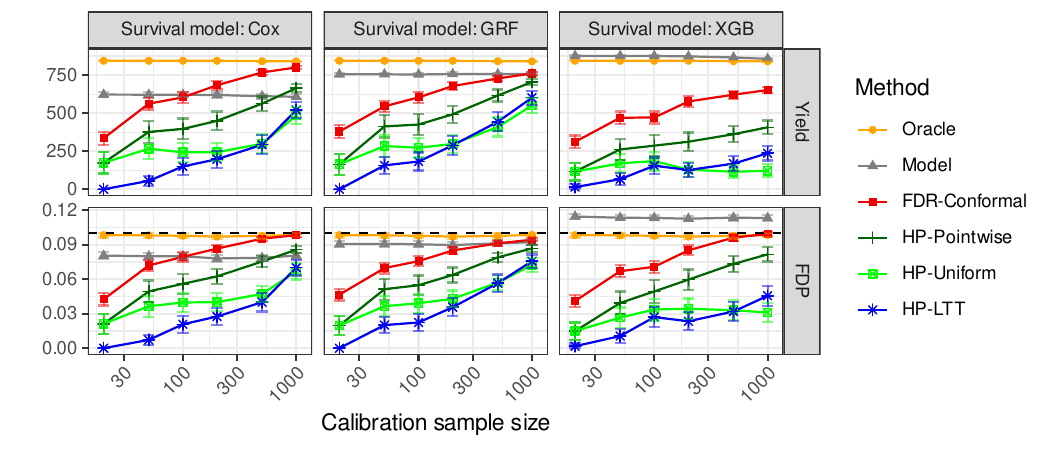}
\caption{
Average performance of low-risk screening methods on semi-synthetic data, as in Figure~\ref{fig:1}, at horizon $t_0=2$ months. The results are shown as a function of the calibration sample size, for different survival models. Error bars represent two standard errors.
}
\label{fig:4}
\end{figure*}

Figure~\ref{fig:4} illustrates how the yield of all methods tends to increase with the number of calibration samples. 
These results correspond to experiments similar to those in Figure~\ref{fig:1},
at $t_0 = 2$ and comparing three different survival models: Cox, generalized random
forest (GRF), and gradient boosting (XGB).
In this setting, FDR-Conformal performs quite similarly to the ideal oracle when
the calibration sample size is large enough, while HP-LTT suffers the steepest
decline in yield when calibration data are scarce, falling below HP-Uniform, as 
anticipated in Section~\ref{subsec:ltt}.

Additional results following the same layout of Figure~\ref{fig:1} but using a
smaller calibration sample size ($n_{\text{cal}} = 100$) are provided in
Figures~\ref*{fig:semi-synthetic-n100-grf-delta0.1}--\ref*{fig:semi-synthetic-n100-grf-delta}
and Table~\ref*{tab:semi-synthetic_grf_n100} in Supplement
Section~\ref*{app:numerical-sim}. These results highlight that variability
increases and the number of selected patients decreases as $n_{\text{cal}}$
becomes small. Despite this loss in precision, all procedures keep the FDP below
the target on average.

\subsubsection{The Advantage of AIPCW for Double Robustness}  \label{sec:exp-aipcw-dr}

In the semi-synthetic settings considered so far, the censoring mechanism is
relatively easy to estimate, and therefore all calibration methods perform very
similarly regardless of whether they use AIPCW (the default) or IPCW
pseudo-outcomes; e.g., see Figure~\ref{fig:semi-synthetic-dr-easy}.
To highlight the advantage of double robustness, we conduct additional
experiments in a more challenging adversarial setting, where censoring
depends sharply on a single key covariate and falls most heavily on the patients at
highest risk; see Supplement Section~\ref*{app:synthetic-dr} for a more detailed 
description of this setup.

\begin{figure*}[!htbp]
\centering
\includegraphics[width=0.8\textwidth]{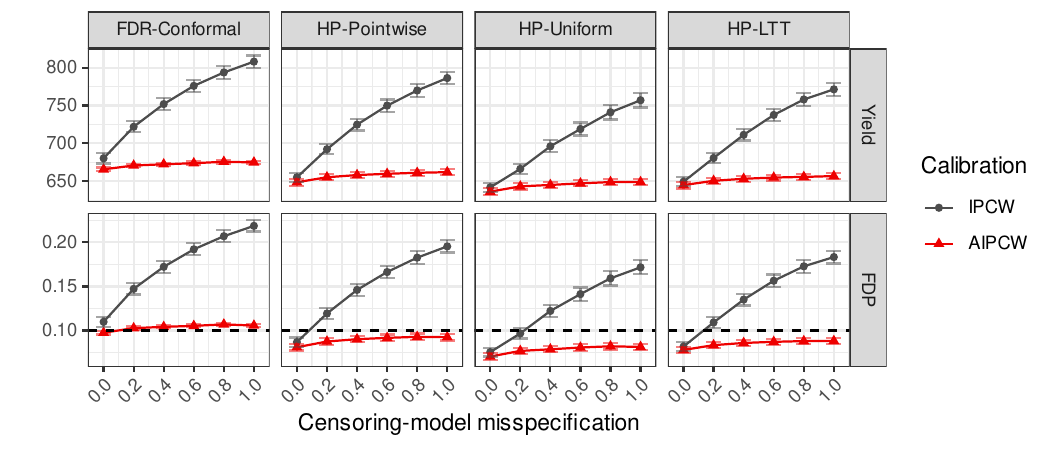}
\caption{
Empirical demonstration of the double robustness of calibration methods based
on AIPCW pseudo-outcomes, in an adversarial semi-synthetic setting constructed
so that misspecifying the censoring model is costly. Each panel contrasts one
method applied with AIPCW against the same method applied with IPCW
pseudo-outcomes, at horizon $t_0=2$ months. The horizontal axis represents the
fraction of training samples used to fit the censoring model that come from a
misspecified distribution: at $0$ the censoring model is correctly specified
and the two pseudo-outcomes are nearly indistinguishable, while at $1$ it is
badly misspecified and IPCW becomes invalid. AIPCW remains valid throughout,
because the survival model is correctly specified. All methods use the same
generalized random forest survival model. Other details as in
Figure~\ref{fig:4}.
}
\label{fig:5}
\end{figure*}

Figure~\ref{fig:5} compares the methods as a function of the mixing proportion
controlling the misspecification of the censoring model $\hat{G}$. The survival model is
fitted as usual on independent training data from the target distribution, while $\hat{G}$ is fitted on a separate
sample drawn from a mixture of the target distribution and an alternative one in
which the sharp dependence of the censoring time on the key covariate is removed; the mixing
proportion is the weight on that alternative distribution.
The results show that IPCW becomes anti-conservative as $\hat{G}$ degrades, 
while AIPCW remains valid.

\subsubsection{The Importance of Tuning the Time Shift Parameter}  \label{sec:exp-gamma}

We now examine empirically how the time shift $\gamma$ affects the performance
of the FDR-Conformal method of Section~\ref{sec:fdr}, which until now has been
tuned on the training data as described in Supplement Section~\ref*{app:tuning-gamma}. 
Figure~\ref{fig:gamma-tuning} compares the tuned
method against the same method applied with a range of fixed values,
$\gamma \in \{0, 1, 10, 100, 1000\}$, as a function of the calibration sample
size and otherwise in the setting of Figure~\ref{fig:4}.

At $\gamma = 0$ the method makes no selections once the calibration sample is
large, as anticipated by the theoretical analysis presented in Supplement Section~\ref{app:tuning-gamma}. 
At all sample sizes, yield tends to increase with $\gamma$ up to about $\gamma = 100$ and then falls again,
underscoring the importance of empirical tuning. The tuned method performs near optimally throughout.
Whenever $\gamma$ is well chosen, the yield grows with the calibration sample
size and the realized FDP approaches the 10\% target from below.

\begin{figure*}[!htbp]
\centering
\includegraphics[width=0.8\textwidth]{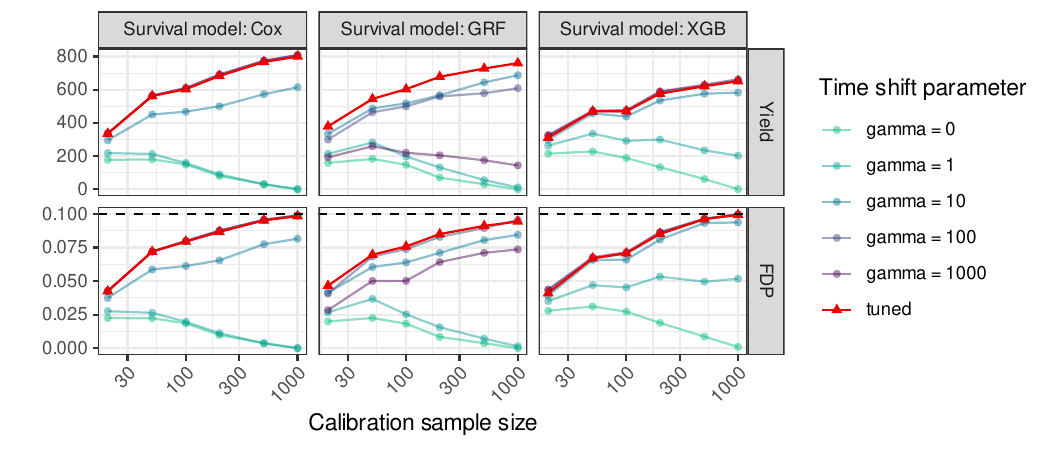}
\caption{
Effect of the time shift $\gamma$ on the performance of FDR-Conformal, as a
function of the calibration sample size. Faded lines show fixed values of
$\gamma$, shaded from green to purple as $\gamma$ increases; the tuned method
is drawn in red. All methods use AIPCW pseudo-outcomes. Other details as in
Figure~\ref{fig:4}.
}
\label{fig:gamma-tuning}
\end{figure*}

\section{Application to FHRD data}

We next apply the described calibration methods directly to the FHRD data described in Section~\ref{sec:data}, using the true observed times and censoring indicators. We randomly divide the 7{,}000 observations into a training set of 5{,}000, a calibration set of 1{,}000, and a test set of 1{,}000, and we average all reported results over 100 independent splits. Because the event times are only partially observed, we report surrogate lower and upper bounds of the event rates obtained by treating censored observations optimistically or pessimistically, respectively.
Aside from this adjustment, we follow the same analysis protocol used for the semi-synthetic experiments in Section~\ref{subsec:analysis-protocol}. In particular, we use a generalized random forest model to estimate the censoring distribution and evaluate three survival models (i.e., gradient boosting, the Cox model, and a generalized random forest) for estimating the event-time distribution.

Table~\ref{tab:real_xgb} summarizes the results obtained with a gradient boosting survival model; a visual summary is provided by Figure~\ref*{fig:real-n1000-xgb-delta0.1} in Supplement Section~\ref*{app:numerical-real}.
Overall, these findings mirror those from the semi-synthetic experiments. At the shorter horizons, FDR-Conformal and all HP procedures keep the FDP near or below the target 10\% level on average; at 9 months, where selections become rare, the event rate among selected patients rises above 10\% for FDR-Conformal and HP-Pointwise. FDR-Conformal consistently selects more patients compared to the HP calibration approaches. At longer horizons, selections become rarer and variability increases across all methods.
In sharp contrast to the calibrated procedures, the uncalibrated model benchmark selects too many patients, resulting in event rates well above the desired 10\% level, particularly at longer horizons.

\begin{table*}[!htb]
\centering
\caption{Summary of low-risk screening results on FHRD data.
Because of censoring, event rates cannot be evaluated exactly; instead, we report deterministic lower (LB) and upper (UB) bounds for both the FDP and the event rate among selected patients given at least one selection (Ev.rate).
We also report the fraction of trials with at least one selected patient, $P(|\hat{\mathcal{S}}|>0)$.
Each entry shows the mean and, in parentheses, two standard errors across repeated splits; a dash marks a conditional entry that is not reported because selections occur in fewer than 10\% of splits.
Values highlighted in red indicate event rates significantly above the 10\% target, while boldface denotes the highest yield among methods with valid event rates.}
\label{tab:real_xgb}
{\footnotesize
\input{tables_main/data_1_ntrain_5000_ncal_1000_xgb_aipcw.txt}
}
\end{table*}

Tables~\ref*{tab:real_cox}--\ref*{tab:real_grf} and Figures~\ref*{fig:real-n1000-cox-delta0.1}--\ref*{fig:real-n1000-grf-delta0.1} in Supplement Section~\ref*{app:numerical-real} show analogous results for the Cox model and the generalized random forest. The trends align with those in Table~\ref*{tab:real_xgb} and Figure~\ref*{fig:real-n1000-xgb-delta0.1}: the calibrated methods keep the FDP near the target level on average, with FDR-Conformal typically achieving the highest yield.
In these two settings, however, the uncalibrated model leads to a conservative rather than overly liberal screening rule.
Figures~\ref*{fig:real-n1000-xgb-delta}--\ref*{fig:real-n1000-grf-delta} in Supplement Section~\ref*{app:numerical-real} visualize the effect of changing the significance level $\delta$ used by HP-LTT, with results analogous to those reported in Figure~\ref{fig:2}.
Finally, Table~\ref{tab:real_xgb_ipcw} repeats Table~\ref{tab:real_xgb} with IPCW in place of AIPCW pseudo-outcomes. The results are nearly identical, consistent with Section~\ref{sec:experiments}, since the censoring distribution is easily estimated on these data.

\section{Discussion} \label{sec:discussion}

The methods discussed in this paper extend naturally to other problems. Firstly, they can be applied symmetrically to calibrate high-risk screening rules, by simply inverting the score inequality \citep{sesiavsvetnik2025}.
Moreover, they can be adapted to handle covariate shift via additional reweighting \citep{tibshirani2019conformal,park2021pac,jin2023model}. 
In the future, it would be interesting to extend the FDR-Conformal method using e-values \citep{wang2022false,vovk2023confidence,lee2024boosting}, a direction that could lead to lower algorithmic variability caused by random sample splitting \citep{NEURIPS2023_cec8ad77} and the possibility of incorporating additional constraints into the patient subset selection problem \citep{nair2025diversifying}.

\section*{Acknowledgments}
The authors thank Dr.~Peining Tao and Dr.~Michael Johnson for preparing the analytical dataset used in this study, and Dr.~Mehmet Burcu for suggesting the problem of predicting short-term survival of oncology patients.
The authors also thank two anonymous referees for their helpful feedback on an earlier version of this manuscript.

\section*{Data Availability Statement}
Software implementing the methods and experiments is available at \url{https://github.com/msesia/screening_calibration}, along with usage examples. The data analyzed in this paper were obtained from the Flatiron Health Research Database (\url{https://flatiron.com/database-characterization}) and are not publicly available.


%% file: tables_main/data_1_ntrain_5000_ncal_1000_xgb_aipcw.txt
\begin{tabular}[t]{lrrrrrr}
\toprule
Method & Yield & FDP (LB) & FDP (UB) & Ev.rate (LB) & Ev.rate (UB) & P($|S|>0$)\\
\midrule
\addlinespace[0.3em]
\multicolumn{7}{l}{\textbf{Horizon: 2 months}}\\
\hspace{1em}Model & 833 (2.99) & 0.09 (0.00) & 0.12 (0.00) & 0.09 (0.00) & 0.12 (0.00) & 1.00 (0.00)\\
\hspace{1em}FDR-Conformal & \textbf{874} (5.10) & 0.10 (0.00) & 0.12 (0.00) & 0.10 (0.00) & 0.12 (0.00) & 1.00 (0.00)\\
\hspace{1em}HP-Pointwise & 788 (13.80) & 0.09 (0.00) & 0.11 (0.00) & 0.09 (0.00) & 0.11 (0.00) & 1.00 (0.00)\\
\hspace{1em}HP-Uniform & 705 (22.74) & 0.08 (0.00) & 0.10 (0.00) & 0.08 (0.00) & 0.10 (0.00) & 1.00 (0.00)\\
\hspace{1em}HP-LTT & 740 (18.48) & 0.08 (0.00) & 0.11 (0.00) & 0.08 (0.00) & 0.11 (0.00) & 1.00 (0.00)\\
\addlinespace[0.3em]
\multicolumn{7}{l}{\textbf{Horizon: 3 months}}\\
\hspace{1em}Model & 711 (3.30) & \textcolor{red}{0.11 (0.00)} & 0.15 (0.00) & \textcolor{red}{0.11 (0.00)} & 0.15 (0.00) & 1.00 (0.00)\\
\hspace{1em}FDR-Conformal & \textbf{570} (9.28) & 0.10 (0.00) & 0.14 (0.00) & 0.10 (0.00) & 0.14 (0.00) & 1.00 (0.00)\\
\hspace{1em}HP-Pointwise & 446 (20.48) & 0.08 (0.00) & 0.12 (0.00) & 0.08 (0.00) & 0.12 (0.00) & 1.00 (0.00)\\
\hspace{1em}HP-Uniform & 195 (34.42) & 0.05 (0.01) & 0.08 (0.01) & 0.05 (0.01) & 0.10 (0.01) & 0.85 (0.07)\\
\hspace{1em}HP-LTT & 346 (32.46) & 0.07 (0.01) & 0.10 (0.01) & 0.08 (0.00) & 0.12 (0.00) & 0.87 (0.07)\\
\addlinespace[0.3em]
\multicolumn{7}{l}{\textbf{Horizon: 6 months}}\\
\hspace{1em}Model & 501 (3.54) & \textcolor{red}{0.17 (0.00)} & 0.24 (0.00) & \textcolor{red}{0.17 (0.00)} & 0.24 (0.00) & 1.00 (0.00)\\
\hspace{1em}FDR-Conformal & \textbf{140} (8.19) & 0.08 (0.00) & 0.15 (0.01) & 0.10 (0.00) & 0.18 (0.00) & 0.86 (0.03)\\
\hspace{1em}HP-Pointwise & 76 (11.94) & 0.06 (0.01) & 0.12 (0.02) & 0.08 (0.01) & 0.16 (0.01) & 0.75 (0.09)\\
\hspace{1em}HP-Uniform & 9 (3.59) & 0.02 (0.01) & 0.04 (0.01) & 0.05 (0.01) & 0.14 (0.02) & 0.29 (0.09)\\
\hspace{1em}HP-LTT & 17 (9.59) & 0.01 (0.01) & 0.02 (0.01) & 0.09 (0.01) & 0.17 (0.02) & 0.12 (0.07)\\
\addlinespace[0.3em]
\multicolumn{7}{l}{\textbf{Horizon: 9 months}}\\
\hspace{1em}Model & 392 (3.50) & \textcolor{red}{0.23 (0.00)} & 0.34 (0.00) & \textcolor{red}{0.23 (0.00)} & 0.34 (0.00) & 1.00 (0.00)\\
\hspace{1em}FDR-Conformal & 17 (3.21) & 0.04 (0.01) & 0.08 (0.01) & \textcolor{red}{0.12 (0.01)} & 0.26 (0.01) & 0.30 (0.05)\\
\hspace{1em}HP-Pointwise & 5 (2.85) & 0.02 (0.01) & 0.04 (0.02) & \textcolor{red}{0.13 (0.03)} & 0.25 (0.03) & 0.15 (0.07)\\
\hspace{1em}HP-Uniform & \textbf{2} (1.15) & 0.01 (0.01) & 0.02 (0.01) & -- & -- & 0.07 (0.05)\\
\hspace{1em}HP-LTT & 0 (0.00) & 0.00 (0.00) & 0.00 (0.00) & -- & -- & 0.00 (0.00)\\
\bottomrule
\end{tabular}

%% file: appendix.tex
\section{Additional Methodological Details}

\subsection{Calibration via High-Probability Risk Control} \label{app:high-probability}

\subsubsection{Pointwise Risk Estimation} \label{app:high-probability-estimation}

For a fixed threshold $\lambda$, recall that the population-level \emph{selected-set risk} is defined as
\[
r(\lambda) = \frac{\theta(\lambda)}{\mu(\lambda)},
\qquad
\theta(\lambda) = \EV{\I{T \le t_0}A_\lambda(X)},
\qquad
\mu(\lambda) = \mathbb{E}[A_\lambda(X)],
\]
where $A_\lambda(X)$ indicates whether a patient with covariates $X$ would be selected by the screening rule at threshold $\lambda$.
Using the calibration data, we estimate these quantities as
\begin{equation} \label{eq:generic-estimator}
  \hat r(\lambda) = \frac{\hat\theta(\lambda)}{\hat\mu(\lambda)},
  \qquad
  \hat\theta(\lambda) = \frac{1}{n}\sum_{i=1}^n A_\lambda(X_i)\,\hat\Psi_i,
  \qquad
  \hat\mu(\lambda) = \frac{1}{n}\sum_{i=1}^n A_\lambda(X_i),
\end{equation}
where $\hat\Psi_i$ is a per-subject pseudo-outcome not depending on $\lambda$.
All estimators considered here are of this form and differ only in the choice of $\hat\Psi_i$:
\begin{align} \label{eq:pseudo-outcomes}
  \hat\Psi_i^{\,\mathrm{et}}
    &= \frac{\mathbb I(\tilde T_i \le t_0,\, E_i=1)}{\hat G(\tilde T_i\mid X_i)},
  &&\text{(event-time IPCW, \eqref{eq:ipcw-et})} \nonumber \\
  \hat\Psi_i^{\,\mathrm{ft}}
    &= 1 - \frac{\mathbb I(\tilde T_i \ge t_0)}{\hat G(t_0\mid X_i)},
  &&\text{(fixed-time IPCW)} \\
  \hat\Psi_i^{\,\mathrm{dr}}
    &= \hat\Psi^{\mathrm{dr}}(O_i, t_0) \text{ from } \eqref{eq:aipcw-pseudo}.
  &&\text{(AIPCW)} \nonumber
\end{align}
The derivations below are stated for a generic $\hat\Psi_i$ and apply to all three.

\subsubsection{Asymptotic Bounds via Delta Method} \label{app:ipcw-delta-var}

We review here how to construct an approximate $(1-\delta)$ upper confidence bound for the true risk $r(\lambda)$ using the delta method.

Define the function $g(\theta,\mu) = \theta / \mu$, so that $\hat r(\lambda) = g(\hat\theta, \hat\mu)$.
The gradient of $g$ is
\[
\nabla g(\theta,\mu) =
\begin{pmatrix}
1/\mu \\[3pt]
-\theta/\mu^2
\end{pmatrix}.
\]
By the multivariate delta method, the asymptotic variance of $\hat r(\lambda)$ is approximated by
\[
\hat\sigma^2(\lambda)
=
\nabla g(\hat v)^\top
\hat\Sigma(\lambda)
\nabla g(\hat v),
\]
where $\hat v = (\hat\theta, \hat\mu)^\top$ and
\[
\hat\Sigma(\lambda)
=
\begin{pmatrix}
\hat\Sigma_{\theta\theta}(\lambda) & \hat\Sigma_{\theta\mu}(\lambda) \\[3pt]
\hat\Sigma_{\mu\theta}(\lambda) & \hat\Sigma_{\mu\mu}(\lambda)
\end{pmatrix}
\]
denotes the empirical covariance matrix of $(\hat\theta(\lambda), \hat\mu(\lambda))$.
Each block in $\hat\Sigma(\lambda)$ is estimated from individual sample contributions:
\[
\hat\Sigma_{ab}(\lambda)
=
\frac{1}{n}
\sum_{i=1}^n
\big(\psi_i^{(a)}(\lambda)-\bar\psi^{(a)}(\lambda)\big)
\big(\psi_i^{(b)}(\lambda)-\bar\psi^{(b)}(\lambda)\big),
\qquad a,b\in\{\theta,\mu\},
\]
where $\psi_i^{(\theta)}$ and $\psi_i^{(\mu)}$ represent each subject's contribution to the estimators of $\theta$ and $\mu$, centered by their sample means:
\[
\psi_i^{(\theta)}(\lambda)
=
A_\lambda(X_i)\,\hat\Psi_i - \hat\theta(\lambda),
\qquad
\psi_i^{(\mu)}(\lambda) = A_\lambda(X_i)-\hat\mu(\lambda).
\]

Expanding the expression for the asymptotic variance of $\hat r(\lambda)$ gives:
\[
\hat\sigma^2(\lambda)
=
\frac{\hat\Sigma_{\theta\theta}(\lambda)}{\hat\mu^2(\lambda)}
-2\,\frac{\hat\theta(\lambda)}{\hat\mu^3(\lambda)}\,\hat\Sigma_{\theta\mu}(\lambda)
+\frac{\hat\theta^2(\lambda)}{\hat\mu^4(\lambda)}\,\hat\Sigma_{\mu\mu}(\lambda),
\]
or equivalently, as a single sample variance,
\[
\hat\sigma^2(\lambda)
=
\frac{1}{n}
\sum_{i=1}^n
\Bigg(
\frac{\psi_i^{(\theta)}(\lambda)}{\hat\mu(\lambda)}
-
\frac{\hat\theta(\lambda)}{\hat\mu^2(\lambda)}\,\psi_i^{(\mu)}(\lambda)
\Bigg)^{\!2}.
\]
This last expression simplifies further. Since $\hat\theta(\lambda) = \hat r(\lambda)\hat\mu(\lambda)$, the two centering constants cancel:
\[
\frac{\psi_i^{(\theta)}(\lambda)}{\hat\mu(\lambda)}
-\frac{\hat\theta(\lambda)}{\hat\mu^2(\lambda)}\,\psi_i^{(\mu)}(\lambda)
=
\frac{A_\lambda(X_i)\hat\Psi_i - \hat\theta(\lambda)}{\hat\mu(\lambda)}
-\frac{\hat\theta(\lambda)\{A_\lambda(X_i)-\hat\mu(\lambda)\}}{\hat\mu^2(\lambda)}
=
\frac{A_\lambda(X_i)\big\{\hat\Psi_i - \hat r(\lambda)\big\}}{\hat\mu(\lambda)},
\]
which is the influence function $\phi_i(\lambda)$ defined in~\eqref{eq:influence-generic}. Hence
\begin{equation} \label{eq:sigma-generic}
  \hat\sigma^2(\lambda)
  = \frac{1}{n}\sum_{i=1}^n \phi_i(\lambda)^2
  = \frac{1}{n\,\hat\mu^2(\lambda)}\sum_{i=1}^n A_\lambda(X_i)\big\{\hat\Psi_i - \hat r(\lambda)\big\}^2,
\end{equation}
so that only selected subjects contribute, and the estimator enters solely through $\hat\Psi_i$. Then, substituting the
expressions for the pseudo-outcomes from~\eqref{eq:pseudo-outcomes} allows evaluating the corresponding variance estimate. The same influence contributions $\phi_i(\lambda)$ are used to construct the Gaussian multiplier band in Supplement Section~\ref{app:uniform}.

Finally, the $(1-\delta)$ one-sided upper confidence bound for the selected-set risk is
\[
\mathrm{UCB}^{\mathrm{pt}}(\lambda;\delta)
=
\hat r(\lambda)
+
z_{1-\delta}\,\frac{\hat\sigma(\lambda)}{\sqrt{n}},
\]
where $z_{1-\delta}$ is the $(1-\delta)$ quantile of the standard normal distribution.

In practice, we apply this method with mild winsorization of the weights, capping the top 1\% at the 99th percentile; for $\hat\Psi_i^{\,\mathrm{dr}}$ the same cap is applied to all three terms of~\eqref{eq:aipcw-pseudo}.
When $\hat{\mu}(\lambda)$ is extremely small (e.g., $< 15/n$), we revert to the more conservative finite-sample upper bound described in Appendix~\ref{app:ipcw-fs}, which is computed with $\hat\Psi_i^{\,\mathrm{et}}$ in all cases.

\FloatBarrier

\subsubsection{Asymptotic Bounds via Nonparametric Bootstrap} \label{app:bootstrap-ub}

As an alternative to the delta method, one may construct a nonparametric bootstrap upper confidence bound for $r(\lambda)$ by resampling the calibration data as outlined by Algorithm~\ref{alg:bootstrap-ub}.
The bootstrap also relies on large-sample asymptotics; therefore, if $\hat{\mu}(\lambda)$ is very small (e.g., $< 15/n$), we recommend reverting to the more conservative finite-sample upper bound described in Appendix~\ref{app:ipcw-fs}.

\begin{algorithm}[H]
\caption{Nonparametric bootstrap pointwise upper confidence bound.}
\label{alg:bootstrap-ub}
\begin{algorithmic}[1]
\INPUT Pre-trained screening rule $A_\lambda : \mathcal{X} \mapsto \{0,1\}$, calibration data $\mathcal D_{\mathrm{cal}}=\{(X_i,\tilde T_i,E_i)\}_{i=1}^n$; fitted survival and censoring models $\hat S(\cdot\mid X)$, $\hat G(\cdot\mid X)$; number of bootstrap replicates $B$; significance level $\delta$; pseudo-outcome flavor ${\tt et}$, ${\tt ft}$ or ${\tt dr}$.
\STATE Compute the appropriate $\hat\Psi_i$ as in~\eqref{eq:pseudo-outcomes} for $i \in [n]$, with winsorization.
\FOR{$b = 1,\dots,B$}
  \STATE Resample $n$ pairs $\{(X_i^{\ast(b)}, \hat\Psi_i^{\ast(b)})\}_{i=1}^n$ with replacement from $\{(X_i,\hat\Psi_i)\}_{i=1}^n$.
  \STATE Recompute $\hat\theta^{\ast(b)}(\lambda)$ and $\hat\mu^{\ast(b)}(\lambda)$ as in~\eqref{eq:generic-estimator}.
  \STATE Form the bootstrap replicate
  $\hat r^{\ast(b)}(\lambda)
  = \hat\theta^{\ast(b)}(\lambda)/\hat\mu^{\ast(b)}(\lambda)$.
\ENDFOR
\STATE Compute
\begin{align} \label{eq:boot-ucb}
\mathrm{UCB}^{\mathrm{boot}}(\lambda;\delta)
=
\text{Quantile}_{1-\delta}\!\big(\{\hat r^{\ast(b)}(\lambda)\}_{b=1}^B\big).
\end{align}
\STATE If $n_{\mathrm{sel}}(\lambda)=\sum_i A_\lambda(X_i)$ is small (e.g., $<15$),
replace $\mathrm{UCB}^{\mathrm{boot}}(\lambda;\delta)$ with a finite-sample bound.
\OUTPUT $(1-\delta)$ pointwise upper confidence bound $\mathrm{UCB}^{\mathrm{boot}}(\lambda;\delta)$ for $r(\lambda)$.
\end{algorithmic}
\end{algorithm}

Note that the pseudo-outcomes are computed once, in step 1, from the models fitted on the independent training sample, and are then resampled together with the corresponding covariates; the survival and censoring models are not refitted within the bootstrap loop. This reflects the conditional perspective adopted throughout, in which $\hat S$ and $\hat G$ are treated as fixed.

\FloatBarrier

\subsubsection{Finite-Sample Bounds via Empirical Bernstein Inequality} \label{app:ipcw-fs}

When the number of selected samples is very small, asymptotic approximations such as the
delta-method or bootstrap can no longer be justified.
In such cases, we can construct a finite-sample upper confidence bound for $r(\lambda) = \theta(\lambda) / \mu(\lambda)$ by combining
an empirical-Bernstein concentration inequality for the numerator $\theta(\lambda)$ with an exact Clopper-Pearson lower bound for the selection fraction $\mu(\lambda)$. For simplicity, we focus on the IPCW-ET approach.

With the notation from Section~\ref{app:high-probability-estimation}, write $Z_i(\lambda) = A_\lambda(X_i)\, \hat\Psi_i^{\,\mathrm{et}}$, so that $\hat\theta(\lambda) = \frac{1}{n}\sum_{i=1}^n Z_i$. By Assumption (A2), $\hat\Psi_i^{\,\mathrm{et}} \leq M := 1/c$ almost-surely.
In practice, the constant $M$ can be estimated empirically using the maximum of the relevant weights in the calibration sample, after winsorization.
The sample variance of the weighted contributions is given by
\[
\hat v(\lambda)
=
\frac{1}{n-1}
\sum_{i=1}^n
\big(Z_i(\lambda) - \hat\theta(\lambda)\big)^2.
\]
The empirical-Bernstein inequality of \citet{maurer2009empirical} then implies that
\begin{align} \label{eq:bernstein-theta}
\theta(\lambda)
\leq
\hat\theta(\lambda)
+
\sqrt{\frac{2\,\hat v(\lambda)\,\log(4/\delta)}{n}}
+
\frac{7M\,\log(4/\delta)}{3\,(n-1)}
\qquad \text{with probability $\geq 1-\frac{\delta}{2}$}.
\end{align}

For the denominator, note that $n\hat\mu(\lambda) = \sum_{i=1}^n A_\lambda(X_i)$
follows a $\mathrm{Binomial}(n, \mu(\lambda))$ distribution.
An exact $(1-\delta/2)$ lower confidence bound for $\mu(\lambda)$ is therefore
\begin{align} \label{eq:cp-mu}
\hat\mu_{\mathrm{low}}(\lambda;\delta/2)
=
\Psi_{\mathrm{low}}\!\big(n,\,n\hat\mu(\lambda),\,\delta/2\big),
\end{align}
where $\Psi_{\mathrm{low}}(n,k,\delta)$ is defined implicitly by
\[
\texttt{pbinom}\!\left(k-1;\,n,\,\Psi_{\mathrm{low}}(n,k,\delta)\right)
= 1-\delta,
\]
and \texttt{pbinom} denotes the cumulative distribution function of the binomial distribution.

Combining the numerator bound~\eqref{eq:bernstein-theta}
and the denominator bound~\eqref{eq:cp-mu} yields a finite-sample
$(1-\delta)$ one-sided upper confidence bound for the selected-set risk:
\begin{align} \label{eq:ucb-fs}
\mathrm{UCB}^{\mathrm{fs}}(\lambda;\delta)
=
\frac{\hat\theta_{\mathrm{upp}}(\lambda;\delta/2)}
     {\hat\mu_{\mathrm{low}}(\lambda;\delta/2)},
\end{align}
where $\hat\theta_{\mathrm{upp}}(\lambda;\delta/2)$
denotes the empirical-Bernstein upper bound in~\eqref{eq:bernstein-theta}.
In practice, we recommend applying $\mathrm{UCB}^{\mathrm{fs}}$
whenever $n_{\mathrm{sel}}(\lambda) = \sum_i A_\lambda(X_i) < 15$.

\FloatBarrier

\subsubsection{Greedy Pointwise Calibration} \label{app:hp-greedy}

\begin{algorithm}[H]
\caption{Greedy pointwise calibration}
\label{alg:greedy}
\begin{algorithmic}[1]
\INPUT Pre-trained screening rule $A_\lambda : \mathcal{X} \mapsto \{0,1\}$ with parameter $\lambda$, calibration data $\mathcal D_{\mathrm{cal}}=\{(X_i,\tilde T_i,E_i)\}_{i=1}^n$; survival model $\hat S$; censoring model $\hat G$; target risk level $\alpha\in(0,1)$; significance level $\delta\in(0,1)$; grid $\Lambda=\{\lambda_1,\ldots,\lambda_L\}\subset[0,1]$; winsorization level $\tau$ (e.g., $1\%$); flavor $\texttt{dr} \in \{\textsc{false}, \textsc{true}\}$.
\STATE Compute winsorized IPCW weights $\hat w_i = 1/\hat G(\tilde T_i \mid X_i)$.
\IF{$\texttt{dr} = \textsc{false}$}
  \STATE For $i \in [n]$, compute pseudo-outcomes $\hat\Psi_i$ with the IPCW form in~\eqref{eq:ipcw-et}.
\ELSE
  \STATE For $i \in [n]$, compute pseudo-outcomes $\hat\Psi_i$ with the AIPCW form in~\eqref{eq:aipcw-pseudo}.
\ENDIF
  \STATE Compute $\hat\theta(\lambda)=\frac{1}{n}\sum_i A_\lambda(X_i)\hat\Psi_i$, $\hat\mu(\lambda)=\frac{1}{n}\sum_i A_\lambda(X_i)$ and $\hat r(\lambda) = \hat\theta(\lambda) / \hat\mu(\lambda)$ for all $\lambda\in\Lambda$.
  \STATE Compute $\mathrm{UCB}^{\mathrm{pt}}(\lambda;\delta)$ (e.g., delta–method or finite-sample if $n_{\text{sel}}(\lambda)=\sum_i A_\lambda(X_i) \leq 15$).
\STATE Define the feasible set $\mathcal F \leftarrow \{\lambda\in\Lambda:\ \mathrm{UCB}^{\mathrm{pt}}(\lambda;\delta)\le \alpha\}$.
\STATE If $\mathcal F=\emptyset$, set $\hat\lambda\leftarrow +\infty$. Otherwise choose $\hat\lambda \leftarrow \min\{\lambda\in\mathcal F\}$.
\OUTPUT Calibrated screening rule $A_{\hat\lambda}$.
\end{algorithmic}
\end{algorithm}

\FloatBarrier

\subsubsection{Uniform Calibration} \label{app:uniform}

The simplest way to obtain a UCB for $r(\lambda)$ that holds simultaneously with high probability across all $\lambda \in \Lambda$ is to apply the pointwise estimation methods from Section~\ref{subsec:pointwise-estimation} with a Bonferroni correction.
Specifically, the pointwise significance level $\delta$ is replaced by the more conservative value $\delta / L$, where $L = |\Lambda|$.
The calibrated threshold is then chosen as
\[
\hat\lambda_{\text{Bonferroni}} =
\min\ \big\{\ \lambda\in\Lambda  :\
\mathrm{UCB}^{\mathrm{pt}}(\lambda;\delta/|\Lambda|)\le \alpha\ \big\},
\]
where $\mathrm{UCB}^{\mathrm{pt}}(\lambda;\delta)$ represents a level-$\delta$ pointwise upper confidence bound for $r(\lambda)$, computed with any of the methods from Section~\ref{subsec:pointwise-estimation}.
This Bonferroni correction guarantees $\mathbb{P}[r(\hat\lambda_{\text{Bonferroni}}) > \alpha] \leq \delta$, but tends to be overly conservative in practice because it does not account for the often strong dependencies between the estimation problems corresponding to different thresholds $\lambda \in \Lambda$.

A less conservative alternative is the Gaussian multiplier (perturbation) approach \citep{chernozhukov2013gaussian}, which constructs a joint confidence band by simulating correlated random fluctuations of the estimated risk across all $\lambda \in \Lambda = \{\lambda_1, \ldots, \lambda_L\}$.
This method builds on the same influence-function linearization that underlies the delta-method variance estimator (see Appendix~\ref{app:ipcw-delta-var} for additional details): for each individual $i$, the contribution $\phi_i(\lambda_l)$ from~\eqref{eq:influence-generic} approximates the influence of observation $i$ on $\hat{r}(\lambda_l)$, in the sense that
\[
  \sqrt n \,\{\hat r(\lambda)-r(\lambda)\}
  \;\approx\;
  \frac{1}{\sqrt n}\sum_{i=1}^{n}\phi_i(\lambda) .
\]

To approximate the joint sampling variability of $\{\hat{r}(\lambda_l)\}_{l=1}^L$, we repeatedly simulate random perturbations of these influence contributions.
Let $\xi_i^{(b)} \stackrel{\text{i.i.d.}}{\sim} \mathcal{N}(0,1)$, sampled independently across replicates $b = 1, \ldots, B$ and individuals $i = 1, \ldots, n$, and form the studentized multiplier process
\begin{equation} \label{eq:mult-boot}
Z^{(b)}(\lambda)
=
\frac{\frac{1}{\sqrt{n}}\sum_{i=1}^n \xi_i^{(b)}\,\phi_i(\lambda)}
     {\widehat{\mathrm{sd}}(\lambda)} ,
\end{equation}
where $\widehat{\mathrm{sd}}(\lambda)$ estimates the standard deviation of the
influence contributions at threshold $\lambda$.
For numerical stability, to prevent random fluctuations from making the denominator too small, we take $\widehat{\mathrm{sd}}(\lambda)$ to be a 95\% upper confidence bound for that
standard deviation rather than the usual empirical standard deviation
$\hat\sigma(\lambda)$ from~\eqref{eq:sigma-generic}.

Since only an upper bound on $r(\lambda)$ is required, we record the one-sided
maximum fluctuation across thresholds,
\[
  T^{(b)} = \max_{l \le L} Z^{(b)}(\lambda_l) ,
\]
and let $q_{1-\delta}$ denote the empirical $(1-\delta)$ quantile of
$\{T^{(b)}\}_{b=1}^{B}$.
This yields the simultaneous upper confidence bound
\[
\mathrm{UCB}^{\mathrm{GM}}(\lambda_l;\delta)
= \hat r(\lambda_l) + q_{1-\delta}\,
  \frac{\widehat{\mathrm{sd}}(\lambda_l)}{\sqrt{n}},
\qquad l=1,\ldots,L,
\]
truncated to $[0,1]$, and the calibrated threshold is chosen as
\begin{align*}
  \hat\lambda_{\text{GM}} = \min \big\{\lambda\in\Lambda : \mathrm{UCB}^{\mathrm{GM}}(\lambda;\delta)\le \alpha \big\}.
\end{align*}
Because all thresholds share the same simulated perturbations, this approach
automatically captures the correlation between neighboring $\hat{r}(\lambda)$
values, while studentization prevents the few noisiest thresholds from dictating
the width of the band everywhere.
As a result, these uniform bounds are often substantially tighter than
Bonferroni bounds when $\hat{r}(\lambda)$ varies smoothly with $\lambda$.

Finally, for thresholds at which fewer than 15 patients are selected the
asymptotic approximation underlying the Gaussian multiplier bootstrap is not
justified. We therefore build the band only on
$\Lambda_+ := \{\lambda\in\Lambda : n\hat\mu(\lambda)\ge 15\}$. Elsewhere one
would in principle revert to the finite-sample bound of
Appendix~\ref{app:ipcw-fs}, as in the pointwise setting of
Appendix~\ref{app:ipcw-delta-var}; but that bound would here have to be corrected
for multiplicity across the grid, which makes it vacuous. We therefore simply set
$\mathrm{UCB}^{\mathrm{uni}}(\lambda;\delta)=1$ for $\lambda\notin\Lambda_+$, so
that such thresholds are never selected. Algorithm~\ref{alg:uniform} summarizes this approach.

\begin{algorithm}[H]
\caption{Uniform calibration}
\label{alg:uniform}
\begin{algorithmic}[1]
\INPUT Pre-trained screening rule $A_\lambda : \mathcal{X}\!\mapsto\!\{0,1\}$ with parameter $\lambda$, calibration data $\mathcal D_{\mathrm{cal}}=\{(X_i,\tilde T_i,E_i)\}_{i=1}^n$; survival model $\hat S$; censoring model $\hat G$; target risk level $\alpha$; significance level $\delta$; grid $\Lambda=\{\lambda_1,\ldots,\lambda_L\}$; band type ${\tt band}\in\{\texttt{bonferroni},\texttt{multiplier}\}$; number of draws $B$; winsorization level $\tau$; flavor $\texttt{dr} \in \{\textsc{false}, \textsc{true}\}$.
\STATE Compute winsorized IPCW weights $\hat w_i = 1/\hat G(\tilde T_i \mid X_i)$.
\IF{$\texttt{dr} = \textsc{false}$}
  \STATE For $i \in [n]$, compute pseudo-outcomes $\hat\Psi_i$ with the IPCW form in~\eqref{eq:ipcw-et}.
\ELSE
  \STATE For $i \in [n]$, compute pseudo-outcomes $\hat\Psi_i$ with the AIPCW form in~\eqref{eq:aipcw-pseudo}.
\ENDIF
\STATE Compute $\hat\theta(\lambda_l)=\frac{1}{n}\sum_i A_{\lambda_l}(X_i)\hat\Psi_i$, $\hat\mu(\lambda_l)=\frac{1}{n}\sum_i A_{\lambda_l}(X_i)$, and $\hat r(\lambda_l)=\hat\theta(\lambda_l)/\hat\mu(\lambda_l)$ for all $\lambda_l\in\Lambda$.
\STATE Split the grid into $\Lambda_+ \leftarrow \{\lambda_l\in\Lambda : n\hat\mu(\lambda_l)\ge 15\}$ and $\Lambda_-\leftarrow\Lambda\setminus\Lambda_+$.
\STATE Construct uniform upper confidence bounds on $\Lambda_+$:
\begin{itemize}
\item \texttt{bonferroni}: $\mathrm{UCB}^{\mathrm{uni}}(\lambda_l;\delta)\leftarrow \mathrm{UCB}^{\mathrm{pt}}(\lambda_l;\delta/|\Lambda_+|)$.
\item \texttt{multiplier}: compute the influence contributions $\phi_i(\lambda_l)$ of~\eqref{eq:influence-generic} evaluated at $\hat\Psi_i$, and the studentization factors $\widehat{\mathrm{sd}}(\lambda_l)$ (95\% upper confidence bounds for the standard deviation of $\phi_i(\lambda_l)$); draw $\xi_i^{(b)}\!\sim\!\mathcal N(0,1)$ for $b\in[B]$ and compute $Z^{(b)}(\lambda_l)=\{\sqrt n\,\widehat{\mathrm{sd}}(\lambda_l)\}^{-1}\sum_i \xi_i^{(b)}\phi_i(\lambda_l)$; set $q_{1-\delta}$ to the $(1-\delta)$ quantile of $T^{(b)}=\max_{\lambda_l\in\Lambda_+} Z^{(b)}(\lambda_l)$, and $\mathrm{UCB}^{\mathrm{uni}}(\lambda_l;\delta)=\hat r(\lambda_l)+q_{1-\delta}\,\widehat{\mathrm{sd}}(\lambda_l)/\sqrt{n}$.
\end{itemize}
\STATE For $\lambda_l\in\Lambda_-$, set $\mathrm{UCB}^{\mathrm{uni}}(\lambda_l;\delta)\leftarrow 1$, the vacuous bound.
\STATE Define the feasible set $\mathcal F \leftarrow \{\lambda_l \in\Lambda:\ \mathrm{UCB}^{\mathrm{uni}}(\lambda_l;\delta)\le \alpha\}$.
\STATE If $\mathcal F=\emptyset$, set $\hat\lambda\leftarrow +\infty$; otherwise $\hat\lambda \leftarrow \min \{\lambda\in\mathcal F\}$.
\OUTPUT Calibrated screening rule $A_{\hat\lambda}$.
\end{algorithmic}
\end{algorithm}

\FloatBarrier
\clearpage

\subsubsection{Learn-then-Test (LTT) Calibration} \label{app:ltt}

\begin{algorithm}[H]
\caption{Learn--Then--Test (LTT) calibration}
\label{alg:ltt}
\begin{algorithmic}[1]
\INPUT Pre-trained screening rule $A_\lambda:\mathcal{X}\!\mapsto\!\{0,1\}$; calibration data $\mathcal D_{\mathrm{cal}}=\{(X_i,\tilde T_i,E_i)\}_{i=1}^n$; survival model $\hat S$; censoring model $\hat G$; target risk level $\alpha$; significance level $\delta$; winsorization level $\tau$; flavor $\texttt{dr} \in \{\textsc{false}, \textsc{true}\}$.
\STATE Compute calibration scores $Z_i=\hat S(t_0\mid X_i)$, and winsorized IPCW weights $\hat w_i = 1/\hat G(\tilde T_i \mid X_i)$.
\IF{$\texttt{dr} = \textsc{false}$}
  \STATE For $i \in [n]$, compute pseudo-outcomes $\hat\Psi_i$ with the IPCW form in~\eqref{eq:ipcw-et}.
\ELSE
  \STATE For $i \in [n]$, compute pseudo-outcomes $\hat\Psi_i$ with the AIPCW form in~\eqref{eq:aipcw-pseudo}.
\ENDIF
\STATE Define anchor thresholds $\lambda^{(1)}_0=1-\alpha$ and $\lambda^{(2)}_0=1-\alpha/2$.
For each anchor $k \in \{1,2\}$, construct sequence $\Lambda^{(k)}=\{\lambda^{(k)}_0>\lambda^{(k)}_1>\cdots>\lambda^{(k)}_L\}$ by setting $\lambda^{(k)}_j$ to successive lower quantiles of $\{Z_i\}$ below $\lambda^{(k)}_0$ (e.g., 95th, 92nd, 90th, $\ldots$ percentiles).
\STATE Set per-path significance level $\delta_\star=\delta/2$.
\FOR{$k=1,2$}
  \STATE Initialize $\hat\lambda_{\mathrm{LTT}}^{(k)}\leftarrow +\infty$.
  \FOR{$\lambda\in\Lambda^{(k)}$ (in decreasing order)}
    \STATE Compute pointwise upper confidence bound $\mathrm{UCB}^{\mathrm{pt}}(\lambda;\delta_\star)$ for $r(\lambda)$ from $\{(A_\lambda(X_i),\hat\Psi_i)\}_{i=1}^n$.
    \IF{$\mathrm{UCB}^{\mathrm{pt}}(\lambda;\delta_\star)\le\alpha$}
      \STATE Set $\hat\lambda_{\mathrm{LTT}}^{(k)}\leftarrow\lambda$.
    \ELSE
      \STATE \textbf{break}
    \ENDIF
  \ENDFOR
\ENDFOR
\STATE Set $\hat\lambda_{\mathrm{LTT}}=\min\{\hat\lambda_{\mathrm{LTT}}^{(1)},\hat\lambda_{\mathrm{LTT}}^{(2)}\}$.
\OUTPUT Calibrated screening rule $A_{\hat\lambda_{\mathrm{LTT}}}$.
\end{algorithmic}
\end{algorithm}

\FloatBarrier

\subsection{Efficient Computation of the Augmented Pseudo-Outcomes}
\label{app:aipcw-implementation}

This section explains how to evaluate the AIPCW pseudo-outcomes
$\hat\Psi^{\mathrm{dr}}(O_i, t)$ of~\eqref{eq:aipcw-pseudo} in practice, at
either the fixed horizon $t_0$ required by the risk estimator~\eqref{eq:aipcw} or
the individual-specific horizons $b_z(X_i)$ required by the augmented conformal
$p$-values of Section~\ref{sec:fdr-augmented}. The latter case requires more
care: $b_z(X_i)$ varies with both the calibration individual $i$ and the
candidate threshold $z$, so a naive implementation would re-evaluate the integral
in~\eqref{eq:aipcw-pseudo} from scratch for every pair. We show below that a
single $O(nK)$ preprocessing pass over the calibration data reduces each
subsequent evaluation to $O(1)$, where $K$ is the number of time grid points at which
the fitted models are evaluated.
The total cost is $O(nK + n \log n)$ for the risk
estimator, which needs the $n$ pseudo-outcomes at the single horizon $t_0$, and
$O(nK + nm + m \log m)$ for the conformal $p$-values.

We consider fitted models $\hat S(\cdot\mid x)$ and $\hat G(\cdot\mid x)$ that
are right-continuous step functions on a time grid $0 = u_0 < u_1 < \cdots < u_K = t_0$.
Both $\hat{S}$ and $\hat G$ are winsorized.
Write $\hat G_k(x) = \hat G(u_k\mid x)$, $\hat S_k(x) = \hat S(u_k\mid x)$, and
\begin{equation} \label{eq:discrete-hazard}
  \Delta\hat\Lambda^{C}_k(x) = 1 - \frac{\hat G_k(x)}{\hat G_{k-1}(x)},
  \qquad
  \hat Q_k(x, t) = \frac{\hat S_k(x) - \hat S(t \mid x)}{\hat S_k(x)},
  \qquad k = 1,\dots,K,
\end{equation}
so that $\hat Q(u, t \mid x) = \hat Q_{k-1}(x,t)$ for $u \in (u_{k-1}, u_k]$, by
the left-limit convention in~\eqref{eq:qhat}.

On the grid, $\hat Q$ is thus evaluated through $\hat S(u^{-}\mid x)$, whereas
$\hat G$ is evaluated at $u$ itself, so that the atom of $\hat\Lambda^{C}$ at
$u_k$ carries the weight $\hat G_k(x)^{-1}$.
 Letting
$k_i(t) = \min\{k : \tilde T_i \wedge t \le u_k\}$ index the grid interval
containing $\tilde T_i \wedge t$, and
$\ell_i(t) = \#\{k \ge 1 : u_k \le \tilde T_i \wedge t\}$ the number of atoms of
$\hat\Lambda^{C}$ in $(0, \tilde T_i \wedge t]$, the expression for
$\hat\Psi^{\mathrm{dr}}(O_i, t)$ from~\eqref{eq:aipcw-pseudo} becomes
\begin{equation} \label{eq:aipcw-discrete}
  \hat\Psi^{\mathrm{dr}}(O_i, t)
  =
  \underbrace{\frac{E_i\,\mathbb I(\tilde T_i \le t)}{\hat G(\tilde T_i\mid X_i)}}_{\text{(a)}}
  +
  \underbrace{(1-E_i)\,\mathbb I(\tilde T_i \le t)\,
    \frac{\hat Q_{k_i(t)-1}(X_i, t)}{\hat G(\tilde T_i \mid X_i)}}_{\text{(b)}}
  \;-\;
  \underbrace{\sum_{k=1}^{\ell_i(t)} \frac{\hat Q_{k-1}(X_i, t)}{\hat G_{k}(X_i)}\,\Delta\hat\Lambda^{C}_k(X_i)}_{\text{(c)}} .
\end{equation}
Note that $\ell_i(t) = k_i(t) - 1$ unless $\tilde T_i \wedge t = u_{k_i(t)}$
exactly, in which case $\ell_i(t) = k_i(t)$; the latter occurs for every subject
with $\tilde T_i \ge t_0$ when $t = t_0$.

Evaluating~\eqref{eq:aipcw-discrete} directly requires $O(K)$ operations per
pair $(i,t)$, because of the sum in (c).
However, multiple evaluations can be made more efficient.
Since $\hat Q_k(x,t)$ is \emph{affine} in $\hat S(t\mid x)$ by~\eqref{eq:discrete-hazard},
the term (c) splits into
\begin{align*}
  \sum_{k=1}^{\ell_i(t)} \frac{\hat Q_{k-1}(X_i, t)}{\hat G_{k}(X_i)}\,\Delta\hat\Lambda^{C}_k(X_i)
  & = \sum_{k=1}^{\ell_i(t)} \frac{\Delta\hat\Lambda^{C}_k(X_i)}{\hat G_{k}(X_i)}
    - \hat S(t \mid X_i) \sum_{k=1}^{\ell_i(t)} \frac{\Delta\hat\Lambda^{C}_k(X_i)}{\hat S_{k-1}(X_i) \, \hat G_{k}(X_i)} \\
  & = A_{\ell_i(t)}(X_i) - \hat S(t \mid X_i) \, B_{\ell_i(t)}(X_i),
\end{align*}
where the prefix sums, for $\ell = 0,\dots,K$,
\begin{equation} \label{eq:prefix-sums}
  A_\ell(x) := \sum_{k=1}^{\ell} \frac{\Delta\hat\Lambda^{C}_k(x)}{\hat G_{k}(x)}
             = \frac{1}{\hat G_\ell(x)} - 1 ,
  \qquad
  B_\ell(x) := \sum_{k=1}^{\ell}
    \frac{\Delta\hat\Lambda^{C}_k(x)}{\hat G_{k}(x)\,\hat S_{k-1}(x)},
\end{equation}
do not depend on $t$. The closed form for $A_\ell$ follows because
$\Delta\hat\Lambda^{C}_k/\hat G_k = \hat G_k^{-1} - \hat G_{k-1}^{-1}$
telescopes, with $\hat G_0 \equiv 1$.

Consequently the whole computation proceeds as follows. In a single $O(nK)$ pass
over the calibration data we form $\Delta\hat\Lambda^{C}_k(X_i)$ and the prefix
sums~\eqref{eq:prefix-sums} for every $i$ and $\ell$. Thereafter each
evaluation of~\eqref{eq:aipcw-discrete} requires only constant numbers of table lookups and arithmetic operations,
so computing the $p$-values at all $m$
candidate thresholds costs $O(nm)$ rather than $O(nmK)$. The index $\ell_i(b_z(X_i))$
is likewise obtained without search: since $b_z(x)$ is non-increasing in $z$,
sweeping the candidate thresholds in decreasing order---pre-sorting them costs $O(m \log m)$---lets each pointer advance
monotonically, for a total of $O(K)$ steps per calibration individual.
Thus, the overall cost is $O(nK + nm + m \log m)$ for the conformal $p$-values.

At the fixed horizon $t = t_0$ the computation is even simpler: the pseudo-outcomes
$\hat\Psi^{\mathrm{dr}}(O_i,t_0)$ do not depend on the screening threshold
$\lambda$, so the risk estimate over the whole threshold grid is obtained by
sorting subjects by $\hat z(X_i)$ at cost $O(n \log n)$,
computing the $n$ pseudo-outcomes at cost $O(nK)$, and forming a single cumulative sum.
Thus, the overall cost is $O(nK + n \log n)$ for the risk estimator.

\FloatBarrier

\subsection{Calibration via FDR Control and Conformal Inference} \label{app:tuning-gamma}

\subsubsection{The Role of the Time Shift Parameter}  \label{app:gamma-role}

This section explains why the choice of $\gamma$ in the
nonconformity score function $R(x,t) = \hat{S}(t + \gamma \mid x) \I{t \le t_0}$
from~\eqref{eq:conformal-score-R} is crucial for the power of the
FDR--conformal screening method. In particular, we show that with $\gamma = 0$ the procedure
tends to select either every subject or none, depending only on whether the mean risk
of the whole cohort meets the target.
By contrast, a (large) positive time shift $\gamma>0$ is precisely what allows it to select low-risk subgroups. 

To isolate the role of $\gamma$ from other complications like estimation errors, finite-sample variation, and censoring, 
in this section we study an idealized problem where the survival model is perfectly well specified ($\hat{S} = S$) 
and the IPCW ($\hat{p}(z)$) and AIPCW ($\hat{p}_{\mathrm{dr}}(z)$) conformal p-values  from Section~\ref{sec:fdr}
are replaced by their population oracle counterpart $p^\star$ defined in~\eqref{eq:def-pstar}:
\[
  p^\star(z) := \P{R(X,T)\ge z},
\]
where $(X,T)$ is an independent draw from the population. Proposition~\ref{prop:eps} in Supplement Section~\ref{app:fdr-risk} tells us that the IPCW ($\hat{p}(z)$) and AIPCW ($\hat{p}_{\mathrm{dr}}(z)$) conformal p-values  from Section~\ref{sec:fdr} asymptotically behave like $p^\star(z)$ in the large-sample limit, under suitable consistency assumptions.

With the score function $R$ defined as in~\eqref{eq:conformal-score-R} for $\hat S = S$, the oracle p-value for a cohort subject with (fixed) covariates $x_{n+j}$ becomes
\begin{align*}
  p^\star_{j,\gamma}
  & := p^\star\bigl( S(t_0 + \gamma \mid x_{n+j}) \bigr) \\
  & = \P{S(T + \gamma \mid X)\,\I{T \le t_0} \geq S(t_0 + \gamma \mid x_{n+j})} \\
  & = \P{S(T + \gamma \mid X) \geq S(t_0 + \gamma \mid x_{n+j}), T \le t_0} \\
  & = \P{S(T + \gamma \mid X) \geq S(t_0 + \gamma \mid x_{n+j}), S(T \mid X) \geq S(t_0 \mid X)},
\end{align*}
since $S$ is non-increasing.
To simplify further, define the random variable $U := S(T\mid X)$ and the conditional survival probability
\begin{align*}
  & V_\gamma(x,t) := \frac{S(t+\gamma\mid x)}{S(t\mid x)}.
\end{align*}
By the probability integral transform, $U \sim \text{Uniform}([0,1])$ and $U \indep X$.
In other words, knowing $U$ tells us nothing about the risk of a patient with covariates $X$.
Then,
\begin{align} \label{eq:p-star-U}
  p^\star_{j,\gamma}
  & = \P{U V_\gamma(X,T) \geq S(t_0 \mid x_{n+j}) V_\gamma(x_{n+j},t_0), U \geq S(t_0\mid X)}.
\end{align}

\paragraph{Behavior with $\gamma=0$.}
Using the independence of $U$ and $X$, the expression for $p^\star_{j,\gamma}$ simplifies to
\begin{align*}
  p^\star_{j,0}
  & = \P{U \geq \max\{S(t_0 \mid x_{n+j}), S(t_0\mid X) \}} \\
  & = \EV{1 - \max\{ S(t_0 \mid x_{n+j}), S(t_0\mid X) \} } \\
  & = \EV{\min\{ 1-S(t_0 \mid x_{n+j}), 1-S(t_0\mid X) \} }.
\end{align*}

Write the risk at $t_0$ of the $j$-th cohort subject as $r_j = 1-S(t_0 \mid x_{n+j})$, and the corresponding risk of
a random subject as $r(X) = 1-S(t_0 \mid X)$. Define also $A_j(X) = \I{r_j \leq r(X)}$ with $\pi_j = \P{r_j \leq r(X)}$ representing
the probability that a random subject from the population
is at higher risk than the $j$-th cohort subject.
 Then,
\begin{align}\label{eq:pstar-gamma-0}
  p^\star_{j,0}
  & = \EV{\min\{ r_j, r(X) \} }
   = r_j \pi_j + \EV{r(X) \mid A_j(X) = 0 } (1-\pi_j).
 \end{align}
Since $r(X) < r_j$ whenever $A_j(X) = 0$, this implies
\begin{equation}\label{eq:pstar-bounds}
  \pi_j\, r_j \leq p^\star_{j,0} \leq r_j .
\end{equation}
Therefore, the p-values $p^\star_{j,0}$ are capped from below and can never be very small, 
essentially making the BH procedure all-or-nothing and generally ineffective, as explained
in more detail below.

To make this intuition more precise, consider a special case where 
the risk is uniformly distributed on $(a,b)$ across the population, and the cohort is an independent
sample from it. Then $\pi_j = (b - r_j)/(b-a)$ and
$\EV{r(X) \mid A_j(X) = 0} = (a + r_j)/2$, so
\begin{equation}\label{eq:pstar-uniform}
  p^\star_{j,0}
  = \frac{r_j\,(b - r_j)}{b-a} + \frac{r_j - a}{b-a}\cdot\frac{a + r_j}{2}
  = \frac{2 r_j b - r_j^2 - a^2}{2(b-a)} .
\end{equation}
Now order the cohort by risk and write $v = k/m$ for the fraction selected, so that
the $k$-th lowest-risk cohort subject has $r_{(k)} \approx a + v(b-a)$. Substituting
this into~\eqref{eq:pstar-uniform} and simplifying,
we find that the p-value for this subject is
\begin{equation}\label{eq:pstar-uk}
  p^\star_{(k),0} = a + v(b-a)\left(1 - \frac{v}{2}\right).
\end{equation}
Applying BH at level $\alpha$ selects $k$ subjects only if
$p^\star_{(k)} \le \alpha k/m = \alpha v$, that is, only if
\begin{equation}\label{eq:bh-uniform}
  \varphi(v) \;:=\; \frac{a}{v} + (b-a)\left(1 - \frac{v}{2}\right) - \alpha
  \leq 0 .
\end{equation}
Since $\varphi'(v) = -a/v^2 - (b-a)/2 < 0$, the requirement
in~\eqref{eq:bh-uniform} is strictly easier to satisfy the larger the selected
set. It is therefore least demanding at $v = 1$, where
\begin{equation*}
  \varphi(1) = a + \frac{b-a}{2} - \alpha = \frac{a+b}{2} - \alpha ,
\end{equation*}
so the criterion reduces to the requirement that the mean risk of the
entire cohort be at most $\alpha$.

The procedure is thus all-or-nothing. If the mean risk of the entire cohort is below $\alpha$,
BH selects every subject; otherwise, it selects none. No intermediate outcomes can occur.
In particular, with $\gamma=0$, BH can never isolate a low-risk
subgroup from a cohort whose overall risk exceeds $\alpha$, which is the main task
screening is meant to perform. 

The root cause of this degenerate behavior is the near-proportionality between $p^\star_{j,0}$ and $r_j$
in~\eqref{eq:pstar-bounds}, which causes the p-value $p^\star_{(k),0}$ to rise across the ordered cohort 
nearly in parallel with the BH threshold $\alpha k/m$, so no intermediate $k$ is
ever favoured over selecting everyone. 
This undesirable behavior is precisely what the time shift parameter $\gamma$ is meant to break.
As shown below, $\gamma>0$ allows the $p$-values of low-risk subjects to fall faster than the BH threshold,
enabling interior selections.

\paragraph{The general case of $\gamma>0$.}
Returning to the general expression for $p^\star_{j,\gamma}$ in~\eqref{eq:p-star-U}, the first requirement in
$p^\star_{j,\gamma}$ reads
\begin{equation*}
  U \geq S(t_0 \mid x_{n+j})\, \frac{V_\gamma(x_{n+j},t_0)}{V_\gamma(X,T)} .
\end{equation*}
The time shift therefore inflates the first threshold that $U$ must clear by the
ratio of two conditional survival probabilities: that of the cohort subject
over the window $(t_0, t_0+\gamma]$, and that of the calibration subject over
$(T, T+\gamma]$. 
This ratio will tend to be larger than one if the random calibration subject
has higher risk than the cohort subject, and smaller than one otherwise.
This pushes the p-values down for low-risk cohort subjects and up for high-risk subjects,
as desired.

To visualize this idea more concretely, consider an exponential survival model
with $S(t \mid x) = e^{-h(x) t}$, so that $V_\gamma(x,t) = e^{-h(x) \gamma}$ does not depend on $t$. 

The first requirement in $p^\star_{j,\gamma}$ then becomes $U \geq (1-r_j) \, e^{-\gamma [h(x_{n+j})-h(X)]}$.
Combining this with $U \ge 1 - r(X)$ and using the independence of $U$ and $X$, the expression for $p^\star_{j,\gamma}$ simplifies to
\begin{align} \label{eq:pstar-gamma-exp}
\begin{split}
  p^\star_{j,\gamma}
  & = \P{U \geq \max\{(1-r_j) \, e^{-\gamma [h(x_{n+j})-h(X)]}, 1-r(X) \}} \\
  & = \EV{ \left( 1 - \max\{(1-r_j) \, e^{-\gamma [h(x_{n+j})-h(X)]}, 1-r(X) \} \right)_+}.
\end{split}
\end{align}
In the limit of $\gamma \to \infty$, the factor $e^{-\gamma [h(x_{n+j})-h(X)]}$ diverges if $A_j(X)=1$, so the threshold exceeds one and the positive part in~\eqref{eq:pstar-gamma-exp} vanishes. On the other hand, if $A_j(X)=0$, the factor tends to zero, the second argument of the maximum binds, and the contribution is $r(X)$. Therefore,
\begin{equation}\label{eq:pstar-gamma-inf}
  p^\star_{j,\infty}
  = \EV{ r(X)\, \I{A_j(X) = 0} }
  = \bigl(1 - \pi_j\bigr)\, \EV{ r(X) \mid A_j(X) = 0 } .
\end{equation}
Comparing this with the corresponding expression~\eqref{eq:pstar-gamma-0} at $\gamma = 0$ shows that the shift has
removed the term $\pi_j r_j$, which is what created the undesired parallelism with the BH threshold.

Returning to the uniform example, where $\pi_j = (b - r_j)/(b-a)$ and
$\EV{r(X) \mid A_j(X) = 0} = (a + r_j)/2$, we have
\begin{equation}\label{eq:pstar-uniform-inf}
  p^\star_{j,\infty}
  = \frac{r^2_j-a^2}{2(b-a)}.
\end{equation}
As before, order the cohort by risk and write $v = k/m$ for the fraction selected, so that
the $k$-th lowest-risk cohort subject has $r_{(k)} \approx a + v(b-a)$. Substituting
this into~\eqref{eq:pstar-uniform-inf} and simplifying,
we find that the p-value for this subject is
\begin{equation}\label{eq:pstar-uk-inf}
  p^\star_{(k),\infty} = v \left(a + v\frac{b-a}{2}\right) = v\, \bar{r}_{(k)},
\end{equation}
where $\bar{r}_{(k)}$ is the average risk of the $k$ lowest-risk cohort subjects.
Then, the BH requirement $p^\star_{(k)} \le \alpha v$ reduces to
\begin{equation}\label{eq:bh-gamma-inf}
  \bar{r}_{(k)} \leq \alpha .
\end{equation} 
In the uniform case, solving for $v$ tells us that BH will select the fraction 
\begin{equation*}
  v^\star = \min\left\{ 1,\; \frac{2(\alpha - a)}{b - a} \right\}
\end{equation*}
of lowest-risk patients. This contrasts sharply with the all-or-nothing behavior induced by $\gamma=0$.
For example, if $\alpha=0.1$, $a = 0.02$ and $b=0.3$, BH selects approximately 57\% of cohort patients with $\gamma = \infty$ and nobody with $\gamma = 0$.

That being said, sending $\gamma \to \infty$ is not optimal in general because the shift is beneficial only insofar as the ranking induced
by $S(\cdot + \gamma \mid x)$ agrees with the ranking of risk at $t_0$. 
The two coincide under the exponential model but not in general.
Moreover, a very large value of $\gamma$ requires extrapolating $\hat{S}(\cdot + \gamma \mid x)$
beyond the range observed in the training data, where it may be unreliable.
This is why in practice we recommend tuning $\gamma$ using independent data, as detailed next.

\clearpage

\subsubsection{Tuning the Time Shift Parameter} \label{app:tuning-gamma-alg}

The time shift parameter $\gamma$ used by Algorithm~\ref{tab:alg-fdr} to compute $Z_i = \hat{S}(\tilde{T}_i + \gamma \mid X_i)$ and $Z_{n+j} = \hat{S}(t_0 + \gamma \mid X_{n+j})$ should be tuned on a tuning data set \emph{independent} of the calibration data in order to maximize power without invalidating any guarantees.

In practice, we tune $\gamma$ as described below and summarized by Algorithm~\ref{alg:tuning-gamma}.
First, define a grid $\Gamma$ of candidate values (e.g., $\{0\}$ together with several quantiles of the observed event times).
Then, repeatedly split the training data (assumed independent of the calibration data) into three subsets: a tuning-training set for model fitting, a tuning-calibration set for conformal calibration, and a tuning-test set for evaluation.
On each split, fit smaller survival and censoring models using the tuning-training subset, then apply Algorithm~\ref{tab:alg-fdr} for each candidate value of $\gamma$ using the tuning-calibration and tuning-test subsets in place of the true calibration and test sets.
Measure performance on the tuning-test set either by
(i) the yield at level $\alpha$, or
(ii) in lower-power settings, by the smoother Fisher combination statistic $-2\sum_j \log \hat{p}_j$, where $\hat{p}_j$ denote the conformal $p$-values for the tuning-test set.  In our experiments, we adopt the Fisher criterion by default because we have observed that it performs better in low-yield settings and similarly otherwise, and we set the stability threshold for the optional abstention rule to $\nu=0.9$ when estimating $\mathbb{P}(|\hat{\mathcal{S}}|>0)$ via bootstrap.
Then, we average the performance score across several random splits, and select
\[
\gamma^\star = \arg\max_{\gamma\in\Gamma} \text{AverageScore}(\gamma).
\]
Finally, fix $\gamma = \gamma^\star$ and run Algorithm~\ref{tab:alg-fdr} on the actual calibration and test sets.
For computational efficiency, we can precompute $\hat S(\cdot\mid X)$ on a fine time grid and use interpolation to evaluate $\hat S(\tilde T_i+\gamma\mid X_i)$ and $\hat S(t_0+\gamma\mid X_i)$ for all $\gamma\in\Gamma$, avoiding repeated model calls.

\begin{algorithm}[!h]
\caption{Cross--validation to tune the time shift parameter $\gamma$ for Algorithm~\ref{tab:alg-fdr}}
\label{alg:tuning-gamma}
\begin{algorithmic}[1]
\INPUT Tuning dataset $\mathcal D_{\mathrm{tune}}=\{(X_i,\tilde T_i,E_i)\}_{i=1}^N$ (independent of final calibration data); pre-fitted survival model $\hat S$ and censoring model $\hat G$; horizon $t_0$; target level $\alpha$; candidate grid $\Gamma$; number of repetitions $R$; split fractions $(\pi_{\mathrm{cal}},\pi_{\mathrm{test}})$; evaluation criterion $\in\{\texttt{Yield},\texttt{Fisher}\}$.
\FOR{$r=1$ to $R$}
  \STATE Randomly split $\mathcal D_{\mathrm{tune}}$ into disjoint subsets
  $D_{\mathrm{tune}}^{\mathrm{train}}$,
  $D_{\mathrm{tune}}^{\mathrm{cal}}$, and
  $D_{\mathrm{tune}}^{\mathrm{test}}$
  according to fractions
  $(1-\pi_{\mathrm{cal}}-\pi_{\mathrm{test}},\pi_{\mathrm{cal}},\pi_{\mathrm{test}})$.
  \STATE Optionally refit smaller models
  $\hat S_r,\hat G_r$ on $D_{\mathrm{tune}}^{\mathrm{train}}$
  (otherwise use $\hat S,\hat G$).
  \FOR{each $\gamma\in\Gamma$}
    \STATE Apply Algorithm~\ref{tab:alg-fdr} using calibration set
    $D_{\mathrm{tune}}^{\mathrm{cal}}$, test set
    $D_{\mathrm{tune}}^{\mathrm{test}}$, and time shift $\gamma$, obtaining conformal $p$--values $\{p_j\}$.
    \STATE Compute the performance score $J_r(\gamma)$:
    number of BH rejections at level $\alpha$,
    or Fisher statistic $J_r(\gamma)=-2\sum_j \log p_j$.
  \ENDFOR
\ENDFOR
\STATE Aggregate scores $\bar J(\gamma)=R^{-1}\sum_{r=1}^R J_r(\gamma)$ .
\STATE Select $\gamma^\star=\arg\max_{\gamma\in\Gamma}\bar J(\gamma)$
(break ties by smallest $\gamma$).
\OUTPUT Tuned time shift $\gamma^\star$.
\end{algorithmic}
\end{algorithm}

\clearpage

\section{Mathematical Proofs and Auxiliary Theoretical Results} \label{appendix:proofs}

\subsection{Pointwise Risk Estimation}

\begin{proof}[Proof of Proposition~\ref{prop:double-robustness}]

We condition throughout on the training data used to fit $\hat S$ and $\hat G$,
treating these as fixed functions. Recall that $T$ and $C$ are continuously
distributed given $X$, so ties have probability zero; no continuity is required of
$\hat S$ or $\hat G$, which in practice are step functions, in which case the
integrals below reduce to finite sums.

Conditioning on $X$ and using the tower property, the bias can be equivalently written as
\begin{align*}
  \EV{ \hat \theta_{\mathrm{dr}}(\lambda) } - \theta(\lambda)
   & = \EV{ A_\lambda(X) \hat\Psi^{\mathrm{dr}}(O, t_0)  } - \EV{ \I{T\le t_0}A_\lambda(X) } \\
   & = \EV{ A_\lambda(X) \left(  \EV{ \hat\Psi^{\mathrm{dr}}(O, t_0)  \mid X } - [ 1 - S(t_0\mid X) ]  \right) }.
\end{align*}
Therefore, it suffices to prove that
\begin{align} \label{eq:conditional-claim}
\begin{split}
  & \mathbb E\big[\hat\Psi^{\mathrm{dr}}(O, t_0)  \mid X = x\big] - \{1 - S(t_0\mid x)\} \\
  & \qquad = \int_0^{t_0} \frac{G(u\mid x)}{\hat G(u\mid x)}\, S(u\mid x)\,
    \big(\hat Q - Q\big)(u, t_0 \mid x)\,
    \big(d\Lambda^{C} - d\hat\Lambda^{C}\big)(u \mid x)
  \end{split}
\end{align}
for almost every $x$.

By conditional independence and continuity, $\mathbb P(C > u \mid T = u, X=x) = G(u \mid x)$, so the first term of~\eqref{eq:aipcw-pseudo} has conditional mean
\begin{equation*}
  \mathbb E\!\left[\frac{\mathbb I(T \le t_0,\, C > T)}{\hat G(T \mid X)} \mid X=x\right]
  = \int_0^{t_0} \frac{G(u \mid x)}{\hat G(u \mid x)}\, f_T(u \mid x)\, du
  = -\int_0^{t_0} \frac{G(u\mid x)}{\hat G(u\mid x)} \, dS(u \mid x) .
\end{equation*}

The second term of~\eqref{eq:aipcw-pseudo} has conditional mean
\begin{align*}
  \EV{\frac{\mathbb I(\tilde T \le t_0, T > C)\,\hat Q(\tilde T, t_0 \mid X)}{\hat G(\tilde T \mid X)} \mid X=x}
  & = \int_0^{t_0} \frac{S(u \mid x) \hat Q(u, t_0\mid x)}{\hat G(u \mid x)}\, f_C(u \mid x)\, du \\
  & = \int_0^{t_0} S(u \mid x) \hat Q(u, t_0\mid x) \frac{G(u\mid x)}{\hat G(u\mid x)} \, d\Lambda^{C}(u),
\end{align*}
where the second equality uses the identity $f_C(u)\,du = -dG(u) = G(u)\,d\Lambda^{C}(u)$.

The third term of~\eqref{eq:aipcw-pseudo} has conditional mean
\begin{align*}
  \EV{ \int_{0}^{\tilde T \wedge t_0} \frac{\hat Q(u, t_0\mid X)}{\hat G(u \mid X)}\, d\hat\Lambda^{C}(u \mid X) \mid X=x}
  & = \EV{ \int_{0}^{t_0} \I{\tilde{T} \geq u} \frac{\hat Q(u, t_0\mid x)}{\hat G(u \mid x)}\, d\hat\Lambda^{C}(u \mid x) \mid X=x} \\
  & = \int_{0}^{t_0} \P{\tilde{T} \geq u \mid X=x} \frac{\hat Q(u, t_0\mid x)}{\hat G(u \mid x)}\, d\hat\Lambda^{C}(u \mid x) \\
  & = \int_{0}^{t_0} S(u \mid x) \hat Q(u, t_0\mid x) \frac{G(u\mid x)}{\hat G(u\mid x)} \, d\hat\Lambda^{C}(u \mid x),
\end{align*}
since $\mathbb P(\tilde T \ge u \mid X = x) = S(u\mid x)G(u\mid x)$ by continuity of $T$ and $C$.

Collecting the three pieces, we obtain
\begin{align} \label{eq:collected}
\begin{split}
  & \EV{\hat\Psi^{\mathrm{dr}}(O, t_0)  \mid X=x} - [ 1 - S(t_0\mid x) ] \\
  & \qquad = -\int_0^{t_0} \frac{G(u\mid x)}{\hat G(u\mid x)}\, dS(u \mid x) - [ 1 - S(t_0\mid x) ] \\
   & \qquad \qquad + \int_0^{t_0} \hat Q(u, t_0\mid x)\, S(u \mid x)\, \frac{G(u\mid x)}{\hat G(u\mid x)}\, [d\Lambda^{C}(u \mid x) - d\hat\Lambda^{C}(u \mid x)].
 \end{split}
\end{align}

Note that $dG(u \mid x) = -G(u \mid x)\,d\Lambda^{C}(u \mid x)$ by continuity of
$C$, while $d\,\hat G(u\mid x)^{-1} = \hat G(u \mid x)^{-1} d\hat\Lambda^{C}(u \mid x)$
holds whether or not $\hat G(\cdot \mid x)$ has atoms: at a jump point $u_k$,
$\hat G_k^{-1} - \hat G_{k-1}^{-1} = \Delta\hat\Lambda^{C}_k(x)/\hat G_k(x)$ with
$\Delta\hat\Lambda^{C}_k(x) = 1 - \hat G_k(x)/\hat G_{k-1}(x)$ as in
Section~\ref{app:aipcw-implementation}. Since $G$ is continuous there is no covariation term,
and the product rule gives
\[
  d \frac{G(u\mid x)}{\hat G(u\mid x)}
  = \frac{G(u\mid x)}{\hat G(u\mid x)} \,[ d\hat\Lambda^{C}(u \mid x) - d\Lambda^{C}(u \mid x)],
\]
with $G(0\mid x)/\hat G(0\mid x) = 1$, so that
\[
\frac{G(u\mid x)}{\hat G(u\mid x)} - 1 = \int_0^{u} \frac{G(s\mid x)}{\hat G(s\mid x)}
  \,[ d\hat\Lambda^{C}(s \mid x) - d\Lambda^{C}(s \mid x) ].
\]
Here $\hat G$ is evaluated at $u$ rather than $u^{-}$, matching the weights
in~\eqref{eq:aipcw-pseudo}.

This allows us to rewrite the first two terms on the right-hand-side of~\eqref{eq:collected} as:
\begin{align*}
  & -\int_0^{t_0} \frac{G(u\mid x)}{\hat G(u\mid x)}\, dS(u \mid x) - [ 1 - S(t_0\mid x) ] \\
  & \qquad \qquad = -\int_0^{t_0} \left[\frac{G(u\mid x)}{\hat G(u\mid x)} - 1\right]\, dS(u \mid x) \\
  & \qquad \qquad = -\int_0^{t_0} \int_0^{u} \frac{G(s \mid x)}{\hat{G}(s \mid x)}\,[ d\hat\Lambda^{C}(s \mid x) - d\Lambda^{C}(s \mid x) ] \, dS(u \mid x) \\
  & \qquad \qquad = -\int_0^{t_0} \frac{G(s \mid x)}{\hat{G}(s \mid x)}\,[ d\hat\Lambda^{C}(s \mid x) - d\Lambda^{C}(s \mid x) ] \int_{s}^{t_0}  \, dS(u \mid x) \\
  & \qquad \qquad = \int_0^{t_0} \frac{G(s \mid x)}{\hat{G}(s \mid x)} [S(s \mid x) - S(t_0 \mid x) ] [ d\hat\Lambda^{C}(s \mid x) - d\Lambda^{C}(s \mid x) ] \\
  & \qquad \qquad = - \int_0^{t_0} Q(s,t_0 \mid x) \frac{G(s \mid x)}{\hat{G}(s \mid x)} S(s \mid x) [ d\Lambda^{C}(s \mid x) - d\hat\Lambda^{C}(s \mid x)  ],
\end{align*}
where the third equality uses Fubini's theorem. Note that $Q$ involves no left
limit, $S$ being continuous, whereas $\hat Q$ in~\eqref{eq:qhat} is defined
through $\hat S(u^{-}\mid x)$, since $\hat S$ may jump.

Replacing this into~\eqref{eq:collected}, we arrive at
\begin{align*}
  & \EV{\hat\Psi^{\mathrm{dr}}(O, t_0)  \mid X=x} - [ 1 - S(t_0\mid x) ] \\
  & \qquad =\int_0^{t_0}  S(u \mid x)\, \frac{G(u\mid x)}{\hat G(u\mid x)} \, [\hat Q(u, t_0\mid x) - Q(u, t_0\mid x)] [d\Lambda^{C}(u \mid x) - d\hat\Lambda^{C}(u \mid x)] .
\end{align*}
The desired result is then obtained by averaging over $X$.

Finally, $\hat G = G$ implies $G/\hat{G} \equiv 1$ and $d\hat\Lambda^{C} = d\Lambda^{C}$, so the right-hand side above vanishes; and $\hat S = S$ implies $\hat Q = Q$, so it vanishes as well.

\end{proof}

\clearpage

\subsection{Calibration via FDR Control and Conformal Inference} \label{app:fdr-risk}

\subsubsection{Auxiliary Definitions and Lemmas}

With $R(x,t) := \hat S(t + \gamma \mid x)\,\I{t \le t_0}$ from~\eqref{eq:conformal-score-R}, define the function $p^\star : [0,1]\to [0,1]$ as
\begin{equation} \label{eq:def-pstar}
  p^\star(z) := \P{R(X,T)\ge z} = \P{T \leq b_z(X)},
\end{equation}
which represents the ideal population target for the conformal p-value functions in~\eqref{eq:pvalue-as-score} and~\eqref{eq:aipcw-conformal-p}.
For any function $f : [0,1]\to\mathbb R$, define
\begin{align*}
  & \|f\|_\infty := \sup_{z \in (0,1]} |f(z)|.
\end{align*}
Accordingly, for any $\Dcal$-measurable function $\hat{p} : [0,1]\to\mathbb R$, implicitly indexed by $n = |\Dcal|$, define
\begin{align} \label{eq:varepsilon-n}
  & \varepsilon_n(\hat{p}) := \| \hat{p} -p^\star \|_\infty.
\end{align}

\begin{definition}[$p$-Value Function] \label{def:p-val-func}
A (data-dependent) function $\hat p:[0,1]\to [0,1]$ is said to be a {\em $p$-value function} if it is $\mathcal D_{\mathrm{cal}}$-measurable and non-increasing in $z$.
\end{definition}

\begin{lemma} \label{lem:superunif}
Let $V$ be a real random variable and $F(v):=\P{V\ge v}$. Then
$\P{F(V)\le\beta}\le\beta$ for every $\beta\in[0,1]$.
\end{lemma}

\begin{proof}[Proof of Lemma~\ref{lem:superunif}]
Since $F$ is non-increasing, $A:=\{v:F(v)\le\beta\}$ is closed upward: $v\in A$
and $v'\ge v$ imply $v'\in A$, because $F(v')\le F(v)\le\beta$. Hence either $A$ is
empty, in which case $\P{V\in A}=0\le\beta$, or $A$ is a ray with $v_\beta:=\inf A$. In
the latter case, either $v_\beta\in A$ and
$\P{V\in A}=\P{V\ge v_\beta}=F(v_\beta)\le\beta$ by the definition of $A$, or
$A=(v_\beta,\infty)$ and
$\P{V\in A}=\P{V>v_\beta}=\lim_{v\downarrow v_\beta}F(v)\le\beta$, since
$F(v)\le\beta$ for every $v>v_\beta$.
\end{proof}

\begin{lemma} \label{lem:majorant}
Let $f:(0,1]\to\mathbb R$ be bounded and let
$\Pi f(z) := \max\{0, \min\{1,\ \sup_{z'\ge z} f(z')\}\}$ as in~\eqref{eq:aipcw-conformal-p}. Then $\Pi f$ is non-increasing
and takes values in $[0,1]$. Moreover, for any non-increasing
$g:(0,1]\to[0,1]$,
\begin{align*}
  & \|\Pi f - g\|_\infty \leq \|f - g\|_\infty,
  & \|f\|_\infty := \sup_{z \in (0,1]} |f(z)|.
\end{align*}
\end{lemma}

\begin{proof}[Proof of Lemma~\ref{lem:majorant}]
Monotonicity is immediate, since $z\mapsto\sup_{z'\ge z}f(z')$ is non-increasing
and the minimum with a constant preserves this. For the second claim, define
$\delta:=\|f-g\|_\infty \geq 0$. For any $z,z' \in (0,1]$ with $z \leq z'$, we have
$f(z')\le g(z')+\delta\le g(z)+\delta$ by monotonicity of $g$, so
$\sup_{z'\ge z}f(z')\le g(z)+\delta$ and hence $\Pi f(z) = \max\{0, \min\{1, \sup_{z'\ge z} f(z')\}\} \leq \max\{0, g(z)+\delta\} = g(z)+\delta$.
Conversely, $\Pi f(z) = \max\{0, \min\{1, \sup_{z'\ge z} f(z')\}\} \geq \max\{0, \min\{1, f(z)\}\} \geq \max\{0, \min\{1, g(z) - \delta\}\} = \max\{0, g(z) - \delta\} \geq g(z) - \delta$, using $g(z) \leq 1$ and $\delta \geq 0$.
Combining the two bounds gives $|\Pi f(z)-g(z)|\le\delta$ for every $z \in (0,1]$.
\end{proof}

\subsubsection{BH with Approximately Super-Uniform Conformal $p$-Values}

\begin{lemma} \label{lem:cond-superunif}
Under Assumptions \textup{(A1)--(A3)}, consider $Z_{n+j} = \hat{S}(t_0 + \gamma \mid X_{n+j})$ as in~\eqref{eq:def-scores-test}, for $j \in [m]$.
Then, for any $p$-value function $\hat p$ satisfying Definition~\ref{def:p-val-func}, almost surely,
\[
  \P{\hat p(Z_{n+j})\le\tau,\ H^{0}_{j}=1 \mid \mathcal D_{\mathrm{cal}}}
  \leq \tau+\varepsilon_n(\hat p)
  \qquad\text{for every }\tau\in(0,1) .
\]
\end{lemma}

\begin{proof}[Proof of Lemma~\ref{lem:cond-superunif}]
Fix $j \in [m]$, set $W := R(X_{n+j},T_{n+j}) = \hat S(T_{n+j} + \gamma \mid X_{n+j})\,\I{T_{n+j} \le t_0} \in[0,1]$, and define
$\beta := \tau+\varepsilon_n(\hat p)$, a constant conditional on $\mathcal D_{\mathrm{cal}}$.
We may assume $\beta\le1$, the claim being trivial otherwise.

On $\{H^{0}_{j}=1\} = \{T_{n+j}\le t_0\}$, we have $W=\hat S(T_{n+j}+\gamma\mid X_{n+j}) \geq \hat S(t_0+\gamma\mid X_{n+j})=Z_{n+j}$, since
$T_{n+j}\le t_0$ and $\hat S(\cdot\mid x)$ is non-increasing.
This implies $\hat p(W)\le\hat p(Z_{n+j})$ because $\hat p$ is non-increasing, and therefore
\[
  \{\hat p(Z_{n+j})\le\tau\}\cap\{H^{0}_{j}=1\}
  \;\subseteq\; \{\hat p(W)\le\tau\}
  \;\subseteq\; \{p^\star(W)\le\beta\},
\]
where the last inclusion follows from the definitions of $\beta$ and $\varepsilon_n(\hat{p})$.
Therefore,
\[
  \P{\hat p(Z_{n+j})\le\tau,\ H^{0}_{j}=1 \mid \mathcal D_{\mathrm{cal}}}
  \leq \P{p^\star(W)\le \tau+\varepsilon_n(\hat p) \mid \mathcal D_{\mathrm{cal}}}.
\]
By Assumption (A3) the pair $(X_{n+j},T_{n+j})$ is independent of
$\mathcal D_{\mathrm{cal}}$ and drawn from the population, so conditionally on
$\mathcal D_{\mathrm{cal}}$ the variable $W$ has the same law as $R(X,T)$, whose
upper tail function is precisely $p^\star$ from~\eqref{eq:def-pstar}. Applying
Lemma~\ref{lem:superunif} conditionally on $\mathcal D_{\mathrm{cal}}$ completes
the proof.
\end{proof}

\begin{theorem} \label{thm:fdr-slack}
Under Assumptions \textup{(A1)--(A3)}, for any $p$-value function
$\hat p$ satisfying Definition~\ref{def:p-val-func}, the BH procedure at level $\alpha$ applied to
$(p_1,\ldots,p_m)$ with $p_j := \hat{p}(\hat{S}(t_0 + \gamma \mid X_{n+j}) )$ satisfies
\[
  \mathrm{FDR} \leq \alpha +
  \EV{\varepsilon_n(\hat p)\,\frac{m}{\max(R,1)}},
\]
where $R$ is the number of rejections.
\end{theorem}

\begin{proof}[Proof of Theorem~\ref{thm:fdr-slack}]
Define $A_j:=\I{j\in\hat{\mathcal{S}}}$, where $\hat{\mathcal{S}}$ is the rejection set, so that $R =\sum_{j=1}^m A_j = |\hat{\mathcal{S}}|$.
For $j\in[m]$ let $\tilde R^{(j)}$ denote the number of BH rejections at level
$\alpha$ obtained when $p_j$ is replaced by $0$; it is a function of
$\{p_\ell\}_{\ell\ne j}$ alone. Lowering a $p$-value can only increase the
number of rejections, and setting it to zero forces $j$ to be rejected, so
\begin{equation} \label{eq:Rtilde-lower}
  \tilde R^{(j)} \;\ge\; \max(R,1) .
\end{equation}

\emph{Step 1.} On $\{R=k\}$ the BH threshold equals $\alpha k/m$, and replacing
$p_j$ by $0$ leaves the rejection count unchanged on the event that $j$ is
already rejected. Therefore,
\begin{align} \label{eq:fdp-loo}
  \mathrm{FDP}
  &= \frac{\sum_{j=1}^m H^{0}_{j}A_j}{\max(R,1)}
    =\sum_{j=1}^m\sum_{k=1}^m \frac{H^{0}_{j}}{k}\,A_j\,\I{R=k}
    = \sum_{j=1}^m\sum_{k=1}^m \frac{H^{0}_{j}}{k}\, \I{p_j\le\alpha k/m}\,\I{\tilde R^{(j)}=k} .
\end{align}

\emph{Step 2.} Fix $j$ and $k$ and condition on $\mathcal D_{\mathrm{cal}}$.
Since $p$ is $\mathcal D_{\mathrm{cal}}$-measurable and the cohort is
i.i.d.\ and independent of $\mathcal D_{\mathrm{cal}}$ by Assumption~(A3), the pairs
$(p_j,H^{0}_{j})_{j=1}^m$ are i.i.d.\ conditionally on
$\mathcal D_{\mathrm{cal}}$. Hence $H^{0}_{j}\I{p_j\le\alpha k/m}$, a function of
$(p_j,H^{0}_{j})$, and $\I{\tilde R^{(j)}=k}$, a function of
$\{p_\ell\}_{\ell\ne j}$, are conditionally independent, so that by
Lemma~\ref{lem:cond-superunif} with $\tau=\alpha k/m$,
\[
  \EVc{H^{0}_{j}\,\I{p_j\le\alpha k/m}\,\I{\tilde R^{(j)}=k}}{\mathcal D_{\mathrm{cal}}}
  \leq \Big(\frac{\alpha k}{m}+\varepsilon_n(\hat p)\Big)\,
    \P{\tilde R^{(j)}=k\mid\mathcal D_{\mathrm{cal}}} .
\]

\emph{Step 3.} Taking expectations in~\eqref{eq:fdp-loo} and substituting,
\begin{align*}
  \mathrm{FDR}
  &\le \sum_{j=1}^m\sum_{k=1}^m\frac1k
    \EV{\Big(\frac{\alpha k}{m}+\varepsilon_n(\hat p)\Big)
        \P{\tilde R^{(j)}=k\mid\mathcal D_{\mathrm{cal}}}} \\
  &= \frac{\alpha}{m}\sum_{j=1}^m
      \EV{\sum_{k=1}^m\P{\tilde R^{(j)}=k\mid\mathcal D_{\mathrm{cal}}}}
   + \sum_{j=1}^m \EV{\varepsilon_n(\hat p)
      \sum_{k=1}^m\frac1k\,\P{\tilde R^{(j)}=k\mid\mathcal D_{\mathrm{cal}}}} \\
  &\le \alpha + \EV{\varepsilon_n(\hat p)\sum_{j=1}^m\frac{1}{\tilde R^{(j)}}} ,
\end{align*}
where we used $\sum_k\P{\tilde R^{(j)}=k\mid\mathcal D_{\mathrm{cal}}}\le1$ and the
$\mathcal D_{\mathrm{cal}}$-measurability of $\varepsilon_n(\hat p)$.
By~\eqref{eq:Rtilde-lower} the inner sum is at most $m/\max(R,1)$.
\end{proof}

\subsubsection{Asymptotic FDR Analysis} \label{app:fdr-errors}

The following notation is used in this section.
For a measurable function $f$ of a single observation, we write
\[
  \mathbb P_n f := \frac1n\sum_{i=1}^n f(O_i) ,
  \qquad
  P f := \EV{f(O)} ,
\]
for its empirical average over $\Dcal$ and its expectation under the population
law $P$, so that $Pf$ is deterministic and $\mathbb P_nf$ is
$\Dcal$-measurable with $\EV{\mathbb P_nf} = Pf$. We also write
$Pf^2 := \EV{f(O)^2}$.

\begin{proof}[Proof of Theorem~\ref{thm:fdr-asymptotic}]
Throughout we condition on $\mathcal D_{\mathrm{train}}$, so that by
Assumption~(A2) the fitted models $\hat S,\hat G$, and the errors $\mathcal G_n,\mathcal S_n,\mathcal L_n$ are all fixed. Letting $\varepsilon_n(\hat p)$ be as
in~\eqref{eq:varepsilon-n}, and defining $W_{n,m} := m/\max(|\hat{\mathcal{S}}|,1)$, Theorem~\ref{thm:fdr-slack} gives
\begin{align} \label{eq:master}
  \mathrm{FDR} 
  \leq \alpha + \EV{\varepsilon_n(\hat p)\,W_{n,m}}.
\end{align}
Next, we use Proposition~\ref{prop:eps} to bound $\varepsilon_n(\hat p)$.

\emph{Step 1: applicability of Proposition~\ref{prop:eps} with $M=\{2+\log(1/c)\}/c$.}
The IPCW~\eqref{eq:pvalue-as-score} and AIPCW~\eqref{eq:aipcw-conformal-p}
conformal p-value functions can both be written in the form
\begin{equation} \label{eq:pvalue-general}
  \hat p(z) = \Pi \left( \frac{1+ \sum_{i=1}^n \hat{\Psi}_z(O_i)  }{1+n} \right)
  \qquad\text{with}\qquad
  \hat{\Psi}_z(O_i) := \begin{cases}
    E_i\,\hat{w}_i\,\I{R(X_i,\tilde T_i)\ge z} & \text{for IPCW},\\[2pt]
    \hat\Psi^{\mathrm{dr}}(O_i, b_z(X_i)) \text{ from } \eqref{eq:aipcw-pseudo}
      & \text{for AIPCW},
  \end{cases}
\end{equation}
where $\Pi f(z) := \max\{0,\min\{1,\ \sup_{z'\ge z} f(z')\}\}$
as in~\eqref{eq:aipcw-conformal-p}.
\begin{itemize}
\item For IPCW, $\hat{\Psi}_z(O_i)=E_i\,\hat{w}_i\,\I{R(X_i,\tilde T_i)\ge z}$, which is non-negative and
non-increasing in $z$, so we can take $A_z=\hat{\Psi}_z$ and $B_z=0$. By Assumption (A2), the
weights satisfy $E_i\,\hat{w}_i\leq 1/c\le M$ almost surely. Proposition~\ref{prop:eps} thus applies.

\item For AIPCW, $\hat{\Psi}_z(O_i) =\hat\Psi^{\mathrm{dr}}(O_i, b_z(X_i))$ as in~\eqref{eq:aipcw-pseudo}.
The horizon $b_z(X) \leq t_0$ in~\eqref{eq:def-bz} is non-increasing in $z$, so $\hat S(b_z(X)\mid X)$ is
non-decreasing and hence
$\hat Q(u, b_z(X) \mid X)=1-\hat S(b_z(X)\mid X)/\hat S(u^-\mid X)$ is
non-increasing in $z$ for every $u$. Therefore, we can write $\hat\Psi^{\mathrm{dr}}(O, b_z(X)) = A_z(O) - B_z(O)$ with
\begin{align*}
  A_z(O) & := \frac{E\,\I{\tilde T\le b_z(X)}}{\hat G(\tilde T\mid X)}
   + \frac{(1-E)\,\I{\tilde T\le b_z(X)}\,\hat Q(\tilde T, b_z(X)\mid X)}
          {\hat G(\tilde T\mid X)} , \\
  B_z(O) & := \int_0^{\tilde T\wedge b_z(X)}
    \frac{\hat Q(u, b_z(X) \mid X)}{\hat G(u\mid X)}\,d\hat\Lambda^{C}(u\mid X),
\end{align*}
where $A_z$ and $B_z$ are both non-negative and non-increasing in $z$ for every $O$.

To bound $A_z(O)$ from above, note that $\hat Q(u, b_z(X) \mid X) =1-\hat S(b_z(X)\mid X)/\hat S(u^- \mid X)$ satisfies
 $\hat Q(u, b_z(X) \mid X) \in [0,1]$ for all $u \leq b_z(X)$, and therefore by Assumption (A2), $A_z(O) \leq 2/c$.

To bound $B_z(O)$ from above, note that, using the discrete representation of $\hat\Lambda^{C}$ from Section~\ref{sec:conformal-aipcw} and the inequality $1-x \leq -\log x$,
\begin{align*}
  B_z(O)
  & \leq \frac{1}{c} \int_0^{\tilde T\wedge b_z(X)} d\hat\Lambda^{C}(u\mid X) \leq \frac{1}{c} \hat\Lambda^{C}(t_0 \mid X) \\
  & = \frac{1}{c} \sum_{k=1}^{K} \left( 1 - \frac{\hat G(u_k \mid X)}{\hat G(u_{k-1} \mid X)} \right) \\
  & \leq - \frac{1}{c} \sum_{k=1}^{K} \log \frac{\hat G(u_k \mid X)}{\hat G(u_{k-1} \mid X)} \\
  & = - \frac{1}{c} \left( \log \hat G(u_K \mid X) - \log \hat G(u_0 \mid X) \right) \\
  & = - \frac{1}{c} \log \hat G(t_0 \mid X)
    \leq \frac{\log(1/c)}{c}.
\end{align*}
Both $A_z$ and $B_z$ are thus bounded by $M=\{2+\log(1/c)\}/c$, so Proposition~\ref{prop:eps} applies.
\end{itemize}
Therefore, Proposition~\ref{prop:eps} applies in both cases with $M=\{2+\log(1/c)\}/c$ and gives:
\begin{align}\label{eq:eps-decomp}
  \varepsilon_n(\hat p) 
  \leq \underbrace{\sup_{z\in[0,1]}\big|P\hat{\Psi}_z - p^\star(z)\big|}_{Z^*}  + \underbrace{\sup_{z\in(0,1]}\big|\mathbb P_n\hat{\Psi}_z - P\hat{\Psi}_z\big|}_{Z_n} + \frac{1}{1+n},
\end{align}
almost surely, and moreover
\begin{align*}
  & \EV{Z_n} \leq \frac{2 M C}{\sqrt n},
  & \big\|Z_n\big\|_{2} \leq \frac{(2C+1)M}{\sqrt n},
\end{align*}
for a universal constant $C$, where $\|\cdot\|_{2}$ is the $L_2(P)$ norm.
Combining these results with~\eqref{eq:master}, and applying Cauchy–Schwarz, gives
\begin{align} \label{eq:eps-decomp-2}
\begin{split}
  \mathrm{FDR} 
  & \leq \alpha + \left( Z^* + \frac{1}{n+1} \right) \EV{W_{n,m}} + \EV{Z_n \, W_{n,m}} \\
  & \leq \alpha + \left( Z^* + \frac{1}{n+1} \right) \kappa_{n,m} + \|Z_n\|_2 \, \kappa^{(2)}_{n,m} \\
  & \leq \alpha + \left( Z^* + \frac{1}{n+1} \right) \kappa_{n,m} + \frac{(2C+1)M}{\sqrt n} \, \kappa^{(2)}_{n,m}.
\end{split}
\end{align}

\emph{Step 2: bounding $Z^*$.}
\begin{itemize}
\item For IPCW, conditioning on $(T,X)$ and using $T\perp\!\!\!\perp C\mid X$
from Assumption~(A1), and recalling $p^\star(z)=\P{R(X,T)\ge z}$
from~\eqref{eq:def-pstar},
\begin{align*}
  \big| P\hat{\Psi}_z - p^\star(z) \big|
  & = \left| \EV{ \frac{E}{\hat G(\tilde T\mid X)}\,\I{R(X,\tilde T)\ge z}
      - \P{R(X,T)\ge z} } \right| \\
  & = \left| \EV{\left(\frac{G(T\mid X)}{\hat G(T\mid X)}-1\right)
      \I{R(X,T)\ge z}} \right|
    \leq \frac{1}{c}\,\EV{\big|G(T\mid X)-\hat G(T\mid X)\big|}
    \leq \frac{\mathcal G_n}{c},
\end{align*}
uniformly over $z$, the last step by Jensen.

\item For AIPCW, the calculation in the proof of
Proposition~\ref{prop:double-robustness} with $b_z(X)$ in place of $t_0$ gives
\[
  P\hat{\Psi}_z - p^\star(z)
  = \EV{ \int_0^{b_z(X)} S\,\frac{G}{\hat G}\,
      (\hat Q_z - Q_z)\,[d\Lambda^{C} - d\hat\Lambda^{C}] },
\]
with $\hat Q_z = \hat Q(u,b_z(X)\mid X)$, $Q_z = Q(u,b_z(X)\mid X)$, and the
dependence of $S,G,\hat G,d\Lambda^{C},d\hat\Lambda^{C}$ on $u$ and $X$
suppressed. Setting
$\|\hat S-S\|_\infty := \sup_{u\le t_0}|S(u\mid X)-\hat S(u\mid X)|$ and using
$\hat S(b_z(X)\mid X)\le \hat S(u^-\mid X)$ for $u\le b_z(X)$,
\begin{align*}
  |\hat Q_z - Q_z|
  & = \left| \frac{S(b_z(X)\mid X)}{S(u\mid X)}
           - \frac{\hat S(b_z(X)\mid X)}{\hat S(u^-\mid X)} \right| \\
  & \le \frac{\|\hat S-S\|_\infty}{s}
      + \frac{\hat S(b_z(X)\mid X)}{\hat S(u^-\mid X)}\cdot
        \frac{\|\hat S-S\|_\infty}{s}
    \leq \frac{2\,\|\hat S-S\|_\infty}{s},
\end{align*}
so that, by Cauchy--Schwarz,
\[
  \big|P\hat{\Psi}_z - p^\star(z)\big|
  \le \frac{2}{c\,s}\,
     \EV{\ \|\hat S-S\|_{\infty}
       \int_0^{t_0}\big|d\Lambda^{C}-d\hat\Lambda^{C}\big| }
  \le \frac{2}{c\,s}\;\mathcal S_n\,\mathcal L_n .
\]
\end{itemize}

\emph{Conclusion.} Substituting Step 2 into~\eqref{eq:eps-decomp-2}, we obtain
\begin{align*}
  \mathrm{FDR} & \leq \alpha + \left(\frac{\mathcal G_n}{c} + \frac{1}{1+n}\right)\kappa_{n,m} + \frac{(2C+1)M}{\sqrt n} \kappa^{(2)}_{n,m}, 
   && \text{under IPCW,} \\
  \mathrm{FDR} & \leq \alpha + \left(\frac{2\mathcal S_n\mathcal L_n}{cs} + \frac{1}{1+n}\right)\kappa_{n,m} + \frac{(2C+1)M}{\sqrt n} \kappa^{(2)}_{n,m},
   && \text{under AIPCW}.
\end{align*}
\end{proof}

\begin{proposition} \label{prop:eps}
For $z \in [0,1]$, let $\hat p(z) = \Pi\left([1+\sum_{i=1}^n \hat{\Psi}_z(O_i)]/(1+n)\right)$,
and suppose that $\hat{\Psi}_z = A_z - B_z$ almost surely, where the maps
$z\mapsto A_z(O)$ and $z\mapsto B_z(O)$ are non-increasing with
$0\le A_z(O)\le M$ and $0\le B_z(O)\le M$ almost surely, for some finite constant $M$.
Then, under Assumptions \textup{(A1)} and \textup{(A2)},
\begin{align*}
  \varepsilon_n(\hat p) 
  \leq \underbrace{\sup_{z\in[0,1]}\big|P\hat{\Psi}_z - p^\star(z)\big|}_{Z^*}  + \underbrace{\sup_{z\in(0,1]}\big|\mathbb P_n\hat{\Psi}_z- P\hat{\Psi}_z\big|}_{Z_n} + \frac{1}{1+n},
\end{align*}
almost surely, and moreover
\begin{align*}
  & \EV{Z_n} \leq \frac{2 M C}{\sqrt n},
  & \big\|Z_n\big\|_{2} \leq \frac{(2C+1)M}{\sqrt n},
\end{align*}
for a universal constant $C$, where $\|\cdot\|_{2}$ is the $L_2(P)$ norm.
\end{proposition}

\begin{proof}[Proof of Proposition~\ref{prop:eps}]

\emph{Step 1: almost-sure bound.}
Both $A_z$ and $B_z$ take values in $[0,M]$, so $|\hat\Psi_z|\le M$ almost surely.
Since $p^\star$ is non-increasing with values in $[0,1]$, Lemma~\ref{lem:majorant}
applies with $g=p^\star$, and from the definition of $\hat p$
in~\eqref{eq:pvalue-general},
\begin{align*}
  \varepsilon_n(\hat{p})
  & = \sup_{z \in (0,1]} \left| \hat{p}(z) -p^\star(z) \right|
   = \sup_{z \in (0,1]} \left| \Pi \left( \frac{1 + n \mathbb P_n\hat{\Psi}_z}{1+n} \right) -p^\star(z) \right| \\
  & \leq \sup_{z \in (0,1]} \left| \frac{1 + n \mathbb P_n\hat{\Psi}_z}{1+n} -p^\star(z) \right|
   = \sup_{z \in (0,1]} \left| \frac{n}{1+n}\big[ \mathbb P_n\hat{\Psi}_z - p^\star(z)\big] + \frac{1-p^\star(z)}{1+n} \right| \\
  & \leq \sup_{z \in (0,1]}\big|\mathbb P_n\hat{\Psi}_z-p^\star(z)\big| + \frac{1}{1+n}
    \leq Z^* + Z_n + \frac{1}{1+n}.
\end{align*}
This proves the first part of the result. Next, we bound $Z_n$.

\emph{Step 2: expected deviation of a monotonically indexed family.}
Let $\{H_z\}_{z\in[0,1]}$ be any family with $z\mapsto H_z(O)$ non-increasing
and $0\le H_z\le M$ almost surely, and set $\mathcal H:=\{H_z/M : z\in[0,1]\}$,
a class of functions of $O$ with values in $[0,1]$, totally ordered pointwise
by the index $z$.

Write $\psi(z):=\EV{H_z/M}$, non-increasing with values in $[0,1]$. Fix
$\varepsilon\in(0,1)$, let $N:=\lceil \varepsilon^{-2}\rceil$, and set
$z_k := \inf\{z : \psi(z)\le 1-k/N\}$ for $k=0,\dots,N$, with
$z_0=0$ and $z_N=1$. Every $z\in[0,1]$ lies in some $[z_k,z_{k+1}]$, and there
$H_z/M$ is contained pointwise in the bracket $[H_{z_{k+1}}/M,\,H_{z_k}/M]$,
whose squared $L_2(P)$ size satisfies, since $(H_{z_k}-H_{z_{k+1}})/M\in[0,1]$,
\[
  \EV{\Big(\tfrac{H_{z_k}-H_{z_{k+1}}}{M}\Big)^{2}}
  \leq \EV{\tfrac{H_{z_k}-H_{z_{k+1}}}{M}}
  \;=\; \psi(z_k)-\psi(z_{k+1}) \leq \tfrac1N \leq \varepsilon^{2}.
\]
Hence the bracketing number satisfies
$N_{[\,]}(\varepsilon,\mathcal H,L_2(P)) \le N \le \varepsilon^{-2}+1$, so that
$\log N_{[\,]}(\varepsilon,\mathcal H,L_2(P)) \le \log(1+1/\varepsilon^2) \leq \log 2 + 4\log(\varepsilon^{-1/2}) \leq 1 + 4 \varepsilon^{-1/2}$
and the bracketing entropy integral is finite:
\begin{align*}
  J_{[\,]}\big(1,\mathcal H,L_2(P)\big)
  := \int_0^{1}\sqrt{1+\log N_{[\,]}\big(\varepsilon,\mathcal H,L_2(P)\big)}\,d\varepsilon
  \leq \int_0^{1}\sqrt{2+4\varepsilon^{-1/2}}\,d\varepsilon
  \leq \sqrt2 + 2\int_0^{1}\varepsilon^{-1/4}\,d\varepsilon 
  < 5.
\end{align*}
The maximal inequality for classes with finite bracketing entropy integral
\citep[\S2.14.2]{Vaart2023} therefore gives
$\EV{\sup_z|\mathbb P_n(H_z/M)-P(H_z/M)|}\le C/\sqrt n$ for a universal constant
$C$, and multiplying by $M$,
\begin{equation} \label{eq:monotone-expectation}
  \EV{\sup_{z\in[0,1]}\big|\mathbb P_nH_z-PH_z\big|}
  \leq \frac{C\,M}{\sqrt n} ,
\end{equation}
uniformly over all such families $\{H_z\}$.

\emph{Step 3: expectation of $Z_n$.}
Applying~\eqref{eq:monotone-expectation} to $\{A_z\}$ and $\{B_z\}$ separately
and using $\hat{\Psi}_z=A_z-B_z$,
\[
  \EV{Z_n} \leq \EV{\sup_z\big|\mathbb P_nA_z-PA_z\big|}
   + \EV{\sup_z\big|\mathbb P_nB_z-PB_z\big|}
   \leq \frac{2 M C}{\sqrt n} ,
\]
which is the first moment bound.

\emph{Step 4: variance of $Z_n$.}
As a function of the $n$ calibration observations, $Z_n$ has bounded differences:
replacing one observation changes each average $\mathbb P_n\hat{\Psi}_z$ by at
most $2M/n$, since $|\hat{\Psi}_z|\le M$, and hence changes the supremum by at
most $2M/n$. The bounded-differences corollary of the Efron--Stein inequality
\citep[Corollary~3.2]{boucheron2013concentration}
therefore gives
\[
  \operatorname{Var}(Z_n) \leq \frac14\sum_{i=1}^{n}\Big(\frac{2M}{n}\Big)^{2}
  \;=\; \frac{M^{2}}{n},
\]
so that, using the result of Step 3,
\begin{align*}
  \|Z_n\|_{2} 
  & = \sqrt{ (\EV{Z_n})^2 + \operatorname{Var}(Z_n)}
    \leq \EV{Z_n} + \sqrt{\operatorname{Var}(Z_n)} 
   \leq \frac{2MC}{\sqrt n} + \frac{M}{\sqrt n}
    = \frac{(2C+1)M}{\sqrt n} .
\end{align*}

\end{proof}

\clearpage

\subsubsection{Relation Between FDR and Risk Control}

\begin{proof}[Proof of Theorem~\ref{thm:fdr-risk}]

To simplify the notation, let $A_j:=\I{j\in\hat{\mathcal{S}}}$ and $R:=\sum_{j=1}^m A_j = |\hat{\mathcal{S}}|$.

By construction, the BH cutoff equals $\alpha k/m$ on the event $\{R=k\}$, for any $k\in[m]$; hence,
\begin{equation}
\label{eq:Aj-on-Rk}
A_j=\I{p_j\le \alpha k/m}\qquad\text{on }\{R=k\}.
\end{equation}

For each $j$, let $\tilde R^{(j)}$ denote the BH rejection count after replacing $p_j$ by $0$.
It is a known and easily verified fact that, for every $j$ and $k\in[m]$,
\begin{equation}
\label{eq:key-identity}
A_j\,\I{R=k}=A_j\,\I{\tilde R^{(j)}=k}\qquad\text{a.s.}
\end{equation}
By the definition of FDR given in~\eqref{eq:FDR},
\begin{align*}
\mathrm{FDR}
&=\EV{\frac{\sum_{j=1}^m H^{0}_{j}A_j}{\max(R,1)}}
=\sum_{j=1}^m\sum_{k=1}^m \frac{1}{k}\,\EV{H^{0}_{j}A_j\,\I{R=k}}
=\sum_{j=1}^m\sum_{k=1}^m \frac{1}{k}\,\EV{H^{0}_{j}A_j\,\I{\tilde R^{(j)}=k}}.
\end{align*}
Using \eqref{eq:Aj-on-Rk} together with \eqref{eq:key-identity},
\[
A_j\,\I{\tilde R^{(j)}=k}=\I{p_j\le \alpha k/m}\,\I{\tilde R^{(j)}=k}\qquad\text{a.s.}
\]
Hence
\[
\mathrm{FDR}
=\sum_{j=1}^m\sum_{k=1}^m \frac{1}{k}\,
\EV{H^{0}_{j}\,\I{p_j\le \alpha k/m}\,\I{\tilde R^{(j)}=k}}.
\]

Let
\[
\mathcal F_{j,k}:=\sigma\!\left(\mathcal D_{\mathrm{cal}},\{p_\ell:\ell\neq j\},\I{p_j\le \alpha k/m}\right).
\]
Since $\tilde R^{(j)}$ depends only on $\mathcal D_{\mathrm{cal}}$ and $\{p_\ell:\ell\neq j\}$, the indicator $\I{\tilde R^{(j)}=k}$ is $\mathcal F_{j,k}$-measurable.
Applying the tower property,
\begin{align*}
\mathrm{FDR}
&=\sum_{j=1}^m\sum_{k=1}^m \frac{1}{k}\,
\EV{\EV{H^{0}_{j}\,\I{p_j\le \alpha k/m}\mid \mathcal F_{j,k}}\;\I{\tilde R^{(j)}=k}} \\
&=\sum_{j=1}^m\sum_{k=1}^m \frac{1}{k}\,
\EV{\I{p_j\le \alpha k/m}\,\EV{H^{0}_{j}\mid \mathcal F_{j,k},\,p_j\le \alpha k/m}\;\I{\tilde R^{(j)}=k}}.
\end{align*}
By conditional independence of $(p_j,H^{0}_{j})$ from $\{p_\ell:\ell\neq j\}$
given $\mathcal D_{\mathrm{cal}}$,
\[
\EV{H^{0}_{j}\mid \mathcal F_{j,k},\,p_j\le \alpha k/m}
=
\P{H^{0}_{j}=1\mid \mathcal D_{\mathrm{cal}},\,p_j\le \alpha k/m}.
\]
By the assumption that the pairs $(p_j,H^{0}_{j})_{j=1}^m$ are i.i.d.~conditional on $D_{\mathrm{cal}}$, this equals $\bar r(\alpha k/m; \mathcal D_{\mathrm{cal}})$.
Substituting,
\[
\mathrm{FDR}
=\sum_{j=1}^m\sum_{k=1}^m \frac{1}{k}\,
\EV{\I{p_j\le \alpha k/m}\,\bar r(\alpha k/m; \mathcal D_{\mathrm{cal}})\,\I{\tilde R^{(j)}=k}}.
\]
Using again \eqref{eq:Aj-on-Rk} and \eqref{eq:key-identity} to replace
$\I{p_j\le \alpha k/m}\,\I{\tilde R^{(j)}=k}$ with $A_j\,\I{R=k}$, we get
\[
\mathrm{FDR}
=\sum_{j=1}^m\sum_{k=1}^m \frac{1}{k}\,
\EV{A_j\,\bar r(\alpha k/m; \mathcal D_{\mathrm{cal}})\,\I{R=k}}.
\]
Finally, on $\{R=k\}$ we have $\sum_{j=1}^m A_j = k$, and therefore
\begin{align*}
\mathrm{FDR}
& =\sum_{k=1}^m \EV{\bar r(\alpha k/m; \mathcal D_{\mathrm{cal}})\,\I{R=k}} \\
& =\EV{\bar r\!\left(\frac{\alpha R}{m}; \mathcal D_{\mathrm{cal}}\right)\,\I{R>0}} \\
& =\EV{ \EV{ \bar r\!\left(\frac{\alpha R}{m}; \mathcal D_{\mathrm{cal}}\right)\,\I{R>0}
        \;\middle|\; \mathcal D_{\mathrm{cal}}} } \\
& =\EV{ \P{R>0 \mid \mathcal D_{\mathrm{cal}}} \cdot
        \EV{ \bar r\!\left(\frac{\alpha R}{m}; \mathcal D_{\mathrm{cal}}\right)
             \;\middle|\; R>0,\ \mathcal D_{\mathrm{cal}}} } .
\end{align*}

\end{proof}

\clearpage
\section{Additional Numerical Results}

\subsection{Estimating the Risk Exceedance Probability} \label{app:numerical-risk-estim}

This section explains how we estimate the risk exceedance probability $\P{r(A;\Dcal) > \alpha}$ 
in the semi-synthetic data experiments of Section~\ref{sec:experiments}.

Ignoring for simplicity that the covariates in our semi-synthetic experiments
are drawn without replacement from a finite population of size $7{,}000$, we
may treat the training, calibration, and cohort data sets analyzed in each
replicate as i.i.d.\ draws from a population; in fact, conditional on the covariates 
the event and censoring times
are, by construction, independent draws from generative models fitted to the real
FHRD data. Within this i.i.d.~model, consider the $b$-th replicate and condition on
its training data $\mathcal{D}_{\mathrm{tr},b}$ as well as on its calibration data $\mathcal{D}_{\mathrm{cal},b}$. 
The screening rule for this replicate, denoted here as $A_b$, is then fixed,
and the number of cohort subjects it selects is
\begin{equation*}
  N_b \mid \mathcal{D}_{\mathrm{cal},b} \;\sim\; \mathrm{Binom}(m, \mu_b),
  \qquad \mu_b := \P{A_b(X) = 1 \mid \mathcal{D}_{\mathrm{cal},b}}.
\end{equation*}
Conditioning further on $N_b$, the number of events among selected cohort subjects is
\begin{equation*}
  Y_b \mid N_b, \mathcal{D}_{\mathrm{cal},b} \;\sim\; \mathrm{Binom}(N_b, r_b),
  \qquad r_b := r(A_b; \mathcal{D}_{\mathrm{cal},b})
       = \P{T \le t_0 \mid A_b(X)=1, \mathcal{D}_{\mathrm{cal},b}} ,
\end{equation*}
where $(X,T)$ denotes a fresh draw from the population. Writing $P$ for the
joint law of $(\mu_b, r_b)$ induced by the draw of
$\mathcal{D}_{\mathrm{tr},b}, \mathcal{D}_{\mathrm{cal},b}$, the replicates therefore follow the
hierarchical model
\begin{align*}
  (\mu_b, r_b) &\;\overset{\text{i.i.d.}}{\sim}\; P, \\
  N_b \mid \mu_b &\;\sim\; \mathrm{Binom}(m, \mu_b), \\
  Y_b \mid N_b, r_b &\;\sim\; \mathrm{Binom}(N_b, r_b).
\end{align*}

The quantity of interest is the exceedance probability $\P{r_b > \alpha}$ under
the $r$-marginal of $P$. Since $r_b$ varies across replicates, the counts $Y_b$
are overdispersed relative to a binomial with a common success probability, and
the beta-binomial is the standard model for this situation
\citep{skellam1948probability}. We
therefore take the $r$-marginal to be $\mathrm{Beta}(a,b)$, fit $(a,b)$ by
maximum likelihood from the observed pairs $(Y_b, N_b)$ using the
\texttt{betabin} function of the \textsf{R} package \texttt{aod}
\citep{lesnoff2012aod}, and report the estimate
$\P{r > \alpha} = 1 - F_{a,b}(\alpha)$, where $F_{a,b}$ is the beta
distribution function.

\clearpage

\subsection{Experiments with Simulated Data} \label{app:numerical-sim}

\subsubsection{Performance at Different Time Horizons}

\begin{table}[!h]
  \centering
  \caption{Detailed numerical summary of the results in Figure~\ref{fig:1}.
    \emph{Yield} is the average number of patients selected. \emph{FDP} is the
    average false discovery proportion across all replicates, counting zero
    selections as FDP $=0$. The \emph{event rate} is calculated among selected
    patients, averaged over replicates that selected at least one, whose
    frequency is reported as $P(|S|>0)$.
    Each entry shows the mean and, in parentheses, two standard errors across
    replicates.
    The risk exceedance probability $P(r>\alpha)$ is estimated as described in Section~\ref{app:numerical-risk-estim} and 
    reported without a standard error.
    The event rate and risk exceedance probability are not reported if selections occur in fewer than 10\% of replicates, 
    where a dash marks the corresponding entry.
    Entries in red are those where the event rate lies above the target $\alpha$, or
    where the estimate of $P(r>\alpha)$ lies above the nominal $\delta$; the Oracle is never flagged.
    Values in bold indicate the method selecting the most patients at each horizon,
    among those not flagged in red and excluding the Oracle.
  }
\label{tab:semi-synthetic_grf}
{\footnotesize
\input{tables/exceedance_0_ntrain_5000_ncal_1000_aipcw.txt}
}
\end{table}

\FloatBarrier
\clearpage

\subsubsection{The Effect of the HP Significance Level $\delta$}

\begin{table*}[!h]
\centering
\caption{Detailed numerical summary of the results in Figure~\ref{fig:2}, for time horizons $t_0 =2,3$ months. Other details are as in Table~\ref{tab:semi-synthetic_grf}.}
\label{tab:fig2-details-23}
{\footnotesize
\input{tables/exceedance_delta_0_ntrain_5000_ncal_1000_t23_aipcw.txt}
}
\end{table*}

\begin{table*}[!h]
\centering
\caption{Detailed numerical summary of the results in Figure~\ref{fig:2}, for time horizons $t_0 =6,9$ months. Other details are as in Table~\ref{tab:semi-synthetic_grf}.}
\label{tab:fig2-details-69}
{\footnotesize
\input{tables/exceedance_delta_0_ntrain_5000_ncal_1000_t69_aipcw.txt}
}
\end{table*}

\FloatBarrier
\clearpage

\subsubsection{The Effect of Survival Model Misspecification}

\begin{table}[!h]
\centering
\caption{Detailed numerical summary of results corresponding to Figure~\ref{fig:3}.
Other details are as in Table~\ref{tab:semi-synthetic_grf}.
}
\label{tab:semi-synthetic_xgb}
{\footnotesize
\input{tables/exceedance_data_0_ntrain_5000_ncal_1000_xgb_aipcw.txt}
}
\end{table}

\begin{figure}[!h]
\centering
\includegraphics[width=0.9\textwidth]{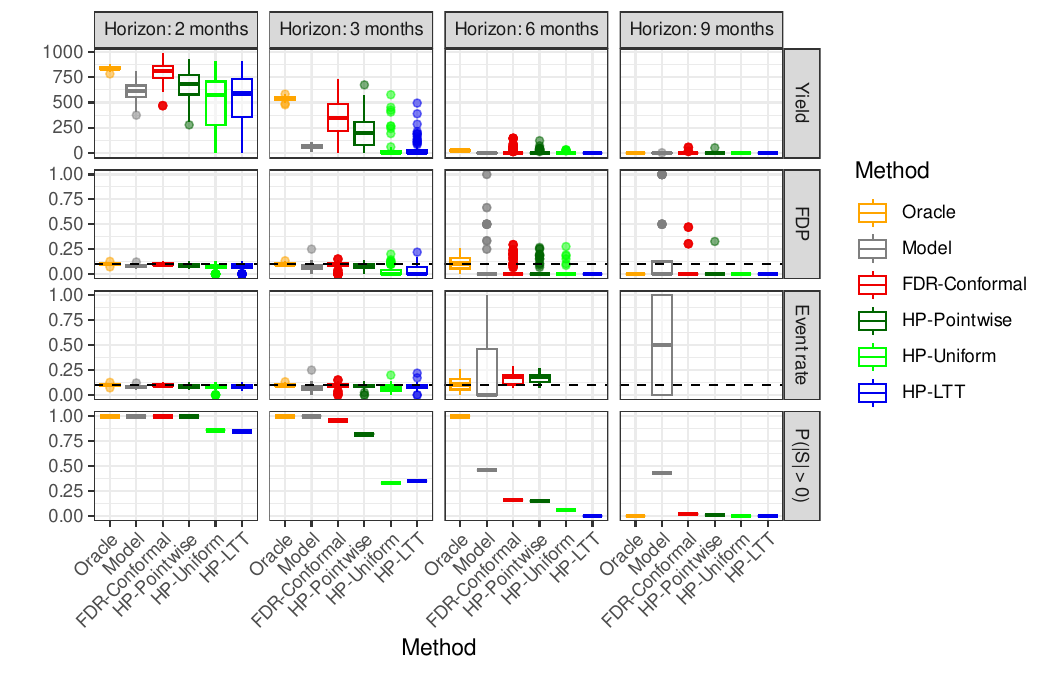}
\caption{
Summary of low-risk screening results obtained by applying different calibration methods to semi-synthetic oncology data simulated using a random forest generative model, at different screening horizons. All methods use the same mis-specified Cox survival model. Other details as in Figure~\ref{fig:1}.
}
\label{fig:semi-synthetic-n1000-cox-delta0.1}
\end{figure}

\begin{table}[!h]
\centering
\caption{Detailed numerical summary of results corresponding to Figure~\ref{fig:semi-synthetic-n1000-cox-delta0.1}.
Other details are as in Table~\ref{tab:semi-synthetic_grf}.
}
\label{tab:semi-synthetic_cox}
{\footnotesize
\input{tables/exceedance_data_0_ntrain_5000_ncal_1000_cox_aipcw.txt}
}
\end{table}

\FloatBarrier
\subsubsection{The Effect of the Calibration Sample Size}

\begin{figure}[!h]
\centering
\includegraphics[width=0.9\textwidth]{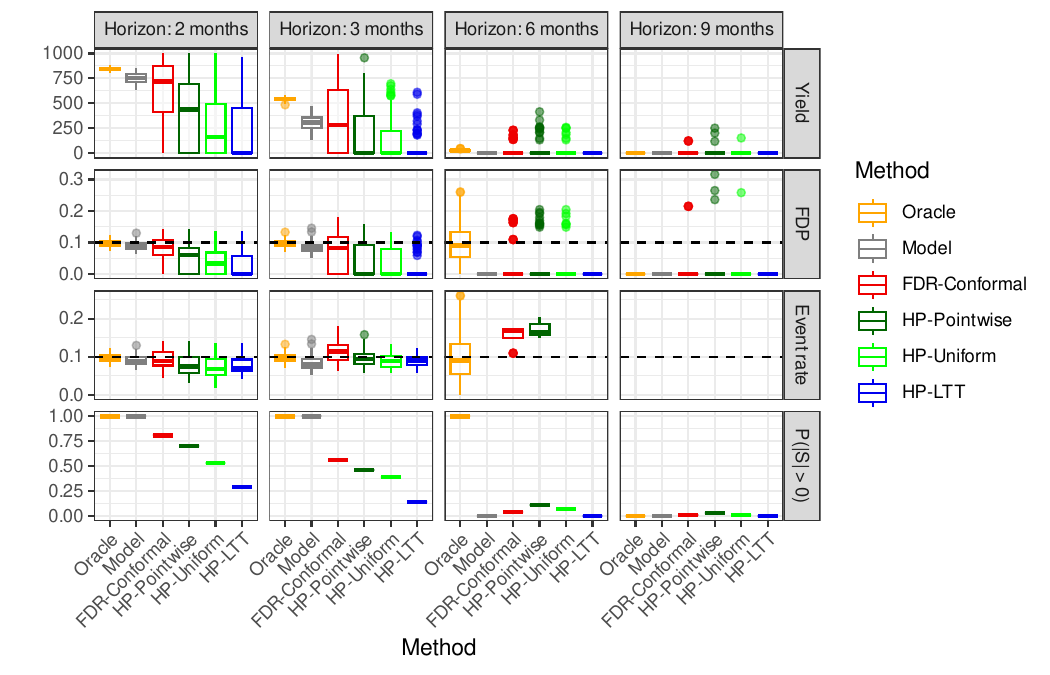}
\caption{
Summary of low-risk screening results obtained by applying different methods to semi-synthetic oncology data simulated from a random forest generative survival model, at different screening horizons. Calibration sample size 100. Other details as in Figure~\ref{fig:1}.}
\label{fig:semi-synthetic-n100-grf-delta0.1}
\end{figure}

\begin{table}[!h]
\centering
\caption{Detailed numerical summary of results corresponding to Figure~\ref{fig:semi-synthetic-n100-grf-delta0.1}.
Other details are as in Table~\ref{tab:semi-synthetic_grf}.
}
\label{tab:semi-synthetic_grf_n100}
{\footnotesize
\input{tables/exceedance_data_0_ntrain_5000_ncal_100_grf_aipcw.txt}
}
\end{table}

\begin{figure}[!htbp]
\centering
\includegraphics[width=0.9\textwidth]{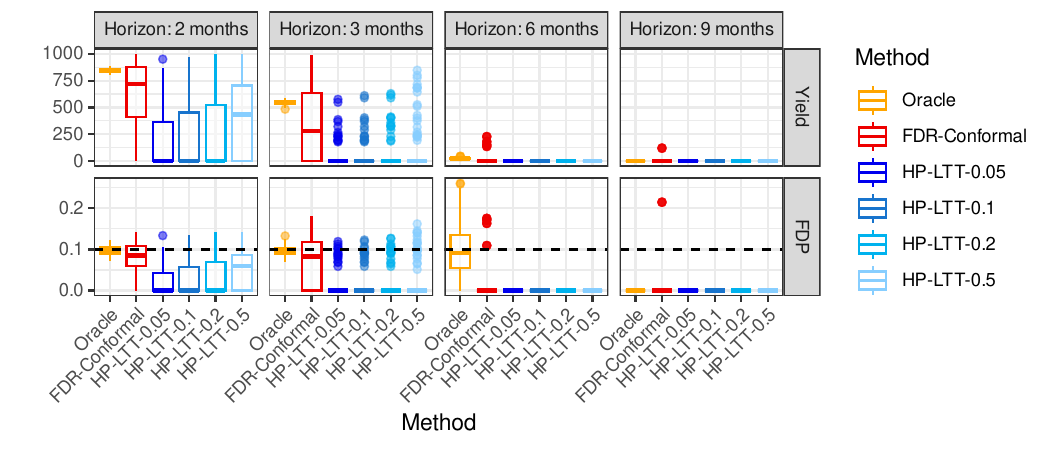}
\caption{
Summary of low-risk screening results on semi-synthetic oncology data, similar to Figure~\ref{fig:1}.
The LTT calibration method is applied at different significance levels $\delta$, ranging from $\delta = 0.05$ (more conservative) to $\delta=0.5$ (more liberal).
Here, the calibration sample size is 100 instead of 1000.
}
\label{fig:semi-synthetic-n100-grf-delta}
\end{figure}

\FloatBarrier
\subsubsection{The Advantage of AIPCW for Double Robustness} \label{app:synthetic-dr}

\begin{figure*}[!htbp]
\centering
\includegraphics[width=0.9\textwidth]{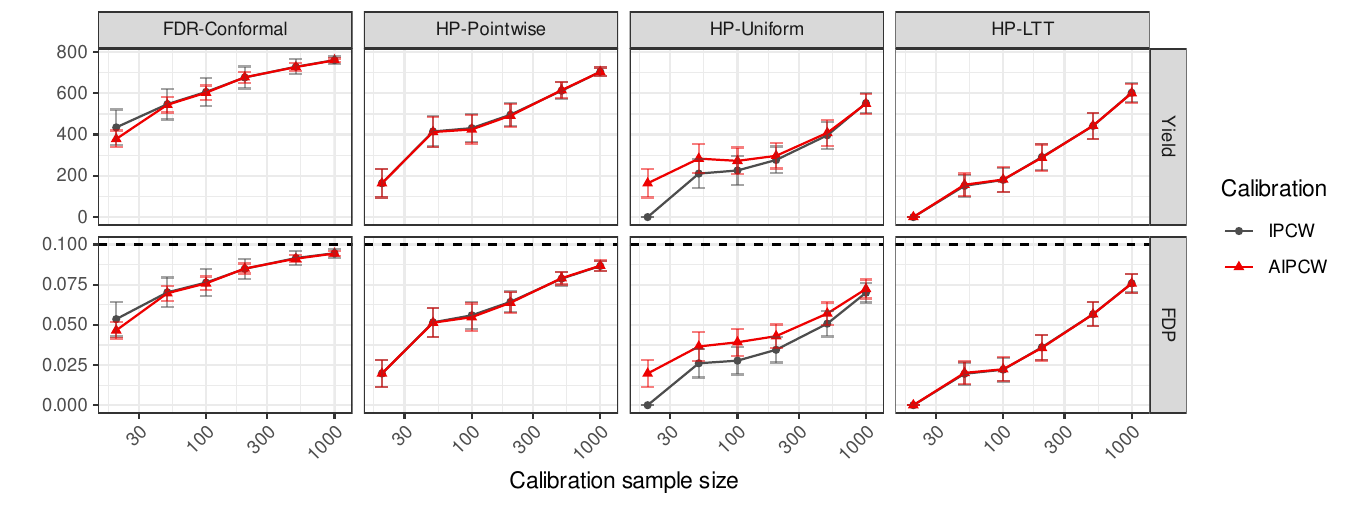}
\caption{
Average performance of low-risk screening methods on semi-synthetic data, as a
function of the calibration sample size, at horizon $t_0=2$ months. Each panel
contrasts one calibration method applied with AIPCW against the same method applied
with event-time IPCW pseudo-outcomes; the two are nearly indistinguishable
here, because the censoring mechanism in these data is easy to estimate and the
augmentation term has little left to correct. All methods use the same
generalized random forest survival model. Other details as in
Figure~\ref{fig:4}.
}
\label{fig:semi-synthetic-dr-easy}
\end{figure*}

The adversarial setting of Figure~\ref{fig:5} is obtained by tilting the
semi-synthetic generator of Figure~\ref{fig:1}. Both conditional
distributions are tilted on the hazard scale: a survival curve satisfies
$S(t \mid x) = \exp\{-H(t \mid x)\}$, so raising it to a power multiplies the
cumulative hazard,
\begin{equation*}
  S_{\mathrm{tilt}}(t \mid x) = S(t \mid x)^{\exp\{f_T(x)\}},
  \qquad
  G_{\mathrm{tilt}}(t \mid x) = G(t \mid x)^{\exp\{f_C(x)\}} ,
\end{equation*}
where $S$ and $G$ are the curves fitted to the real data and $f_T, f_C$ are
log hazard ratios of our choosing.

We take both to depend on a single numerical covariate, $X_{32}$, split at its
median $m_{32}$:
\begin{equation*}
  f_T(x) = -1 + 4 \cdot \mathbf{1}\{x_{32} > m_{32}\},
  \qquad
  f_C(x) = -1 + 6 \cdot \mathbf{1}\{x_{32} > m_{32}\} .
\end{equation*}
Patients above the median therefore have their event hazard multiplied by
$e^{3} \approx 20$ and their censoring hazard by $e^{5} \approx 148$, while
those below have both multiplied by $e^{-1} \approx 0.37$. Censoring varies
roughly $400$-fold across the cohort and falls most heavily on the patients at
highest risk, so a censoring model that ignores this dependence tends to 
produce selections with higher event rate than expected.

The alternative distribution from which the misspecified censoring samples are
drawn uses the same survival tilt $f_T$ but sets $f_C(x) = -1$ for every
patient, removing the dependence on $X_{32}$ entirely. The censoring model is
fitted on a sample in which a fraction $\pi$ of the observations comes from
this alternative distribution and the remainder from the target one; $\pi$ is
the mixing proportion on the horizontal axis of Figure~\ref{fig:5}.

\FloatBarrier
\clearpage

\subsection{Experiments with Real Data} \label{app:numerical-real}

\begin{figure}[!h]
\centering
\includegraphics[width=0.9\textwidth]{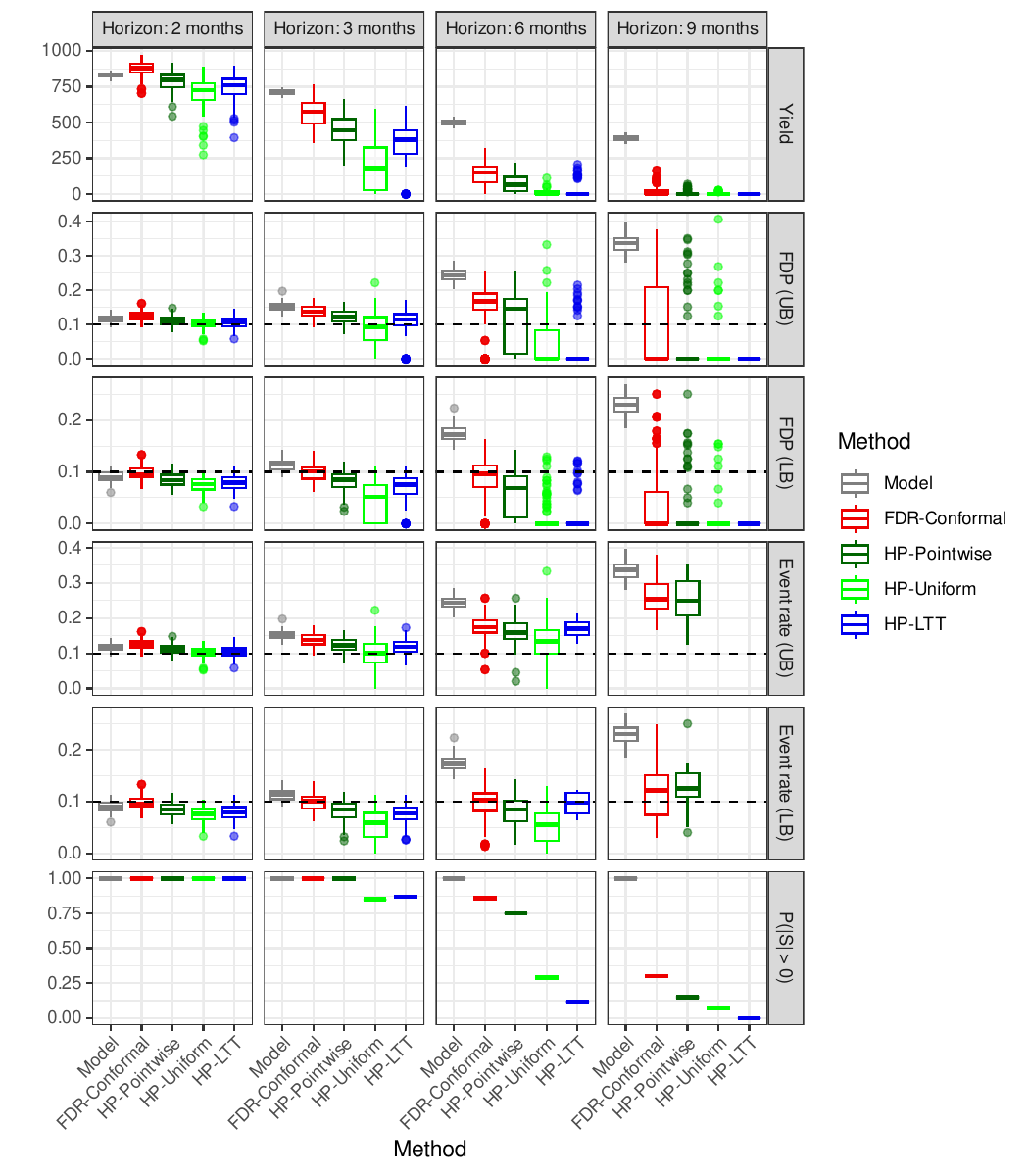}
\caption{Summary of low-risk screening results obtained by applying different methods to real oncology data, at different screening horizons, using a gradient boosting survival model. Other details as in Table~\ref{tab:real_xgb}.}
\label{fig:real-n1000-xgb-delta0.1}
\end{figure}

\begin{table}[!h]
\centering
\caption{Summary of low-risk screening results obtained by applying different methods to real oncology data, at different screening horizons, using a Cox survival model. Other details are as in Table~\ref{tab:real_xgb}.}
\label{tab:real_cox}
{\footnotesize
\input{tables/data_1_ntrain_5000_ncal_1000_cox_aipcw.txt}
}
\end{table}

\begin{figure}[!h]
\centering
\includegraphics[width=0.9\textwidth]{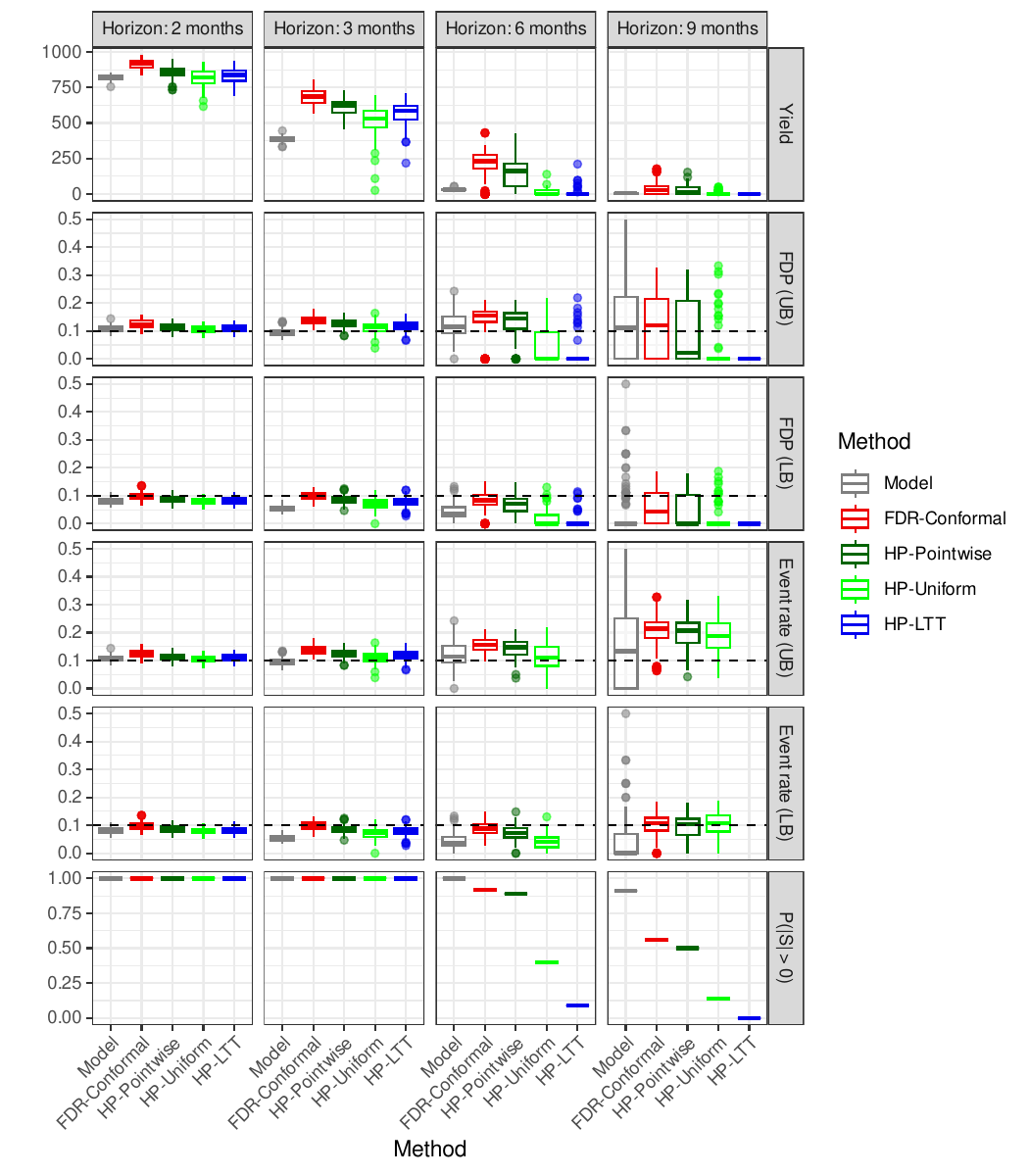}
\caption{Summary of low-risk screening results obtained by applying different methods to real oncology data, at different screening horizons, using a Cox survival model. Other details as in Figure~\ref{fig:real-n1000-xgb-delta0.1}.}
\label{fig:real-n1000-cox-delta0.1}
\end{figure}

\begin{table}[!h]
\centering
\caption{Summary of low-risk screening results obtained by applying different methods to real oncology data, at different screening horizons, using a generalized random forest survival model. Other details are as in Table~\ref{tab:real_xgb}.}
\label{tab:real_grf}
{\footnotesize
\input{tables/data_1_ntrain_5000_ncal_1000_grf_aipcw.txt}
}
\end{table}

\begin{figure}[!h]
\centering
\includegraphics[width=0.9\textwidth]{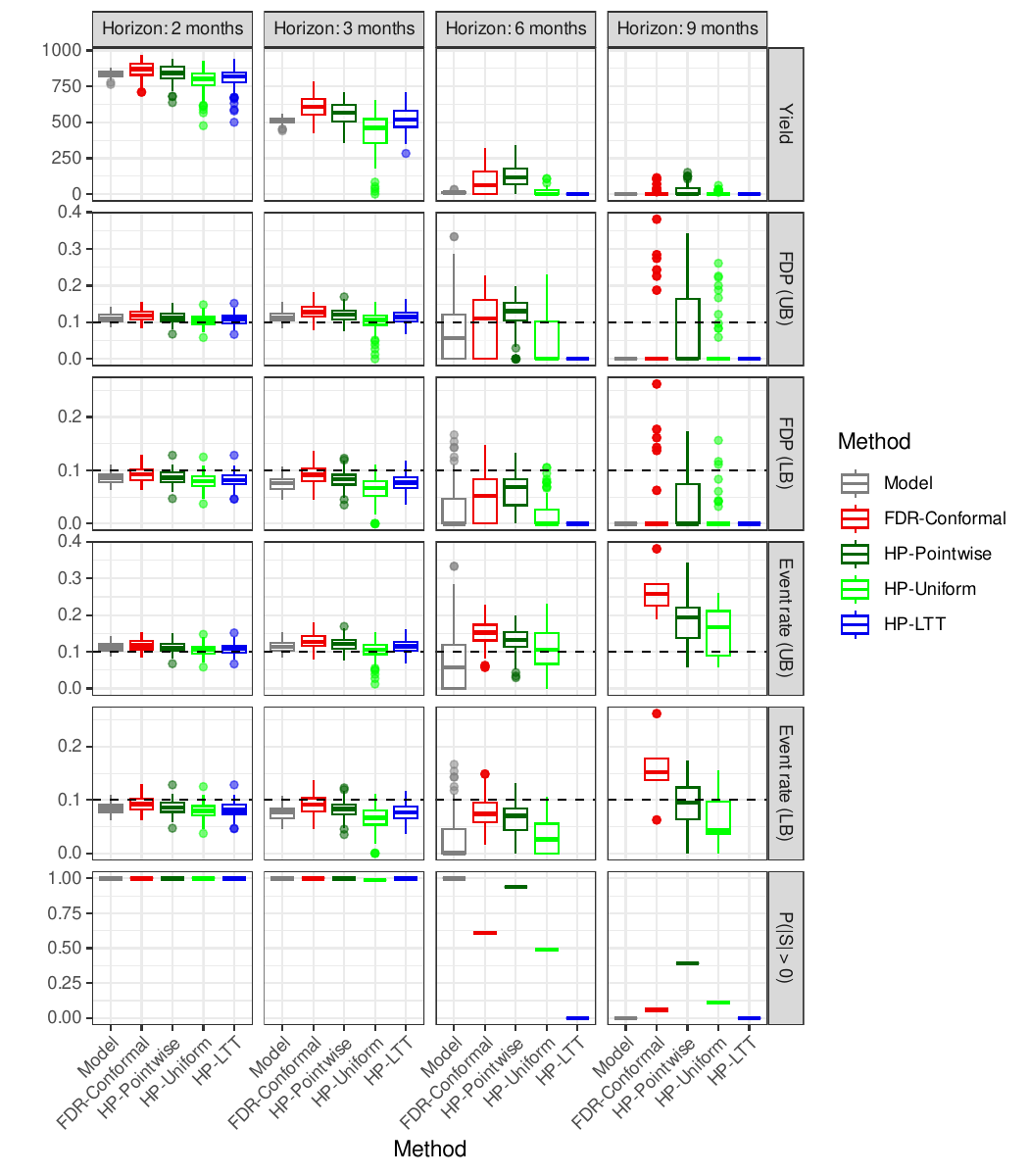}
\caption{Summary of low-risk screening results obtained by applying different methods to real oncology data, at different screening horizons, using a generalized random forest survival model. Other details as in Figure~\ref{fig:real-n1000-xgb-delta0.1}.}
\label{fig:real-n1000-grf-delta0.1}
\end{figure}

\begin{figure}[!h]
\centering
\includegraphics[width=0.9\textwidth]{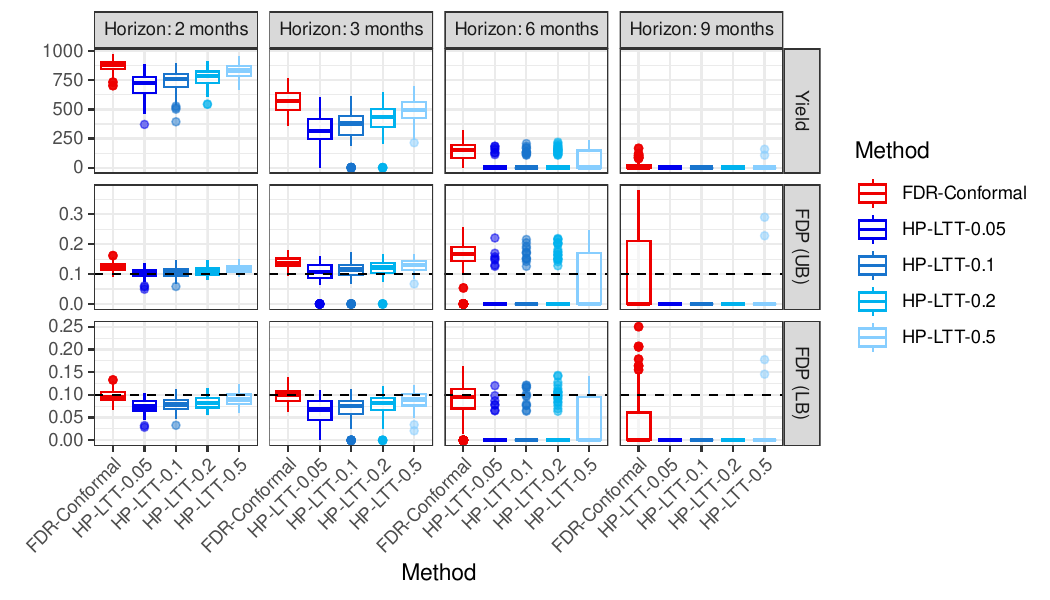}
\caption{
Summary of low-risk screening results on real oncology data, using a gradient boosting survival model as in Figure~\ref{fig:real-n1000-xgb-delta0.1}.
Here, the LTT calibration method is applied at different significance levels $\delta$, ranging from $\delta = 0.05$ (more conservative) to $\delta=0.5$ (more liberal).}
\label{fig:real-n1000-xgb-delta}
\end{figure}

\begin{figure}[!h]
\centering
\includegraphics[width=0.9\textwidth]{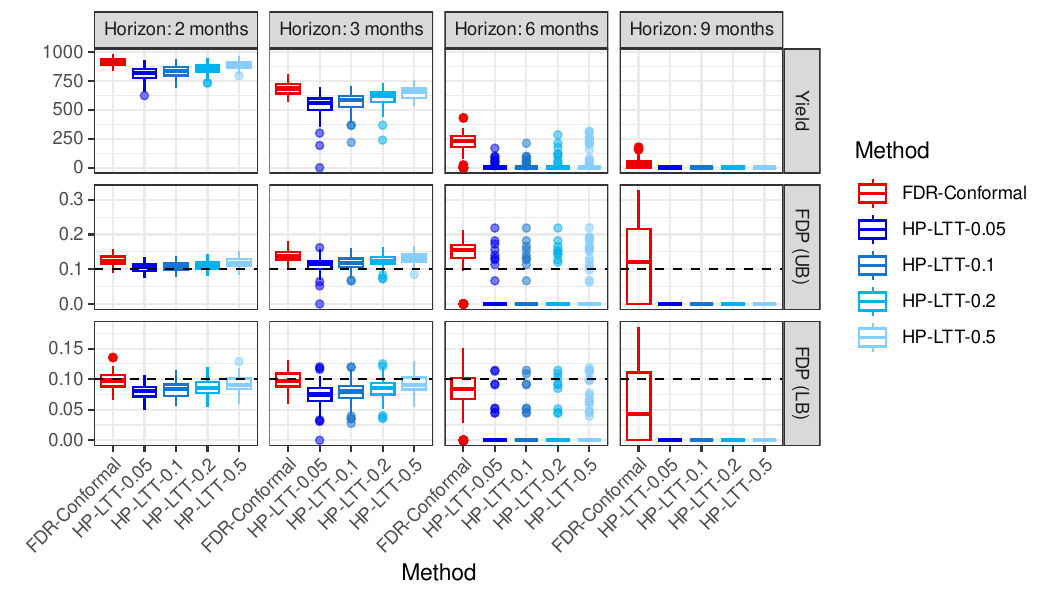}
\caption{
Summary of low-risk screening results on real oncology data, using a Cox survival model as in Figure~\ref{fig:real-n1000-cox-delta0.1}.
Here, the LTT calibration method is applied at different significance levels $\delta$, ranging from $\delta = 0.05$ (more conservative) to $\delta=0.5$ (more liberal).}
\label{fig:real-n1000-cox-delta}
\end{figure}

\begin{figure}[!h]
\centering
\includegraphics[width=0.9\textwidth]{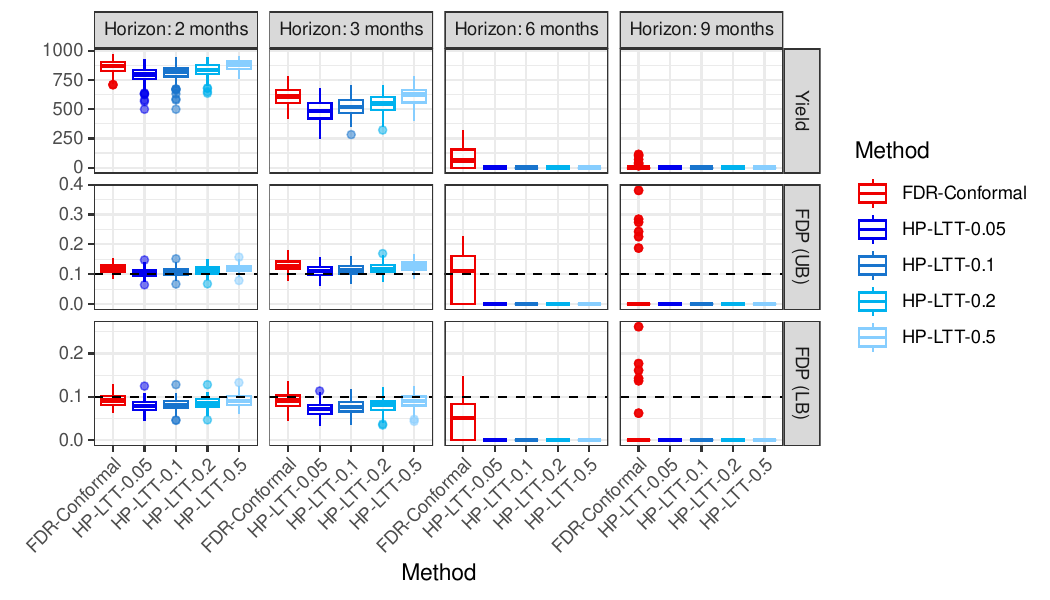}
\caption{
Summary of low-risk screening results on real oncology data, using a generalized random forest survival model as in Figure~\ref{fig:real-n1000-grf-delta0.1}.
Here, the LTT calibration method is applied at different significance levels $\delta$, ranging from $\delta = 0.05$ (more conservative) to $\delta=0.5$ (more liberal).}
\label{fig:real-n1000-grf-delta}
\end{figure}

\begin{table*}[!htb]
\centering
\caption{Summary of low-risk screening results on FHRD data across different screening horizons. All methods use IPCW rather than AIPCW pseudo-outcomes. Other details are as in Table~\ref{tab:real_xgb}.
}
\label{tab:real_xgb_ipcw}
{\footnotesize
\input{tables/data_1_ntrain_5000_ncal_1000_xgb.txt}
}
\end{table*}


%% file: tables/exceedance_0_ntrain_5000_ncal_1000_aipcw.txt
\begin{tabular}[t]{lrrrrr}
\toprule
Method & Yield & FDP & Event rate & $P(r>\alpha)$ & $P(|S|>0)$\\
\midrule
\addlinespace[0.3em]
\multicolumn{6}{l}{\textbf{Horizon: 2 months}}\\
\hspace{1em}Oracle & 840 (3.68) & 0.099 (0.002) & 0.099 (0.002) & 0.00 & 1.00 (0.00)\\
\hspace{1em}Model & 757 (8.65) & 0.092 (0.002) & 0.092 (0.002) & 0.03 & 1.00 (0.00)\\
\hspace{1em}FDR-Conformal & \textbf{760} (8.86) & 0.094 (0.001) & 0.094 (0.001) & 0.29 & 1.00 (0.00)\\
\hspace{1em}HP-Pointwise & 704 (21.12) & 0.087 (0.003) & 0.087 (0.003) & \textcolor{red}{0.14} & 1.00 (0.00)\\
\hspace{1em}HP-Uniform & 548 (49.70) & 0.073 (0.006) & 0.081 (0.004) & 0.04 & 0.90 (0.06)\\
\hspace{1em}HP-LTT & 601 (46.45) & 0.076 (0.006) & 0.084 (0.003) & 0.05 & 0.90 (0.06)\\
\addlinespace[0.3em]
\multicolumn{6}{l}{\textbf{Horizon: 3 months}}\\
\hspace{1em}Oracle & 538 (4.16) & 0.099 (0.003) & 0.099 (0.003) & 0.35 & 1.00 (0.00)\\
\hspace{1em}Model & 325 (14.50) & 0.088 (0.003) & 0.088 (0.003) & 0.00 & 1.00 (0.00)\\
\hspace{1em}FDR-Conformal & \textbf{348} (21.05) & 0.080 (0.004) & 0.098 (0.002) & 0.55 & 0.81 (0.04)\\
\hspace{1em}HP-Pointwise & 303 (34.63) & 0.081 (0.007) & 0.091 (0.004) & \textcolor{red}{0.19} & 0.89 (0.06)\\
\hspace{1em}HP-Uniform & 72 (28.90) & 0.037 (0.010) & 0.088 (0.010) & 0.00 & 0.42 (0.10)\\
\hspace{1em}HP-LTT & 126 (33.20) & 0.039 (0.009) & 0.084 (0.006) & 0.06 & 0.46 (0.10)\\
\addlinespace[0.3em]
\multicolumn{6}{l}{\textbf{Horizon: 6 months}}\\
\hspace{1em}Oracle & 25 (1.23) & 0.109 (0.013) & 0.109 (0.013) & 0.79 & 1.00 (0.00)\\
\hspace{1em}Model & 0 (0.00) & 0.000 (0.000) & -- & -- & 0.00 \vphantom{1} (0.00)\\
\hspace{1em}FDR-Conformal & \textbf{2} (1.35) & 0.007 (0.004) & -- & -- & 0.05 (0.02)\\
\hspace{1em}HP-Pointwise & 4 (2.72) & 0.013 (0.008) & \textcolor{red}{0.119 (0.018)} & \textcolor{red}{1.00} & 0.11 (0.06)\\
\hspace{1em}HP-Uniform & 1 (0.91) & 0.005 (0.005) & -- & -- & 0.04 (0.04)\\
\hspace{1em}HP-LTT & 0 (0.00) & 0.000 (0.000) & -- & -- & 0.00 \vphantom{1} (0.00)\\
\addlinespace[0.3em]
\multicolumn{6}{l}{\textbf{Horizon: 9 months}}\\
\hspace{1em}Oracle & 0 (0.00) & 0.000 (0.000) & -- & -- & 0.00 (0.00)\\
\hspace{1em}Model & 0 (0.00) & 0.000 (0.000) & -- & -- & 0.00 (0.00)\\
\hspace{1em}FDR-Conformal & 0 (0.26) & 0.002 (0.002) & -- & -- & 0.01 (0.01)\\
\hspace{1em}HP-Pointwise & \textbf{1} (0.78) & 0.007 (0.009) & -- & -- & 0.03 (0.03)\\
\hspace{1em}HP-Uniform & 0 (0.28) & 0.001 (0.001) & -- & -- & 0.01 (0.02)\\
\hspace{1em}HP-LTT & 0 (0.00) & 0.000 (0.000) & -- & -- & 0.00 (0.00)\\
\bottomrule
\end{tabular}

%% file: tables/exceedance_delta_0_ntrain_5000_ncal_1000_t23_aipcw.txt
\begin{tabular}[t]{llrrrrr}
\toprule
$\delta$ & Method & Yield & FDP & Event rate & $P(r>\alpha)$ & $P(|S|>0)$\\
\midrule
\addlinespace[0.3em]
\multicolumn{7}{l}{\textbf{Horizon: 2 months}}\\
\hspace{1em}-- & FDR-Conformal & \textbf{760} (8.86) & 0.094 (0.001) & 0.094 (0.001) & 0.29 & 1.00 (0.00)\\
\hspace{1em} & Model & 757 (8.65) & 0.092 (0.002) & 0.092 (0.002) & 0.03 & 1.00 (0.00)\\
\hspace{1em} & Oracle & 840 (3.68) & 0.099 (0.002) & 0.099 (0.002) & 0.00 & 1.00 (0.00)\\
\hspace{1em}0.05 & HP-LTT & 554 (48.36) & 0.072 (0.006) & 0.080 (0.003) & 0.03 & 0.89 (0.06)\\
\hspace{1em} & HP-Pointwise & 647 (30.70) & 0.082 (0.003) & 0.082 (0.003) & \textcolor{red}{0.08} & 1.00 (0.00)\\
\hspace{1em} & HP-Uniform & 490 (51.91) & 0.068 (0.006) & 0.078 (0.004) & 0.02 & 0.88 (0.07)\\
\hspace{1em}0.10 & HP-LTT & 601 (46.45) & 0.076 (0.006) & 0.084 (0.003) & 0.05 & 0.90 (0.06)\\
\hspace{1em} & HP-Pointwise & 704 (21.12) & 0.087 (0.003) & 0.087 (0.003) & \textcolor{red}{0.14} & 1.00 (0.00)\\
\hspace{1em} & HP-Uniform & 548 (49.70) & 0.073 (0.006) & 0.081 (0.004) & 0.04 & 0.90 (0.06)\\
\hspace{1em}0.20 & HP-LTT & 650 (42.96) & 0.081 (0.005) & 0.087 (0.003) & 0.13 & 0.93 (0.05)\\
\hspace{1em} & HP-Pointwise & 751 (18.22) & 0.092 (0.003) & 0.092 (0.003) & \textcolor{red}{0.22} & 1.00 (0.00)\\
\hspace{1em} & HP-Uniform & 627 (43.02) & 0.079 (0.005) & 0.084 (0.004) & 0.10 & 0.94 (0.05)\\
\hspace{1em}0.50 & HP-LTT & 755 (19.58) & 0.092 (0.003) & 0.092 (0.003) & 0.24 & 1.00 (0.00)\\
\hspace{1em} & HP-Pointwise & 821 (15.76) & 0.100 (0.003) & \textcolor{red}{0.100 (0.003)} & 0.50 & 1.00 (0.00)\\
\hspace{1em} & HP-Uniform & 747 (19.47) & 0.091 (0.003) & 0.091 (0.003) & 0.23 & 1.00 (0.00)\\
\addlinespace[0.3em]
\multicolumn{7}{l}{\textbf{Horizon: 3 months}}\\
\hspace{1em}-- & FDR-Conformal & \textbf{348} (21.05) & 0.080 (0.004) & 0.098 (0.002) & 0.55 & 0.81 (0.04)\\
\hspace{1em} & Model & 325 (14.50) & 0.088 (0.003) & 0.088 (0.003) & 0.00 & 1.00 (0.00)\\
\hspace{1em} & Oracle & 538 (4.16) & 0.099 (0.003) & 0.099 (0.003) & 0.35 & 1.00 (0.00)\\
\hspace{1em}0.05 & HP-LTT & 83 (27.85) & 0.027 (0.008) & 0.081 (0.007) & 0.00 & 0.33 (0.09)\\
\hspace{1em} & HP-Pointwise & 231 (35.61) & 0.072 (0.008) & 0.088 (0.005) & \textcolor{red}{0.15} & 0.82 (0.08)\\
\hspace{1em} & HP-Uniform & 38 (20.35) & 0.030 (0.009) & 0.082 (0.012) & 0.00 & 0.36 (0.10)\\
\hspace{1em}0.10 & HP-LTT & 126 (33.20) & 0.039 (0.009) & 0.084 (0.006) & 0.06 & 0.46 (0.10)\\
\hspace{1em} & HP-Pointwise & 303 (34.63) & 0.081 (0.007) & 0.091 (0.004) & \textcolor{red}{0.19} & 0.89 (0.06)\\
\hspace{1em} & HP-Uniform & 72 (28.90) & 0.037 (0.010) & 0.088 (0.010) & 0.00 & 0.42 (0.10)\\
\hspace{1em}0.20 & HP-LTT & 177 (38.56) & 0.049 (0.009) & 0.087 (0.006) & \textcolor{red}{0.22} & 0.56 (0.10)\\
\hspace{1em} & HP-Pointwise & 362 (32.80) & 0.088 (0.006) & 0.093 (0.004) & \textcolor{red}{0.31} & 0.95 (0.04)\\
\hspace{1em} & HP-Uniform & 123 (35.50) & 0.048 (0.010) & 0.090 (0.007) & 0.13 & 0.53 (0.10)\\
\hspace{1em}0.50 & HP-LTT & 272 (41.06) & 0.070 (0.009) & 0.093 (0.005) & 0.34 & 0.75 (0.09)\\
\hspace{1em} & HP-Pointwise & 485 (27.17) & 0.101 (0.004) & \textcolor{red}{0.101 (0.004)} & \textcolor{red}{0.67} & 1.00 (0.00)\\
\hspace{1em} & HP-Uniform & 290 (40.58) & 0.078 (0.008) & 0.096 (0.005) & 0.31 & 0.81 (0.08)\\
\bottomrule
\end{tabular}

%% file: tables/exceedance_delta_0_ntrain_5000_ncal_1000_t69_aipcw.txt
\begin{tabular}[t]{llrrrrr}
\toprule
$\delta$ & Method & Yield & FDP & Event rate & $P(r>\alpha)$ & $P(|S|>0)$\\
\midrule
\addlinespace[0.3em]
\multicolumn{7}{l}{\textbf{Horizon: 6 months}}\\
\hspace{1em}-- & FDR-Conformal & 2 (1.35) & 0.007 (0.004) & -- & -- & 0.05 (0.02)\\
\hspace{1em} & Model & 0 (0.00) & 0.000 (0.000) & -- & -- & 0.00 \vphantom{1} (0.00)\\
\hspace{1em} & Oracle & 25 (1.23) & 0.109 (0.013) & 0.109 (0.013) & 0.79 & 1.00 (0.00)\\
\hspace{1em}0.05 & HP-LTT & 0 (0.00) & 0.000 (0.000) & -- & -- & 0.00 \vphantom{1} (0.00)\\
\hspace{1em} & HP-Pointwise & \textbf{3} (2.24) & 0.011 (0.007) & -- & -- & 0.09 (0.06)\\
\hspace{1em} & HP-Uniform & 1 (0.91) & 0.005 (0.005) & -- & -- & 0.04 \vphantom{1} (0.04)\\
\hspace{1em}0.10 & HP-LTT & 0 (0.00) & 0.000 (0.000) & -- & -- & 0.00 \vphantom{1} (0.00)\\
\hspace{1em} & HP-Pointwise & 4 (2.72) & 0.013 (0.008) & \textcolor{red}{0.119 (0.018)} & \textcolor{red}{1.00} & 0.11 (0.06)\\
\hspace{1em} & HP-Uniform & 1 (0.91) & 0.005 (0.005) & -- & -- & 0.04 (0.04)\\
\hspace{1em}0.20 & HP-LTT & 0 (0.00) & 0.000 (0.000) & -- & -- & 0.00 \vphantom{1} (0.00)\\
\hspace{1em} & HP-Pointwise & 16 (7.37) & 0.049 (0.016) & \textcolor{red}{0.148 (0.023)} & \textcolor{red}{1.00} & 0.33 (0.09)\\
\hspace{1em} & HP-Uniform & 1 (0.93) & 0.005 (0.005) & -- & -- & 0.05 (0.04)\\
\hspace{1em}0.50 & HP-LTT & 0 (0.00) & 0.000 (0.000) & -- & -- & 0.00 \vphantom{1} (0.00)\\
\hspace{1em} & HP-Pointwise & 39 (10.94) & 0.088 (0.019) & \textcolor{red}{0.160 (0.018)} & \textcolor{red}{1.00} & 0.55 (0.10)\\
\hspace{1em} & HP-Uniform & 1 (1.01) & 0.006 (0.005) & -- & -- & 0.06 (0.05)\\
\addlinespace[0.3em]
\multicolumn{7}{l}{\textbf{Horizon: 9 months}}\\
\hspace{1em}-- & FDR-Conformal & 0 (0.26) & 0.002 (0.002) & -- & -- & 0.01 (0.01)\\
\hspace{1em} & Model & 0 (0.00) & 0.000 (0.000) & -- & -- & 0.00 (0.00)\\
\hspace{1em} & Oracle & 0 (0.00) & 0.000 (0.000) & -- & -- & 0.00 (0.00)\\
\hspace{1em}0.05 & HP-LTT & 0 (0.00) & 0.000 (0.000) & -- & -- & 0.00 (0.00)\\
\hspace{1em} & HP-Pointwise & 0 (0.59) & 0.004 (0.006) & -- & -- & 0.02 (0.03)\\
\hspace{1em} & HP-Uniform & 0 (0.28) & 0.001 (0.001) & -- & -- & 0.01 \vphantom{3} (0.02)\\
\hspace{1em}0.10 & HP-LTT & 0 (0.00) & 0.000 (0.000) & -- & -- & 0.00 (0.00)\\
\hspace{1em} & HP-Pointwise & 1 (0.78) & 0.007 (0.009) & -- & -- & 0.03 (0.03)\\
\hspace{1em} & HP-Uniform & 0 (0.28) & 0.001 (0.001) & -- & -- & 0.01 \vphantom{2} (0.02)\\
\hspace{1em}0.20 & HP-LTT & 0 (0.00) & 0.000 (0.000) & -- & -- & 0.00 (0.00)\\
\hspace{1em} & HP-Pointwise & \textbf{2} (1.24) & 0.016 (0.013) & -- & -- & 0.07 (0.05)\\
\hspace{1em} & HP-Uniform & 0 (0.28) & 0.001 (0.001) & -- & -- & 0.01 \vphantom{1} (0.02)\\
\hspace{1em}0.50 & HP-LTT & 0 (0.00) & 0.000 (0.000) & -- & -- & 0.00 (0.00)\\
\hspace{1em} & HP-Pointwise & 5 (2.46) & 0.032 (0.016) & \textcolor{red}{0.199 (0.043)} & \textcolor{red}{1.00} & 0.16 (0.07)\\
\hspace{1em} & HP-Uniform & 0 (0.28) & 0.001 (0.001) & -- & -- & 0.01 (0.02)\\
\bottomrule
\end{tabular}

%% file: tables/exceedance_data_0_ntrain_5000_ncal_1000_xgb_aipcw.txt
\begin{tabular}[t]{lrrrrr}
\toprule
Method & Yield & FDP & Event rate & $P(r>\alpha)$ & $P(|S|>0)$\\
\midrule
\addlinespace[0.3em]
\multicolumn{6}{l}{\textbf{Horizon: 2 months}}\\
\hspace{1em}Oracle & 840 (3.68) & 0.099 (0.002) & 0.099 (0.002) & 0.00 & 1.00 (0.00)\\
\hspace{1em}Model & 856 (5.12) & 0.113 (0.003) & \textcolor{red}{0.113 (0.003)} & 0.98 & 1.00 (0.00)\\
\hspace{1em}FDR-Conformal & 651 (18.03) & 0.099 (0.002) & \textcolor{red}{0.100 (0.002)} & 0.57 & 0.99 (0.01)\\
\hspace{1em}HP-Pointwise & 406 (45.51) & 0.082 (0.006) & 0.089 (0.004) & \textcolor{red}{0.18} & 0.92 (0.05)\\
\hspace{1em}HP-Uniform & 128 (43.04) & 0.032 (0.008) & 0.068 (0.010) & 0.05 & 0.47 (0.10)\\
\hspace{1em}HP-LTT & \textbf{237} (48.16) & 0.046 (0.009) & 0.084 (0.004) & 0.01 & 0.54 (0.10)\\
\addlinespace[0.3em]
\multicolumn{6}{l}{\textbf{Horizon: 3 months}}\\
\hspace{1em}Oracle & 538 (4.16) & 0.099 (0.003) & 0.099 (0.003) & 0.35 & 1.00 (0.00)\\
\hspace{1em}Model & 701 (6.28) & 0.148 (0.003) & \textcolor{red}{0.148 (0.003)} & 1.00 & 1.00 (0.00)\\
\hspace{1em}FDR-Conformal & 179 (14.82) & 0.080 (0.005) & \textcolor{red}{0.107 (0.003)} & 1.00 & 0.75 (0.04)\\
\hspace{1em}HP-Pointwise & 56 (16.70) & 0.048 (0.011) & 0.095 (0.010) & \textcolor{red}{0.55} & 0.51 (0.10)\\
\hspace{1em}HP-Uniform & 8 (4.73) & 0.016 (0.007) & 0.071 (0.016) & 0.00 & 0.22 (0.08)\\
\hspace{1em}HP-LTT & \textbf{11} (8.79) & 0.006 (0.005) & -- & -- & 0.06 (0.05)\\
\addlinespace[0.3em]
\multicolumn{6}{l}{\textbf{Horizon: 6 months}}\\
\hspace{1em}Oracle & 25 (1.23) & 0.109 (0.013) & 0.109 (0.013) & 0.79 & 1.00 (0.00)\\
\hspace{1em}Model & 432 (6.44) & 0.236 (0.004) & \textcolor{red}{0.236 (0.004)} & 1.00 & 1.00 (0.00)\\
\hspace{1em}FDR-Conformal & 4 (1.45) & 0.024 (0.007) & \textcolor{red}{0.198 (0.019)} & 1.00 & 0.12 (0.03)\\
\hspace{1em}HP-Pointwise & \textbf{2} (1.95) & 0.006 (0.007) & -- & -- & 0.03 (0.03)\\
\hspace{1em}HP-Uniform & 1 (0.86) & 0.005 (0.006) & -- & -- & 0.02 (0.03)\\
\hspace{1em}HP-LTT & 1 (1.30) & 0.002 (0.004) & -- & -- & 0.01 (0.02)\\
\addlinespace[0.3em]
\multicolumn{6}{l}{\textbf{Horizon: 9 months}}\\
\hspace{1em}Oracle & 0 (0.00) & 0.000 (0.000) & -- & -- & 0.00 (0.00)\\
\hspace{1em}Model & 301 (5.71) & 0.319 (0.006) & \textcolor{red}{0.319 (0.006)} & 1.00 & 1.00 (0.00)\\
\hspace{1em}FDR-Conformal & \textbf{1} (0.62) & 0.019 (0.008) & -- & -- & 0.06 (0.02)\\
\hspace{1em}HP-Pointwise & 0 (0.00) & 0.000 (0.000) & -- & -- & 0.00 (0.00)\\
\hspace{1em}HP-Uniform & 0 (0.00) & 0.000 (0.000) & -- & -- & 0.00 (0.00)\\
\hspace{1em}HP-LTT & 0 (0.00) & 0.000 (0.000) & -- & -- & 0.00 (0.00)\\
\bottomrule
\end{tabular}

%% file: tables/exceedance_data_0_ntrain_5000_ncal_1000_cox_aipcw.txt
\begin{tabular}[t]{lrrrrr}
\toprule
Method & Yield & FDP & Event rate & $P(r>\alpha)$ & $P(|S|>0)$\\
\midrule
\addlinespace[0.3em]
\multicolumn{6}{l}{\textbf{Horizon: 2 months}}\\
\hspace{1em}Oracle & 840 (3.68) & 0.099 (0.002) & 0.099 (0.002) & 0.00 & 1.00 (0.00)\\
\hspace{1em}Model & 607 (16.51) & 0.081 (0.003) & 0.081 (0.003) & 0.00 & 1.00 (0.00)\\
\hspace{1em}FDR-Conformal & \textbf{800} (9.51) & 0.098 (0.001) & 0.098 (0.001) & 0.44 & 1.00 (0.00)\\
\hspace{1em}HP-Pointwise & 663 (27.26) & 0.086 (0.003) & 0.086 (0.003) & 0.10 & 1.00 (0.00)\\
\hspace{1em}HP-Uniform & 483 (56.93) & 0.066 (0.007) & 0.077 (0.005) & 0.07 & 0.86 (0.07)\\
\hspace{1em}HP-LTT & 520 (53.79) & 0.070 (0.007) & 0.082 (0.004) & 0.09 & 0.85 (0.07)\\
\addlinespace[0.3em]
\multicolumn{6}{l}{\textbf{Horizon: 3 months}}\\
\hspace{1em}Oracle & 538 (4.16) & 0.099 (0.003) & 0.099 (0.003) & 0.35 & 1.00 (0.00)\\
\hspace{1em}Model & 63 (4.66) & 0.072 (0.007) & 0.072 (0.007) & 0.00 & 1.00 (0.00)\\
\hspace{1em}FDR-Conformal & \textbf{351} (19.19) & 0.092 (0.003) & 0.096 (0.003) & 0.58 & 0.96 (0.02)\\
\hspace{1em}HP-Pointwise & 210 (32.60) & 0.073 (0.008) & 0.088 (0.005) & \textcolor{red}{0.28} & 0.82 (0.08)\\
\hspace{1em}HP-Uniform & 36 (20.32) & 0.024 (0.009) & 0.072 (0.017) & \textcolor{red}{0.14} & 0.33 (0.09)\\
\hspace{1em}HP-LTT & 36 (16.19) & 0.030 (0.009) & 0.085 (0.014) & 0.00 & 0.35 (0.10)\\
\addlinespace[0.3em]
\multicolumn{6}{l}{\textbf{Horizon: 6 months}}\\
\hspace{1em}Oracle & 25 (1.23) & 0.109 (0.013) & 0.109 (0.013) & 0.79 & 1.00 (0.00)\\
\hspace{1em}Model & 1 (0.28) & 0.088 (0.043) & \textcolor{red}{0.192 (0.084)} & 0.76 & 0.46 (0.10)\\
\hspace{1em}FDR-Conformal & 9 (2.56) & 0.027 (0.007) & \textcolor{red}{0.168 (0.016)} & 1.00 & 0.16 (0.04)\\
\hspace{1em}HP-Pointwise & 6 (3.52) & 0.025 (0.013) & \textcolor{red}{0.169 (0.032)} & \textcolor{red}{1.00} & 0.15 (0.07)\\
\hspace{1em}HP-Uniform & \textbf{2} (1.30) & 0.010 (0.008) & -- & -- & 0.06 (0.05)\\
\hspace{1em}HP-LTT & 0 (0.00) & 0.000 (0.000) & -- & -- & 0.00 \vphantom{1} (0.00)\\
\addlinespace[0.3em]
\multicolumn{6}{l}{\textbf{Horizon: 9 months}}\\
\hspace{1em}Oracle & 0 (0.00) & 0.000 (0.000) & -- & -- & 0.00 (0.00)\\
\hspace{1em}Model & 1 (0.14) & 0.215 (0.078) & \textcolor{red}{0.500 (0.141)} & 1.00 & 0.43 (0.10)\\
\hspace{1em}FDR-Conformal & \textbf{1} (0.58) & 0.008 (0.006) & -- & -- & 0.02 (0.01)\\
\hspace{1em}HP-Pointwise & 1 (1.04) & 0.003 (0.007) & -- & -- & 0.01 (0.02)\\
\hspace{1em}HP-Uniform & 0 (0.00) & 0.000 (0.000) & -- & -- & 0.00 (0.00)\\
\hspace{1em}HP-LTT & 0 (0.00) & 0.000 (0.000) & -- & -- & 0.00 (0.00)\\
\bottomrule
\end{tabular}

%% file: tables/exceedance_data_0_ntrain_5000_ncal_100_grf_aipcw.txt
\begin{tabular}[t]{lrrrrr}
\toprule
Method & Yield & FDP & Event rate & $P(r>\alpha)$ & $P(|S|>0)$\\
\midrule
\addlinespace[0.3em]
\multicolumn{6}{l}{\textbf{Horizon: 2 months}}\\
\hspace{1em}Oracle & 843 (3.38) & 0.098 (0.002) & 0.098 (0.002) & 0.24 & 1.00 (0.00)\\
\hspace{1em}Model & \textbf{754} (9.91) & 0.090 (0.003) & 0.090 (0.003) & 0.10 & 1.00 (0.00)\\
\hspace{1em}FDR-Conformal & 603 (34.61) & 0.076 (0.004) & 0.094 (0.003) & 0.38 & 0.81 (0.04)\\
\hspace{1em}HP-Pointwise & 425 (68.50) & 0.055 (0.008) & 0.078 (0.006) & \textcolor{red}{0.19} & 0.70 (0.09)\\
\hspace{1em}HP-Uniform & 267 (64.80) & 0.039 (0.008) & 0.073 (0.008) & \textcolor{red}{0.16} & 0.53 (0.10)\\
\hspace{1em}HP-LTT & 182 (59.63) & 0.022 (0.007) & 0.077 (0.008) & \textcolor{red}{0.12} & 0.29 (0.09)\\
\addlinespace[0.3em]
\multicolumn{6}{l}{\textbf{Horizon: 3 months}}\\
\hspace{1em}Oracle & 544 (4.10) & 0.097 (0.002) & 0.097 (0.002) & 0.00 & 1.00 (0.00)\\
\hspace{1em}Model & \textbf{304} (15.14) & 0.083 (0.004) & 0.083 (0.004) & 0.03 & 1.00 (0.00)\\
\hspace{1em}FDR-Conformal & 317 (32.39) & 0.064 (0.006) & \textcolor{red}{0.114 (0.004)} & 0.76 & 0.56 (0.05)\\
\hspace{1em}HP-Pointwise & 202 (51.33) & 0.045 (0.010) & 0.097 (0.007) & \textcolor{red}{0.49} & 0.46 (0.10)\\
\hspace{1em}HP-Uniform & 133 (39.74) & 0.035 (0.009) & 0.090 (0.006) & \textcolor{red}{0.30} & 0.39 (0.10)\\
\hspace{1em}HP-LTT & 44 (24.34) & 0.013 (0.006) & 0.090 (0.010) & \textcolor{red}{0.26} & 0.14 (0.07)\\
\addlinespace[0.3em]
\multicolumn{6}{l}{\textbf{Horizon: 6 months}}\\
\hspace{1em}Oracle & 26 (1.19) & 0.100 (0.012) & 0.100 (0.012) & 0.50 & 1.00 (0.00)\\
\hspace{1em}Model & 0 (0.00) & 0.000 (0.000) & -- & -- & 0.00 \vphantom{1} (0.00)\\
\hspace{1em}FDR-Conformal & 7 (3.50) & 0.006 (0.003) & -- & -- & 0.04 (0.02)\\
\hspace{1em}HP-Pointwise & 27 (16.19) & 0.019 (0.011) & \textcolor{red}{0.171 (0.011)} & \textcolor{red}{1.00} & 0.11 (0.06)\\
\hspace{1em}HP-Uniform & \textbf{14} (10.78) & 0.012 (0.009) & -- & -- & 0.07 (0.05)\\
\hspace{1em}HP-LTT & 0 (0.00) & 0.000 (0.000) & -- & -- & 0.00 \vphantom{1} (0.00)\\
\addlinespace[0.3em]
\multicolumn{6}{l}{\textbf{Horizon: 9 months}}\\
\hspace{1em}Oracle & 0 (0.00) & 0.000 (0.000) & -- & -- & 0.00 (0.00)\\
\hspace{1em}Model & 0 (0.00) & 0.000 (0.000) & -- & -- & 0.00 (0.00)\\
\hspace{1em}FDR-Conformal & 1 (1.21) & 0.002 (0.002) & -- & -- & 0.01 (0.01)\\
\hspace{1em}HP-Pointwise & \textbf{6} (6.76) & 0.008 (0.009) & -- & -- & 0.03 (0.03)\\
\hspace{1em}HP-Uniform & 2 (3.02) & 0.003 (0.005) & -- & -- & 0.01 (0.02)\\
\hspace{1em}HP-LTT & 0 (0.00) & 0.000 (0.000) & -- & -- & 0.00 (0.00)\\
\bottomrule
\end{tabular}

%% file: tables/data_1_ntrain_5000_ncal_1000_cox_aipcw.txt
\begin{tabular}[t]{lrrrrrr}
\toprule
Method & Yield & FDP (LB) & FDP (UB) & Ev.rate (LB) & Ev.rate (UB) & P($|S|>0$)\\
\midrule
\addlinespace[0.3em]
\multicolumn{7}{l}{\textbf{Horizon: 2 months}}\\
\hspace{1em}Model & 821 (4.09) & 0.08 (0.00) & 0.11 (0.00) & 0.08 (0.00) & 0.11 (0.00) & 1.00 (0.00)\\
\hspace{1em}FDR-Conformal & \textbf{916} (3.32) & 0.10 (0.00) & 0.13 (0.00) & 0.10 (0.00) & 0.13 (0.00) & 1.00 (0.00)\\
\hspace{1em}HP-Pointwise & 853 (9.46) & 0.09 (0.00) & 0.11 (0.00) & 0.09 (0.00) & 0.11 (0.00) & 1.00 (0.00)\\
\hspace{1em}HP-Uniform & 813 (12.42) & 0.08 (0.00) & 0.11 (0.00) & 0.08 (0.00) & 0.11 (0.00) & 1.00 (0.00)\\
\hspace{1em}HP-LTT & 832 (10.85) & 0.08 (0.00) & 0.11 (0.00) & 0.08 (0.00) & 0.11 (0.00) & 1.00 (0.00)\\
\addlinespace[0.3em]
\multicolumn{7}{l}{\textbf{Horizon: 3 months}}\\
\hspace{1em}Model & 389 (3.86) & 0.05 (0.00) & 0.10 (0.00) & 0.05 (0.00) & 0.10 (0.00) & 1.00 (0.00)\\
\hspace{1em}FDR-Conformal & \textbf{684} (5.59) & 0.10 (0.00) & 0.14 (0.00) & 0.10 (0.00) & 0.14 (0.00) & 1.00 (0.00)\\
\hspace{1em}HP-Pointwise & 612 (11.98) & 0.09 (0.00) & 0.12 (0.00) & 0.09 (0.00) & 0.12 (0.00) & 1.00 (0.00)\\
\hspace{1em}HP-Uniform & 512 (22.28) & 0.07 (0.00) & 0.11 (0.00) & 0.07 (0.00) & 0.11 (0.00) & 1.00 (0.00)\\
\hspace{1em}HP-LTT & 568 (16.13) & 0.08 (0.00) & 0.12 (0.00) & 0.08 (0.00) & 0.12 (0.00) & 1.00 (0.00)\\
\addlinespace[0.3em]
\multicolumn{7}{l}{\textbf{Horizon: 6 months}}\\
\hspace{1em}Model & 32 (1.62) & 0.04 (0.01) & 0.12 (0.01) & 0.04 (0.01) & 0.12 (0.01) & 1.00 (0.00)\\
\hspace{1em}FDR-Conformal & \textbf{217} (9.29) & 0.08 (0.00) & 0.14 (0.00) & 0.09 (0.00) & 0.16 (0.00) & 0.92 (0.03)\\
\hspace{1em}HP-Pointwise & 146 (18.86) & 0.07 (0.01) & 0.13 (0.01) & 0.07 (0.01) & 0.14 (0.01) & 0.89 (0.06)\\
\hspace{1em}HP-Uniform & 15 (4.50) & 0.02 (0.01) & 0.05 (0.01) & 0.04 (0.01) & 0.11 (0.02) & 0.40 (0.10)\\
\hspace{1em}HP-LTT & 7 (5.45) & 0.01 (0.00) & 0.01 (0.01) & 0.07 (0.02) & 0.14 (0.03) & 0.09 (0.06)\\
\addlinespace[0.3em]
\multicolumn{7}{l}{\textbf{Horizon: 9 months}}\\
\hspace{1em}Model & 6 (0.76) & 0.04 (0.02) & 0.13 (0.03) & 0.04 (0.02) & 0.15 (0.03) & 0.91 (0.06)\\
\hspace{1em}FDR-Conformal & \textbf{36} (4.13) & 0.06 (0.01) & 0.12 (0.01) & 0.10 (0.01) & 0.21 (0.01) & 0.56 (0.05)\\
\hspace{1em}HP-Pointwise & 27 (6.62) & 0.05 (0.01) & 0.10 (0.02) & 0.10 (0.01) & 0.20 (0.02) & 0.50 (0.10)\\
\hspace{1em}HP-Uniform & 4 (2.20) & 0.01 (0.01) & 0.03 (0.01) & 0.10 (0.03) & 0.19 (0.05) & 0.14 (0.07)\\
\hspace{1em}HP-LTT & 0 (0.00) & 0.00 (0.00) & 0.00 (0.00) & -- & -- & 0.00 (0.00)\\
\bottomrule
\end{tabular}

%% file: tables/data_1_ntrain_5000_ncal_1000_grf_aipcw.txt
\begin{tabular}[t]{lrrrrrr}
\toprule
Method & Yield & FDP (LB) & FDP (UB) & Ev.rate (LB) & Ev.rate (UB) & P($|S|>0$)\\
\midrule
\addlinespace[0.3em]
\multicolumn{7}{l}{\textbf{Horizon: 2 months}}\\
\hspace{1em}Model & 835 (4.26) & 0.08 (0.00) & 0.11 (0.00) & 0.08 (0.00) & 0.11 (0.00) & 1.00 (0.00)\\
\hspace{1em}FDR-Conformal & \textbf{864} (5.83) & 0.09 (0.00) & 0.12 (0.00) & 0.09 (0.00) & 0.12 (0.00) & 1.00 (0.00)\\
\hspace{1em}HP-Pointwise & 839 (12.05) & 0.09 (0.00) & 0.11 (0.00) & 0.09 (0.00) & 0.11 (0.00) & 1.00 (0.00)\\
\hspace{1em}HP-Uniform & 789 (16.32) & 0.08 (0.00) & 0.11 (0.00) & 0.08 (0.00) & 0.11 (0.00) & 1.00 (0.00)\\
\hspace{1em}HP-LTT & 806 (15.10) & 0.08 (0.00) & 0.11 (0.00) & 0.08 (0.00) & 0.11 (0.00) & 1.00 (0.00)\\
\addlinespace[0.3em]
\multicolumn{7}{l}{\textbf{Horizon: 3 months}}\\
\hspace{1em}Model & 508 (4.67) & 0.07 (0.00) & 0.11 (0.00) & 0.07 (0.00) & 0.11 (0.00) & 1.00 (0.00)\\
\hspace{1em}FDR-Conformal & \textbf{605} (8.28) & 0.09 (0.00) & 0.13 (0.00) & 0.09 (0.00) & 0.13 (0.00) & 1.00 (0.00)\\
\hspace{1em}HP-Pointwise & 559 (15.76) & 0.08 (0.00) & 0.12 (0.00) & 0.08 (0.00) & 0.12 (0.00) & 1.00 (0.00)\\
\hspace{1em}HP-Uniform & 428 (28.11) & 0.06 (0.00) & 0.10 (0.01) & 0.07 (0.00) & 0.10 (0.01) & 0.99 (0.02)\\
\hspace{1em}HP-LTT & 519 (16.31) & 0.08 (0.00) & 0.11 (0.00) & 0.08 (0.00) & 0.11 (0.00) & 1.00 (0.00)\\
\addlinespace[0.3em]
\multicolumn{7}{l}{\textbf{Horizon: 6 months}}\\
\hspace{1em}Model & 13 (1.26) & 0.03 (0.01) & 0.07 (0.02) & 0.03 (0.01) & 0.07 (0.02) & 1.00 (0.00)\\
\hspace{1em}FDR-Conformal & 85 (8.95) & 0.05 (0.00) & 0.09 (0.01) & 0.08 (0.00) & 0.15 (0.00) & 0.61 (0.05)\\
\hspace{1em}HP-Pointwise & \textbf{128} (15.27) & 0.06 (0.01) & 0.12 (0.01) & 0.07 (0.01) & 0.13 (0.01) & 0.94 (0.05)\\
\hspace{1em}HP-Uniform & 16 (4.44) & 0.02 (0.01) & 0.05 (0.01) & 0.03 (0.01) & 0.10 (0.02) & 0.49 (0.10)\\
\hspace{1em}HP-LTT & 0 (0.00) & 0.00 (0.00) & 0.00 (0.00) & -- & -- & 0.00 \vphantom{1} (0.00)\\
\addlinespace[0.3em]
\multicolumn{7}{l}{\textbf{Horizon: 9 months}}\\
\hspace{1em}Model & 0 (0.00) & 0.00 (0.00) & 0.00 (0.00) & -- & -- & 0.00 (0.00)\\
\hspace{1em}FDR-Conformal & 4 (1.72) & 0.01 (0.00) & 0.02 (0.01) & -- & -- & 0.06 (0.02)\\
\hspace{1em}HP-Pointwise & \textbf{23} (7.11) & 0.04 (0.01) & 0.07 (0.02) & 0.09 (0.01) & 0.19 (0.02) & 0.39 (0.10)\\
\hspace{1em}HP-Uniform & 3 (1.92) & 0.01 (0.01) & 0.02 (0.01) & 0.06 (0.03) & 0.16 (0.04) & 0.11 (0.06)\\
\hspace{1em}HP-LTT & 0 (0.00) & 0.00 (0.00) & 0.00 (0.00) & -- & -- & 0.00 (0.00)\\
\bottomrule
\end{tabular}

%% file: tables/data_1_ntrain_5000_ncal_1000_xgb.txt
\begin{tabular}[t]{lrrrrrr}
\toprule
Method & Yield & FDP (LB) & FDP (UB) & Ev.rate (LB) & Ev.rate (UB) & P($|S|>0$)\\
\midrule
\addlinespace[0.3em]
\multicolumn{7}{l}{\textbf{Horizon: 2 months}}\\
\hspace{1em}Model & 833 (2.99) & 0.09 (0.00) & 0.12 (0.00) & 0.09 (0.00) & 0.12 (0.00) & 1.00 (0.00)\\
\hspace{1em}FDR-Conformal & \textbf{874} (10.11) & 0.10 (0.00) & 0.12 (0.00) & 0.10 (0.00) & 0.12 (0.00) & 1.00 (0.00)\\
\hspace{1em}HP-Pointwise & 786 (13.85) & 0.08 (0.00) & 0.11 (0.00) & 0.08 (0.00) & 0.11 (0.00) & 1.00 (0.00)\\
\hspace{1em}HP-Uniform & 709 (21.01) & 0.08 (0.00) & 0.10 (0.00) & 0.08 (0.00) & 0.10 (0.00) & 1.00 (0.00)\\
\hspace{1em}HP-LTT & 739 (18.59) & 0.08 (0.00) & 0.11 (0.00) & 0.08 (0.00) & 0.11 (0.00) & 1.00 (0.00)\\
\addlinespace[0.3em]
\multicolumn{7}{l}{\textbf{Horizon: 3 months}}\\
\hspace{1em}Model & 711 (3.30) & \textcolor{red}{0.11 (0.00)} & 0.15 (0.00) & \textcolor{red}{0.11 (0.00)} & 0.15 (0.00) & 1.00 (0.00)\\
\hspace{1em}FDR-Conformal & \textbf{568} (18.83) & 0.10 (0.00) & 0.14 (0.00) & 0.10 (0.00) & 0.14 (0.00) & 1.00 (0.00)\\
\hspace{1em}HP-Pointwise & 447 (19.67) & 0.08 (0.00) & 0.12 (0.00) & 0.08 (0.00) & 0.12 (0.00) & 1.00 (0.00)\\
\hspace{1em}HP-Uniform & 184 (35.62) & 0.04 (0.01) & 0.07 (0.01) & 0.06 (0.01) & 0.11 (0.01) & 0.63 (0.10)\\
\hspace{1em}HP-LTT & 344 (32.45) & 0.07 (0.01) & 0.10 (0.01) & 0.08 (0.00) & 0.12 (0.00) & 0.87 (0.07)\\
\addlinespace[0.3em]
\multicolumn{7}{l}{\textbf{Horizon: 6 months}}\\
\hspace{1em}Model & 501 (3.54) & \textcolor{red}{0.17 (0.00)} & 0.24 (0.00) & \textcolor{red}{0.17 (0.00)} & 0.24 (0.00) & 1.00 (0.00)\\
\hspace{1em}FDR-Conformal & \textbf{139} (16.85) & 0.08 (0.01) & 0.15 (0.01) & 0.10 (0.01) & 0.18 (0.01) & 0.85 (0.07)\\
\hspace{1em}HP-Pointwise & 76 (11.96) & 0.06 (0.01) & 0.12 (0.02) & 0.08 (0.01) & 0.16 (0.01) & 0.75 (0.09)\\
\hspace{1em}HP-Uniform & 3 (2.96) & 0.00 (0.00) & 0.00 (0.00) & -- & -- & 0.03 (0.03)\\
\hspace{1em}HP-LTT & 17 (9.59) & 0.01 (0.01) & 0.02 (0.01) & 0.09 (0.01) & 0.17 (0.02) & 0.12 (0.07)\\
\addlinespace[0.3em]
\multicolumn{7}{l}{\textbf{Horizon: 9 months}}\\
\hspace{1em}Model & 392 (3.50) & \textcolor{red}{0.23 (0.00)} & 0.34 (0.00) & \textcolor{red}{0.23 (0.00)} & 0.34 (0.00) & 1.00 (0.00)\\
\hspace{1em}FDR-Conformal & \textbf{18} (6.50) & 0.04 (0.01) & 0.08 (0.02) & 0.12 (0.02) & 0.25 (0.02) & 0.33 (0.09)\\
\hspace{1em}HP-Pointwise & 5 (2.85) & 0.02 (0.01) & 0.04 (0.02) & \textcolor{red}{0.13 (0.03)} & 0.25 (0.03) & 0.15 (0.07)\\
\hspace{1em}HP-Uniform & 0 (0.00) & 0.00 (0.00) & 0.00 (0.00) & -- & -- & 0.00 (0.00)\\
\hspace{1em}HP-LTT & 0 (0.00) & 0.00 (0.00) & 0.00 (0.00) & -- & -- & 0.00 (0.00)\\
\bottomrule
\end{tabular}